\documentclass[superscriptaddress,twocolumn,amsmath,amssymb,prd,floatfix,nofootinbib]{revtex4-2}

\usepackage[dvipdfmx]{graphicx}
\usepackage{dcolumn}
\usepackage[varg]{txfonts}
\usepackage{bm}
\usepackage{multirow}

\usepackage{color}

\newcommand*\Laplace{\mathop{}\!\mathbin\bigtriangleup}

\begin{document}

\title{Integrated perturbation theory for cosmological tensor
  fields. I. Basic formulation
}

\author{Takahiko Matsubara} \email{tmats@post.kek.jp}
\affiliation{%
  Institute of Particle and Nuclear Studies, High Energy
  Accelerator Research Organization (KEK), Oho 1-1, Tsukuba 305-0801,
  Japan}%
\affiliation{%
  The Graduate Institute for Advanced Studies, SOKENDAI,
  Tsukuba 305-0801, Japan}%

\date{\today}

\begin{abstract}
  In order to extract maximal information about cosmology from the
  large-scale structure of the Universe, one needs to use every bit of
  signal that can be observed. Beyond the spatial distributions of
  astronomical objects, the spatial correlations of tensor fields,
  such as galaxy spins and shapes, are ones of promising sources that
  can be accessed in the era of large surveys in the near future. The
  perturbation theory is a powerful tool to analytically describe the
  behaviors and evolutions of correlation statistics on large scales
  for a given cosmology. In this paper, we formulate a nonlinear
  perturbation theory of tensor fields in general, based on the
  formulation of integrated perturbation theory for the scalar-valued
  bias, generalizing it to include the tensor-valued bias. To take
  advantage of rotational symmetry, the formalism is constructed on
  the basis of the irreducible decomposition of tensors, identifying
  physical variables which are invariant under the rotation of the
  coordinates system.
\end{abstract}


\maketitle


\section{Introduction
\label{sec:Introduction}}

The large-scale structure of the Universe is an inexhaustible fountain
of cosmological information and has been playing important roles
in cosmology. The various patterns in the spatial distribution of
galaxies have emerged from initial density fluctuations of the
Universe, whose origin is believed to be generated by quantum
fluctuations in the inflationary phase of the very early Universe
\cite{Lyth:2009zz}. Considerable information on the evolution of the
Universe is also contained in the large-scale structure. For example,
the spatial clustering of galaxies is a promising probe to reveal the
nature of dark energy \cite{SDSS:2005xqv} and the initial condition in
the early Universe such as the primordial spectrum and non-Gaussianity
etc.~\cite{Dalal:2007cu}. While the large-scale structure is used as a
probe of the density field in the relatively late Universe, the
distribution of observable objects, such as galaxies, quasars, and
21\,cm lines, etc., are biased tracers of the underlying matter
distribution of the Universe. The difference between distributions of
galaxies and matter is referred to as ``galaxy bias''
\cite{Desjacques:2016bnm}. The phenomena of biasing are not restricted
to the galaxies, but are unavoidable when we use astronomical objects
as tracers of underlying matter distribution.

The number density of galaxies is the primary target to probe the
properties of large-scale structure. Theoretical predictions for the
spatial distribution of galaxies is investigated by various methods,
including numerical simulations, analytic modeling, and the
perturbation theory. On one hand, the dynamical evolutions of the mass
density fields on relatively large scales are analytically described
by the perturbation theory
\cite{Hunter1964,Tomita1967,Peebles1980,Bernardeau:2001qr}, which is
valid as long as the density fluctuations are small and are in the
quasilinear regime. On the other hand, the dynamical structure
formation on relatively small scales, including the formation of
galaxies and other astronomical objects, is not analytically tractable
because of the full nonlinearity of the problem. One should resort to
numerical simulations and nonlinear modeling to describe and
understand the small-scale structure formation. For example, the halo
model \cite{Mo:1995cs,Mo:1996cn,Cooray:2002dia} statistically predicts
spatial distributions of halos depending on the initial conditions,
using simple assumptions based on the Press-Schechter theory
\cite{Press:1973iz} and its extensions \cite{Bond:1990iw,Sheth:1999mn}
of the nonlinear structure formation.

The biasing of tracers affects not only spatial distributions on small
scales, but also those on large scales. However, the biasing effects
are not so complicated on large scales as those on small scales. For
example, the density contrast $\delta_\mathrm{g}$ of biased tracers on
sufficiently large scales, where the dynamics are described by linear
theory, is proportional to the linear density contrast
$\delta_\mathrm{L}$, and we have a simple relation
$\delta_\mathrm{g} = b\,\delta_\mathrm{L}$, where a proportional
constant $b$ is called the linear bias parameter
\cite{Kaiser:1984sw,Bardeen:1985tr}. While nonlinear dynamics
complicate the situation, the nonlinear perturbation theory to
describe the spatial clustering of biased tracers has been also
developed \cite{Bernardeau:2001qr,Desjacques:2016bnm}. Since the
perturbation theory cannot describe the nonlinear structure formation
which takes place on small scales, it is inevitable to introduce some
unknown parameters or functions whose values strongly depend on the
small-scale nonlinear physics which cannot be predicted from the first
principle. These nuisance parameters are called bias parameters or
bias functions in the nonlinear perturbation theory. When a model of
bias is defined in some way, one can calculate the gravitational
evolution of the spatial distribution of biased tracers by applying
the nonlinear perturbation theory.

There are several different ways of characterizing the bias model in
the nonlinear perturbation theory \cite{Desjacques:2016bnm}. Among
others, the integrated perturbation theory (iPT)
\cite{Matsubara:2011ck,Matsubara:2012nc,Matsubara:2013ofa,Matsubara:2016wth},
which is based on the Lagrangian perturbation theory
\cite{Buchert:1989xx,Moutarde:1991evx} and orthogonal decomposition of
the bias relation \cite{Matsubara:1995wd}, provides a method to
generically include any model of bias formulated in Lagrangian space,
irrespective of whether the bias is local or nonlocal. In this
formalism, the concept of the renormalized bias functions
\cite{Matsubara:2011ck} is introduced. Dynamical evolutions on large
scales which are described by perturbation theory are separated from
complicated nonlinear processes of structure formation on small
scales. The latter complications are all encoded in the renormalized
bias functions in the formalism of iPT. The renormalized bias
functions are given for any models of bias, whether they are local or
nonlocal in general. For example, the halo bias is one of the typical
models of Lagrangian bias, and the renormalized bias functions are
uniquely determined and calculated from the model
\cite{Matsubara:2012nc}.

The number density of biased objects is not the only quantity we can
observe. For example, the position and shape of galaxies are
simultaneously observed in imaging surveys of galaxies, which are
essential in observations of weak lensing fields to probe the nature
of dark matter and dark energy \cite{Bartelmann:1999yn}. Before the
lensing effects, the galaxy shapes are more or less correlated to the
mass density field through, e.g., gravitational tidal fields and other
environmental structures in the underlying density field. The shapes
of galaxies, i.e., sizes and intrinsic alignments
\cite{Catelan:2000vm,Okumura:2008du,Joachimi:2015mma,Kogai:2018nse,Okumura:2019ned,Okumura:2019ozd,Okumura:2020hhr,Taruya:2020tdi,Okumura:2021xgc}
are expected to be promising probes of cosmological information
\cite{Chisari:2013dda} in the near future when unprecedentedly large
imaging surveys are going to take place. For example, the shape
statistics of galaxies offer a new probe of particular features in
primordial non-Gaussianity generated during the inflation in the
presence of higher-spin fields \cite{Arkani-Hamed:2015bza}, whose
features are dubbed ``cosmological collider physics.''

Motivated by recent progress in observational techniques of imaging
surveys, analytical modelings of galaxy shape statistics by the
nonlinear perturbation theory have been actively considered in recent
years
\cite{Blazek:2015lfa,Blazek:2017wbz,Schmitz:2018rfw,Vlah:2019byq,Vlah:2020ovg,Taruya:2021jhg}.
Among others, the authors of Ref.~\cite{Vlah:2019byq} developed a
formalism based on spherical tensors to describe the galaxy shapes,
with which rotation and parity symmetries in the shape statistics are
respected. They use an approach called the effective field theory of
the large-scale structure (EFTofLSS) with the Eulerian perturbation
theory, and galaxy biasing is modeled by introducing a set of
semilocal terms that are generally allowed by conceivable symmetries
of the problem and the coefficients of these terms are free parameters
in the theory.

The galaxy shape statistics are usually characterized by the
second-order moments of galaxy images in the two-dimensional sky,
which correspond to the projection of the three-dimensional
second-order moments of galaxy shapes to the two-dimensional sky. In
principle, the galaxy images contain information about other moments
higher than the second order. The higher-order moments in galaxy
images turn out to be possible probes for the higher-spin fields in
the context of cosmological collider physics through angle-dependent
primordial non-Gaussianity. In particular, Ref.~\cite{Kogai:2020vzz}
showed that the shape statistics of the order-$s$ moment turn out to
be a probe for spin-$s$ fields.

In this paper, inspired by the above developments, we consider a
generalization of the iPT formalism to predict correlations of tensor
fields of rank-$l$ in general with the perturbation theory. In order
to have rotational symmetry apparent, we decompose the tensor fields
into irreducible tensors on the basis of spherical tensors, just
similarly in the pioneering work of
Refs.~\cite{Vlah:2019byq,Vlah:2020ovg}. However, unlike the previous
work, we do not fix the coordinates system to the directions of
wave vectors of Fourier modes, so as to make the theory fully covariant
with respect to the rotational symmetry in three-dimensional space. It
turns out that the generalizing iPT formalism on the basis of
irreducible tensors results in an elegant way of describing the
perturbation theory only with rotationally invariant quantities in the
presence of tensor-valued bias. Various methods developed for the
theory of angular momentum in quantum mechanics, such as the
$3nj$-symbols and their orthogonality relations, sum rules, the
Wigner-Eckart theorem, and so on, are effectively utilized in the
course of calculations for predicting statistical quantities of tensor
fields.

This paper defines a basic formalism of the tensor generalization of
the methods of iPT. Several applications of the fully general
formalism to relatively simple examples of calculating statistics are
also illustrated, i.e., the lowest-order predictions of the power
spectrum of tensor fields in real space and in redshift space,
lowest-order effects of primordial non-Gaussianity in the power
spectrum, and tree-level predictions of the bispectrum of tensor
fields in real space. The main purpose of this paper is to show the
fundamental formulation of tensor-valued iPT, and to give useful
methods and formulas for future applications to calculate statistics
of concrete tensor fields, such as galaxy spins, shapes, etc. Further
developments for calculating loop corrections in the higher-order
perturbation theory are described in a separate paper in the series,
Paper~II \cite{PaperII}, which follows this paper. Since the
three-dimensional tensor fields themselves are not directly
observable, instead, the projected two-dimensional tensors on the sky
are observable. The projection effects for the tensor components are
studied in another separate paper in the series, Paper~III
\cite{PaperIII}. In Papers~I--III, the distant-observer approximation
is assumed, and this approximation is mostly good as long as the
galaxies are located at greater distances from the observer than the
scales of clustering we are interested in. Beyond the distant-observer
approximation, full-sky or wide-angle effects are studied in yet
another separate paper in the series, Paper~IV \cite{PaperIV}.

This paper is organized as follows. In Sec.~\ref{sec:SphericalBasis},
various notations regarding the spherical basis are introduced and
defined, and their fundamental properties are derived and summarized.
In Sec.~\ref{sec:iPTTensor}, the iPT formalism originally introduced
for scalar-valued fields is generalized to those for tensor-valued
fields. In Sec.~\ref{sec:InvariantFunctions}, rotationally invariant
components of renormalized bias functions and higher-order propagators
are identified. Additional symmetries in them are also described.
Explicit forms of several lower-order propagators are calculated and
presented. In Sec.~\ref{sec:TensorSpectrum}, the calculation of the
power spectrum, correlation function, and bispectrum of tensor fields
by our formalism is illustrated and presented in several cases that
are not relatively complicated. In Sec.~\ref{sec:Semilocal}, we define
a class of bias models, which we call semilocal models, in which the
Lagrangian bias of the tensor field is determined only by derivatives
of the gravitational potential at the same position in Lagrangian
space. The relations between the renormalized bias functions in iPT
and bias renormalizations in conventional perturbation theory are
clarified within the scheme of semilocal models of bias. Conclusions
are given in Sec.~\ref{sec:Conclusions}. In
Appendix~\ref{app:SphericalBasis}, explicit expressions of
higher-order spherical tensor basis are derived and given. Formulas
related to spherical harmonics as well as to Wigner's $3j$-, $6j$- and
$9j$-symbols, which are both repeatedly used throughout Papers I--IV,
are summarized, respectively, in
Appendixes~\ref{app:SphericalHarmonics} and \ref{app:3njSymbols}.
  
\section{Decomposition of tensors by spherical basis
  \label{sec:SphericalBasis}
}

In this paper, the irreducible representations of the tensor field in
spherical basis are extensively used, for which nonstandard
conventions of the rotation group are employed for reasons of
notational ease. We define our conventions below in this section.

For a given set of orthonormal basis vectors $\hat{\mathbf{e}}_i$
($i=1,2,3$) in a flat, three-dimensional space, the spherical basis
($\mathbf{e}_0$, $\mathbf{e}_\pm$) is defined by
\cite{Edmonds:1955fi,Khersonskii:1988krb}
\begin{equation}
  \mathbf{e}_0 = \hat{\mathbf{e}}_3, \quad
  \mathbf{e}_\pm = \mp
  \frac{\hat{\mathbf{e}}_1 \pm i\hat{\mathbf{e}}_2}{\sqrt{2}}.
  \label{eq:1}
\end{equation}
It is convenient to introduce the dual basis with an upper index by
taking complex conjugates of the spherical basis:
\begin{equation}
  \mathbf{e}^0 \equiv \mathbf{e}_0^* = \hat{\mathbf{e}}_3, \quad
  \mathbf{e}^\pm \equiv \mathbf{e}_\pm^*
  = \mp \frac{\hat{\mathbf{e}}_1 \mp i\hat{\mathbf{e}}_2}{\sqrt{2}}.
  \label{eq:2}
\end{equation}

The orthogonality relations among these bases are given by
\begin{equation}
  \mathbf{e}_m\cdot\mathbf{e}_{m'} = g^{(1)}_{mm'}, \quad
  \mathbf{e}_m\cdot\mathbf{e}^{m'} = \delta_m^{m'}, \quad
  \mathbf{e}^m\cdot\mathbf{e}^{m'} = g_{(1)}^{mm'},
  \label{eq:3}
\end{equation}
where $m,m'$ are azimuthal indices of the spherical basis which
represent one of $(-,0,+)$. The spherical metric tensors are defined
by
\begin{equation}
  g^{(1)}_{mm'} = g_{(1)}^{mm'} = (-1)^m\delta_{m,-m'},
  \label{eq:4}
\end{equation}
where $\delta_m^{m'}=\delta_{mm'}$ is the Kronecker's delta symbol
which is unity for $m=m'$ and zero for $m\ne m'$. The label ``(1)'' in
$g^{(1)}_{mm'}$ and $g_{(1)}^{mm'}$ suggests that the matrix composed
by these tensors has dimensions $(2l+1)\times (2l+1)$ with $l=1$, and
they are special cases of $g^{(l)}_{mm'}$ and $g_{(l)}^{mm'}$
generally defined in Eqs.~(\ref{eq:27}) or (\ref{eq:28}) below. The
matrices $g_{(1)}^{mm'}$ and $g^{(1)}_{mm'}$ are inverse matrices of
each other,
\begin{equation}
  g_{(1)}^{mm''} g^{(1)}_{m''m'} = \delta^m_{m'},
  \label{eq:5}
\end{equation}
where the repeated index $m''$ is assumed to be summed over, omitting
summation symbols as commonly adopted in the convention of tensor
analysis (Einstein summation convention). Throughout this paper, the
Einstein convention of summation for azimuthal indices is always
assumed, unless otherwise stated. They act as metric tensors of the
spherical basis,
\begin{equation}
  \mathbf{e}^m = g_{(1)}^{mm'} \mathbf{e}_{m'}, \quad
  \mathbf{e}_m = g^{(1)}_{mm'} \mathbf{e}^{m'}.
  \label{eq:6}
\end{equation}
Any vector can be represented by
$\bm{V} \propto V^m\mathbf{e}_m = V_m\mathbf{e}^m$ up to normalization
constant, and thus $\mathbf{e}_m$ is considered as the basis of {\em
  covariant} spherical vectors, and $\mathbf{e}^m$ is considered as
the basis of {\em contravariant} spherical vectors.

One of the advantages of the spherical basis over the Cartesian basis
is that the Cartesian tensors are reduced into irreducible tensors on
the spherical basis under a rotation of coordinates system
\cite{Sakurai:2011zz}. We define the spherical basis of traceless,
symmetric tensors \cite{Desjacques:2010gz,Vlah:2019byq} as\footnote{
  The normalization of the spherical tensors in this paper is somehow
  different from those in Refs.~\cite{Desjacques:2010gz,Vlah:2019byq}.
  Our normalization is chosen so that Eq.~(\ref{eq:11}) below should
  hold. In addition, in the previous literature, the axis
  $\mathbf{e}^0$ of the coordinates system is specifically chosen to
  be parallel to the separation vector $\bm{r}$ of the correlation
  function \cite{Desjacques:2010gz}, or, to the wave vector $\bm{k}$ of
  Fourier modes \cite{Vlah:2019byq}, while our axis can be arbitrarily
  chosen, so that fully rotational symmetry is explicitly kept in our
  formalism, not choosing a special coordinates system.}
\begin{equation}
  \mathsf{Y}^{(0)} = 1, \quad
  \mathsf{Y}^{(m)}_i = {\mathrm{e}^m}_i
  \label{eq:7}
\end{equation}
for tensors of rank-0 and rank-1, and
\begin{align}
  &
    \mathsf{Y}^{(0)}_{ij} = \sqrt{\frac{3}{2}}
    \left(
    {\mathrm{e}^0}_i\,{\mathrm{e}^0}_j - \frac{1}{3}\delta_{ij}
    \right),
  \label{eq:8}\\
  &
    \mathsf{Y}^{(\pm 1)}_{ij} = \sqrt{2}\,
    {\mathrm{e}^0}_{(i}\,{\mathrm{e}^\pm}_{j)},
  \label{eq:9}\\
  &
    \mathsf{Y}^{(\pm 2)}_{ij} =
    {\mathrm{e}^\pm}_i\,{\mathrm{e}^\pm}_j
  \label{eq:10}
\end{align}
for tensors of rank-2, where ${\mathrm{e}^m}_i = [\mathbf{e}^m]_i$ is
the Cartesian components of the spherical basis, and round brackets in
the indices of the right-hand side (rhs) of Eq.~(\ref{eq:9}) indicate
symmetrization with respect to the indices inside the brackets, e.g.,
$x_{(i}y_{j)} = (x_iy_j + x_jy_i)/2$. These bases satisfy
orthogonality relations,
\begin{equation}
    \mathsf{Y}^{(0)}\mathsf{Y}^{(0)} = 1, \quad
    \mathsf{Y}^{(m)}_i \mathsf{Y}^{(m')}_i = g_{(1)}^{mm'},\quad
    \mathsf{Y}^{(m)}_{ij} \mathsf{Y}^{(m')}_{ij} = g_{(2)}^{mm'},
  \label{eq:11}
\end{equation}
where, in the last equation, $g_{(2)}^{mm'} = (-1)^m \delta_{m,-m'}$
is the $5\times 5$ spherical metric, and the indices $m,m'$ run over
the integers $-2,-1,0,+1,+2$ in this case. The Einstein summation
convention is also applied for the Cartesian indices in the above, so
that $i$ and $j$ are summed over on the left-hand side (lhs) of the
above equations. Taking complex conjugates of the spherical basis
virtually lowers the azimuthal indices,
\begin{equation}
  \mathsf{Y}^{(0)*} = \mathsf{Y}^{(0)},\quad
  \mathsf{Y}^{(m)*}_i = g^{(1)}_{mm'}\mathsf{Y}^{(m')}_i,\quad
  \mathsf{Y}^{(m)*}_{ij} = g^{(2)}_{mm'}\mathsf{Y}^{(m')}_{ij}.
  \label{eq:12}
\end{equation}
For a unit vector
$\bm{n}=(\sin\theta\cos\phi,\sin\theta\sin\phi,\cos\theta)$ whose
direction is given by spherical coordinates $(\theta,\phi)$, the
spherical tensors are related to the spherical harmonics
$Y_{lm}(\theta,\phi)$ as
\begin{align}
  Y_{00}(\theta,\phi)
  &= \frac{1}{\sqrt{4\pi}} \mathsf{Y}^{(0)*}, \quad
  Y_{1m}(\theta,\phi) = \sqrt{\frac{3}{4\pi}}\, \mathsf{Y}^{(m)*}_i n_i,
  \nonumber \\
  Y_{2m}(\theta,\phi)
  &= \sqrt{\frac{15}{8\pi}}\, \mathsf{Y}^{(m)*}_{ij} n_i n_j.
  \label{eq:13}
\end{align}
where the normalization convention of the spherical harmonics is given
by a standard one,
\begin{equation}
  Y_{lm}(\theta,\phi) =
  \sqrt{\frac{2l+1}{4\pi}}
  \sqrt{\frac{(l-m)!}{(l+m)!}}\,
  P_l^m(\cos\theta)\,e^{im\phi},
  \label{eq:14}
\end{equation}
and
\begin{equation}
  P_l^m(x) =
  \frac{(-1)^m}{2^l\,l!}
  \left( 1-x^2 \right)^{m/2}
  \frac{d^{l+m}}{dx^{l+m}}
  \left( 1-x^2 \right)^l
  \label{eq:15}
\end{equation}
are associated Legendre polynomials that specifically include the
so-called Condon-Shortley phase factor of $(-1)^m$.
Similarly, spherical bases for higher-rank symmetric tensors are
uniquely defined by imposing the similar properties of
Eq.~(\ref{eq:13}),
\begin{equation}
  Y_{lm}(\theta,\phi) = \sqrt{\frac{(2l+1)!!}{4\pi\,l!}}\,
  \mathsf{Y}^{(m)*}_{i_1i_2\cdots i_l} n_{i_1} n_{i_2}\cdots n_{i_l}.
  \label{eq:16}
\end{equation}
General procedures to construct spherical bases of arbitrary rank are
given and explicit forms of the spherical bases of the rank-0 to -4 are
derived in Appendix~\ref{app:SphericalBasis}.

The standard normalization of spherical harmonics turns out to be not
so convenient to represent various equations in this work. Instead,
the spherical harmonics with Racah's normalization, defined by
\begin{equation}
  C_{lm}(\theta,\phi)
  \equiv \sqrt{\frac{4\pi}{2l+1}}\,
    Y_{lm}(\theta,\phi)
  =
  \sqrt{\frac{(l-m)!}{(l+m)!}}\,
  P_l^m(\cos\theta)\,e^{im\phi},
  \label{eq:17}
\end{equation}
turn out to simplify various equations. One of the reasons is that the
last normalization is closely related to the standard normalization of
Legendre polynomials, as we have
$C_{l0}(\theta,\phi) = P_l(\cos\theta)$, where $P_l(x) = P_l^0(x)$ are
the ordinary Legendre polynomials. With the spherical harmonics with
Racah's normalization, a general relation of Eq.~(\ref{eq:16}) is
given by
\begin{equation}
  \mathsf{Y}^{(m)}_{i_1i_2\cdots i_l} n_{i_1} n_{i_2}\cdots n_{i_l}
  = A_l\,C_{lm}^*(\theta,\phi),
  \label{eq:18}
\end{equation}
where we define a factor,
\begin{equation}
  A_l \equiv \sqrt{\frac{l!}{(2l-1)!!}},
  \label{eq:19}
\end{equation}
which repeatedly appears in later formulas. The spherical bases are
irreducible representations of the rotation group SO(3) in the same
way as the spherical harmonics are. We consider a passive rotation of
the Cartesian basis,
\begin{equation}
  \hat{\mathbf{e}}_i \xrightarrow{\mathbb{R}} \hat{\mathbf{e}}_i'
  =\hat{\mathbf{e}}_j R_{ji},
  \label{eq:20}
\end{equation}
where $R_{ji} =\hat{\mathbf{e}}_j \cdot \hat{\mathbf{e}}_i'$ are
components of a real orthogonal matrix which satisfies
$R^\mathrm{T}R = I$, and $I$ is the $3\times 3$ unit matrix. In terms
of the Euler angles $(\alpha,\beta,\gamma)$ with the $y$ convention,
the rotation matrix $R$ is explicitly given by
\begin{equation}
  R(\alpha,\beta,\gamma) =
  \begin{pmatrix}
    c_\alpha c_\beta c_\gamma - s_\alpha s_\gamma &
    - c_\alpha c_\beta s_\gamma - s_\alpha c_\gamma &
    c_\alpha s_\beta \\
    s_\alpha c_\beta c_\gamma + c_\alpha s_\gamma &
    - s_\alpha c_\beta s_\gamma + c_\alpha c_\gamma &
    s_\alpha s_\beta \\
    - s_\beta c_\gamma & s_\beta s_\gamma & c_\beta
  \end{pmatrix},
  \label{eq:21}
\end{equation}
where $c_\alpha = \cos\alpha$, $s_\alpha = \sin\alpha$,
$c_\beta = \cos\beta$, and so forth. Cartesian components of a unit
vector $\bm{n}$, where $|\bm{n}|=1$, transform as
\begin{equation}
  n_i \xrightarrow{\mathbb{R}} n_i' = n_j R_{ji},
  \label{eq:22}
\end{equation}
because we have $n_i \hat{\mathbf{e}}_i = n_i' \hat{\mathbf{e}}_i'$.

Denoting the spherical coordinates of $n_i$ and $n_i'$ as
$(\theta,\phi)$ and $(\theta',\phi')$, respectively, the spherical
harmonics transform as
\begin{equation}
  C_{lm}(\theta',\phi') = C_{lm'}(\theta,\phi)D_{(l)m}^{m'}(R),
  \label{eq:23}
\end{equation}
where $D_{(l)m}^{m'}(R) = D_{(l)m}^{m'}(\alpha,\beta,\gamma)$ is the
Wigner's rotation matrix of the passive rotation $R$ with the Euler
angles $(\alpha,\beta,\gamma)$.
Therefore, combining the properties of Eqs.~(\ref{eq:18}),
(\ref{eq:22}) and (\ref{eq:23}), we find a useful identity for the
spherical basis,
\begin{equation}
    R_{i_1j_1} R_{i_2j_2}\cdots R_{i_lj_l}
    \mathsf{Y}^{(m)}_{j_1j_2\cdots j_l}
  = D^{m}_{(l)m'}(R^{-1}) \mathsf{Y}^{(m')}_{i_1i_2\cdots i_l},
  \label{eq:24}
\end{equation}
where Wigner's matrix of the inverse rotation is given by
\begin{equation}
  D^{m}_{(l)m'}(R^{-1})
  = D^{m}_{(l)m'}(-\gamma,-\beta,-\alpha)
  = D^{m'*}_{(l)m}(R).
  \label{eq:25}
\end{equation}

The orthogonality relations of Eq.~(\ref{eq:11}) are generalized for
higher-rank tensors. In fact, using Eqs.~(\ref{eq:24}) and
(\ref{eq:25}) with $R^\mathrm{T}R = I$, one can show that they are
given by
\begin{equation}
  \mathsf{Y}^{(m)}_{i_1i_2\cdots i_l}
  \mathsf{Y}^{(m')}_{i_1i_2\cdots i_l} = g_{(l)}^{mm'},
  \label{eq:26}
\end{equation}
where $(2l+1) \times (2l+1)$ matrix $g_{(l)}^{mm'}$ is defined just as
Eq.~(\ref{eq:4}) but the indices $m,m'$ run over integers from $-l$
to $+l$.
The inverse matrix is similarly denoted by $g^{(l)}_{mm'}$,
whose elements are the same as $g_{(l)}^{mm'}$. Thus, similar
relations as in Eqs.~(\ref{eq:4}) and (\ref{eq:5}) hold for
$g_{(l)}^{mm'}$ and $g^{(l)}_{mm'}$:
\begin{equation}
  g_{(l)}^{mm'} = g^{(l)}_{mm'} = (-1)^m \delta_{m,-m'}, \quad
  g_{(l)}^{mm''} g^{(l)}_{m''m'} = \delta^m_{m'}.
  \label{eq:27}
\end{equation}
The relations to the $1jm$-symbol introduced by Wigner \cite{Wigner93}
are given by
\begin{equation}
  (-1)^lg^{(l)}_{mm'} =
  \begin{pmatrix}
    l\\
    m\ \ m'
  \end{pmatrix}
  = (-1)^lg_{(l)}^{mm'}.
  \label{eq:28}
\end{equation}
Taking the complex conjugate of the basis virtually lowers the
azimuthal index as in Eq.~(\ref{eq:12}), i.e., we have
\begin{equation}
  \mathsf{Y}^{(m)*}_{i_1i_2\cdots i_l}
  = g^{(l)}_{mm'}\mathsf{Y}^{(m')}_{i_1i_2\cdots i_l}, \quad
  \mathsf{Y}^{(m)}_{i_1i_2\cdots i_l}
  = g_{(l)}^{mm'}\mathsf{Y}^{(m')*}_{i_1i_2\cdots i_l}.
  \label{eq:29}
\end{equation}
The same properties also apply to spherical harmonics:
\begin{equation}
  C_{lm}^{*}(\theta,\phi)
  = g_{(l)}^{mm'} C_{lm'}(\theta,\phi), \quad
  C_{lm}(\theta,\phi)
  = g^{(l)}_{mm'} C_{lm'}^{*}(\theta,\phi).
  \label{eq:30}
\end{equation}
Consequently, the orthogonality relations of Eq.~(\ref{eq:26}) are
also equivalent to
\begin{equation}
  \mathsf{Y}^{(m)*}_{i_1i_2\cdots i_l}
  \mathsf{Y}^{(m')}_{i_1i_2\cdots i_l}
  = \delta_m^{m'}, \quad
  \mathsf{Y}^{(m)*}_{i_1i_2\cdots i_l}
  \mathsf{Y}^{(m')*}_{i_1i_2\cdots i_l}
  = g^{(l)}_{mm'}.
  \label{eq:31}
\end{equation}

We decompose a scalar function $S$ and a vector function $V_i$ into
spherical components as
\begin{equation}
  S = S_{00} \mathsf{Y}^{(0)}, \quad
  V_i = V_{1m} \mathsf{Y}^{(m)}_i.
  \label{eq:32}
\end{equation}
Because of the orthogonality relations of Eq.~(\ref{eq:11}), the
inverted relations of the above are given by
\begin{equation}
  S_{00} = S \mathsf{Y}^{(0)*}, \quad
  V_{1m} = V_i \mathsf{Y}^{(m)*}_i.
  \label{eq:33}
\end{equation}
Since the spherical basis is a traceless tensor, a symmetric rank-2
tensor function $T_{ij}$, which is not necessarily traceless, is
decomposed into spherical components of the trace part and traceless
part as
\begin{equation}
  T_{ij}
  = \frac{1}{3} \delta_{ij} T_{00} \mathsf{Y}^{(0)}
  + \sqrt{\frac{2}{3}}\, T_{2m} \mathsf{Y}^{(m)}_{ij},
  \label{eq:34}
\end{equation}
where
\begin{equation}
  T_{00} = T_{ii} \mathsf{Y}^{(0)*}, \quad
  T_{2m} =
  \sqrt{\frac{3}{2}}\, T_{ij} \mathsf{Y}^{(m)*}_{ij}.
  \label{eq:35}
\end{equation}
The prefactor $\sqrt{2/3}$ in front of the second term on the rhs of
Eq.~(\ref{eq:34}) is again our normalization convention. The general
rules of our normalization convention are shortly explained at the end
of this section.

Higher-rank tensors can be similarly decomposed into spherical
tensors. We first note that any symmetric tensor of rank-$l$ can be
decomposed into \cite{Spencer1970}
\begin{equation}
  T_{i_1i_2\cdots i_l} =
  T^{(l)}_{i_1i_2\cdots i_l}
  + \frac{l(l-1)}{2(2l-1)}
  \delta_{(i_1i_2} T^{(l-2)}_{i_3\cdots i_l)}
  + \cdots,
  \label{eq:36}
\end{equation}
where $T^{(l)}_{i_1i_2\cdots i_l}$ is a symmetric traceless tensor of
rank-$l$, $T^{(l-2)}_{i_3\cdots i_l}$ is traceless part of
$T_{jji_3i_4\cdots i_l}$, and so forth (one should note that the
Einstein summation convention is applied also to Cartesian indices,
and thus $j=1,2,3$ is summed over in the last tensor). The traceless
part of $T_{i_1\cdots i_l}$ is generally given by a formula
\cite{Spencer1970,Applequist1989}
\begin{multline}
  T^{(l)}_{i_1\cdots i_l} =
  \frac{l!}{(2l-1)!!}
  \sum_{k=0}^{[l/2]}
  \frac{(-1)^k(2l-2k-1)!!}{2^k k! (l-2k)!}
  \\ \times
  \delta_{(i_1i_2}\cdots\delta_{i_{2k-1}i_{2k}}
  T^{(l:k)}_{i_{2k+1}\cdots i_l)},
  \label{eq:37}
\end{multline}
where $[l/2]$ is the Gauss symbol, i.e., $[l/2] = l/2$ if $l$ is even
and $[l/2] = (l-1)/2$ if $l$ is odd, and
\begin{align}
  T^{(l:k)}_{i_{2k+1}\cdots i_l}
  &= \delta_{i_1i_2} \delta_{i_3i_4} \cdots \delta_{i_{2k-1}i_{2k}}
    T_{i_1\cdots i_{2k}i_{2k+1}\cdots i_l}
    \nonumber \\
  &=T_{j_1j_1j_2j_2\cdots j_kj_ki_{2k+1}\cdots i_l}
  \label{eq:38}
\end{align}
is the trace of the original tensor taken $k$ times. Tensors of rank-0
and rank-1 correspond to a scalar and a vector, respectively, and are
already traceless as they do not have any trace component from the
first place. The traceless part of the tensor components forms an
irreducible representation of the rotation group SO(3). Decomposition
of higher-rank tensors can be similarly obtained according to the
procedure described in Ref.~\cite{Spencer1970}. Recursively using
Eq.~(\ref{eq:37}), one obtains explicit expressions of
Eq.~(\ref{eq:36}). Up to the fourth rank, they are given by
\begin{align}
  T_{ij}
  &= T^{(2)}_{ij} + \frac{1}{3} \delta_{ij} T^{(0)},
  \label{eq:39}\\
  T_{ijk}
  &= T^{(3)}_{ijk} + \frac{3}{5} \delta_{(ij}T^{(1)}_{k)},
  \label{eq:40}\\
  T_{ijkl}
  &= T^{(4)}_{ijkl} + \frac{6}{7} \delta_{(ij} T^{(2)}_{kl)}
    + \frac{1}{5} \delta_{(ij}\delta_{kl)} T^{(0)}.
  \label{eq:41}
\end{align}

The traceless part of a tensor is decomposed by spherical bases as
\begin{align}
  T^{(l)}_{i_1i_2\cdots i_l}
  &=
    A_l T_{lm} \mathsf{Y}^{(m)}_{i_1i_2\cdots i_l},
  \label{eq:42}\\
  T_{lm}
  &=
    \frac{1}{A_l}
  T^{(l)}_{i_1i_2\cdots i_l} \mathsf{Y}^{(m)*}_{i_1i_2\cdots i_l}.
  \label{eq:43}
\end{align}
The prefactor $A_l$ on the rhs of Eq.~(\ref{eq:42}) is our
normalization convention, where $A_l$ is given by Eq.~(\ref{eq:19}).
The merits of this particular choice of normalization for the
irreducible tensor $T_{lm}$ in general appear in
Sec.~\ref{sec:Semilocal} below. However, the equations presented in
Secs.~\ref{sec:iPTTensor}--\ref{sec:TensorSpectrum} of this paper are
not affected by the choice of normalization $A_l$, because this
normalization factor simultaneously appears on both sides and cancels
to each other in most of the equations.

The traceless part of a traced tensor is similarly decomposed, e.g.,
as
\begin{align}
  T^{(l-2k)}_{i_{2k+1}\cdots i_l}
  &=
    A_{l-2k} T_{l-2k,m} \mathsf{Y}^{(m)}_{i_{2k+1}\cdots i_l},
    \label{eq:44}\\
  T_{l-2k,m}
  &=
    \frac{1}{A_{l-2k}}
    T^{(l)}_{i_{2k+1}\cdots i_l} \mathsf{Y}^{(m)*}_{i_{2k+1}\cdots i_l}.
  \label{eq:45}
\end{align}

Under the passive rotation of the basis, Eq.~(\ref{eq:20}),
Cartesian tensors of rank-$l$ transform as
\begin{equation}
  T^{(l)}_{i_1i_2\cdots i_l} \xrightarrow{\mathbb{R}}
  {T'}^{(l)}_{i_1i_2\cdots i_l} =
  T^{(l)}_{j_1j_2\cdots j_l} R_{j_1i_1}\cdots R_{j_li_l},
  \label{eq:46}
\end{equation}
and the corresponding transformation of spherical tensors is given by
\begin{equation}
  T_{lm} \xrightarrow{\mathbb{R}}
  T'_{lm} =
  T_{lm'} D^{m'}_{(l)m}(R),
  \label{eq:47}
\end{equation}
as derived from the transformation of the spherical base,
Eq.~(\ref{eq:24}). The contravariant components of the spherical
tensors are defined by
\begin{equation}
  T_{l}^{\,m} = g_{(l)}^{mm'} T_{lm'},
  \label{eq:48}
\end{equation}
and transform under the rotation as
\begin{equation}
  T_{l}^{\,m} \xrightarrow{\mathbb{R}}
  {T'}_{l}^{\,m} =
   D^{m}_{(l)m'}(R^{-1}) T_{l}^{\,m'}.
  \label{eq:49}
\end{equation}

\section{\label{sec:iPTTensor}
  The integrated perturbation theory of tensor fields
}

\subsection{\label{subsec:iPTTFormulation}
  Formulating the iPT of tensor fields
}

The integrated perturbation theory (iPT) is a systematic method of
perturbation theory to describe the quasinonlinear evolution of the
large-scale structure. This method is based on the Lagrangian scheme
of the biasing and nonlinear structure formation. The formalism of
this method is explicitly developed to describe the scalar fields,
such as the number density fields of galaxies or other biased objects
in the Universe \cite{Matsubara:2011ck,Matsubara:2013ofa}. In this
section, the formalism is generalized to be able to describe
quasinonlinear evolutions of objects which generally have tensor
values, such as the angular momentum of galaxies (vector),
the second moment of intrinsic alignment of galaxy shapes (rank-2
tensor), or higher moments of intrinsic alignment (rank-3 or higher
tensors), etc.

The fundamental formalism of iPT to calculate spatial correlations of
biased objects is described in Ref.~\cite{Matsubara:2011ck}. Basically
we can follow the formalism with a generalization of assigning the
tensor values to the objects. We denote the observable objects $X$,
where $X$ is the class of objects selected in a given cosmological
survey, such as a certain type of galaxies in galaxy surveys. We
consider each object of $a\in X$ has a tensor value
$F^a_{i_1i_2\cdots}$ in general, and define the tensor field by
\begin{equation}
  F_{X\,i_1i_2\cdots}(\bm{x}) =
  \frac{1}{\bar{n}_X}
  \sum_{a\in X} F^a_{i_1i_2\cdots}
  \delta_\mathrm{D}^3\left(\bm{x} - \bm{x}_a\right),
  \label{eq:50}
\end{equation}
where $\bm{x}_a$ is the location of a particular object $a$,
$\bar{n}_X$ is the mean number density of objects $X$, and
$\delta_\mathrm{D}^3(\bm{x})$ is the three-dimensional Dirac's delta
function. It is crucial to note that the above definition of the
tensor field is weighted by the number density of the objects $X$. In
the above, although the field really depends on the time
$F_{X\,i_1i_2\cdots}(\bm{x},t)$, the argument of time $t$ is omitted
in the notation for simplicity and the same suppression is employed
for other functions in this paper.

In general, any type of tensor can be decomposed into a sum of
symmetric and isotropic tensors \cite{Spencer1970}, where the latter
are of lower rank than the parent tensor. Bearing that in mind, below
we consider the case that tensor field $F^a_{i_1i_2\cdots}$ is a
symmetric tensor, which can be one of a decomposed symmetric tensor
even if the parent tensor is not symmetric. We decompose the symmetric
tensor of each object $F^a_{i_1i_2\cdots}$ into irreducible tensors
$F^a_{lm}$ according to the scheme described in the previous section.
Accordingly, corresponding components of Eq.~(\ref{eq:50}) are given
by
\begin{equation}
  F_{Xlm}(\bm{x}) =
  \frac{1}{\bar{n}_X}
  \sum_{a\in X} F^a_{lm}
  \delta_\mathrm{D}^3\left(\bm{x} - \bm{x}_a\right),
  \label{eq:51}
\end{equation}
As this tensor field is a number density-weighted quantity by
construction, we also consider the unweighted field $G_{Xlm}(\bm{x})$
defined by
\begin{equation}
  F_{Xlm}(\bm{x}) = \left[1 + \delta_X(\bm{x})\right] G_{Xlm}(\bm{x}),
  \label{eq:52}
\end{equation}
where $\delta_X(\bm{x})$ is the number density contrast of the objects
$X$, defined by
\begin{equation}
  1 + \delta_X(\bm{x}) =
  \frac{1}{\bar{n}_X}
  \sum_{a\in X} \delta_\mathrm{D}^3\left(\bm{x} - \bm{x}_a\right).
  \label{eq:53}
\end{equation}
Substituting Eq.~(\ref{eq:53}) into Eq.~(\ref{eq:52}) and comparing
with Eq.~(\ref{eq:51}), the unweighted field satisfies
\begin{equation}
  G_{Xlm}(\bm{x}_a) = F^a_{Xlm}
  \label{eq:54}
\end{equation}
at each position of object $a$, as naturally expected. The function
$G_{Xlm}(\bm{x})$ can arbitrarily be interpolated where $\bm{x} \ne
\bm{x}_a$ for all $a \in X$, because the field $F_{Xlm}(\bm{x})$ is
not supported at those points.

The iPT utilizes the Lagrangian perturbation theory (LPT) for the
dynamical nonlinear evolutions (see Ref.~\cite{Matsubara:2015ipa} for
notations of LPT employed in this paper). The Eulerian coordinates
$\bm{x}$ and the Lagrangian coordinates $\bm{q}$ of a mass element are
related by
\begin{equation}
  \bm{x} = \bm{q} + \bm{\Psi}(\bm{q}),
  \label{eq:55}
\end{equation}
where $\bm{\Psi}(\bm{q})$ is the displacement field. The number
density of the objects in Lagrangian space $n^\mathrm{L}_X(\bm{q})$ is
given by the Eulerian counterpart by a continuity relation,
\begin{equation}
  n_X(\bm{x}) d^3\!x =  n^\mathrm{L}_X(\bm{q}) d^3\!q,
  \label{eq:56}
\end{equation}
i.e., the Lagrangian number density is defined so that the Eulerian
positions of objects are displaced back into the Lagrangian positions,
and the number density in Lagrangian space also depends on the time of
evaluating $n_X(\bm{x})$. The mean number density of objects
$\bar{n}_X$ is the same in Eulerian and Lagrangian spaces by
construction, thus we have
$1+\delta_X(\bm{x}) = [1+\delta^\mathrm{L}_X(\bm{q})]/J(\bm{q})$,
where the Eulerian coordinate $\bm{x}$ on the lhs is a function of
the Lagrangian coordinates $\bm{q}$ according to Eq.~(\ref{eq:55}),
and $J(\bm{q})=|\partial(\bm{x})/\partial(\bm{q})|$ is the Jacobian of
the mapping. The density contrasts between these spaces are therefore
related by
\begin{equation}
  1 + \delta_X(\bm{x}) =
  \int d^3\!q
  \left[1 + \delta^\mathrm{L}_X(\bm{q})\right]
  \delta_\mathrm{D}^3\left[\bm{x} - \bm{q} - \bm{\Psi}(\bm{q})\right].
  \label{eq:57}
\end{equation}

As in the Eulerian space, the tensor field in Lagrangian space is also
naturally defined with the weight of the number density of objects.
The unweighted tensor field $G^\mathrm{L}_{Xlm}(\bm{q})$ in Lagrangian
space is simply defined by taking the same value displaced back from
the Eulerian position of Eq.~(\ref{eq:52}),
\begin{equation}
  G^\mathrm{L}_{Xlm}(\bm{q}) \equiv G_{Xlm}(\bm{x}),
  \label{eq:58}
\end{equation}
where $\bm{x}$ on the rhs is just given by
Eq.~(\ref{eq:55}) and the number density-weighted tensor field in
Lagrangian space, $F^\mathrm{L}_{lm}(\bm{q})$, is defined by
\begin{equation}
  F^\mathrm{L}_{Xlm}(\bm{q})
  = \left[1 + \delta^\mathrm{L}_X(\bm{q})\right]
  G^\mathrm{L}_{Xlm}(\bm{q}).
  \label{eq:59}
\end{equation}
As obviously derived from Eqs.~(\ref{eq:57}) and (\ref{eq:58}), the
number density-weighted tensor fields in Eulerian space and
Lagrangian space is related by
\begin{equation}
  F_{Xlm}(\bm{x}) =
  \int d^3\!q
  F^\mathrm{L}_{Xlm}(\bm{q})
  \delta_\mathrm{D}^3\left[\bm{x} - \bm{q} - \bm{\Psi}(\bm{q})\right].
  \label{eq:60}
\end{equation}

In cosmology, the properties of observable quantities are determined
by the initial condition of density fluctuations in the Universe, and
the growing mode of linear density contrast, $\delta_\mathrm{L}$, is a
representative of initial conditions. Therefore, the tensor field
$F_{Xlm}$ is considered as a functional of the linear density field
$\delta_\mathrm{L}$. The ensemble average of the $n$th functional
derivatives of the evolved field is called multipoint propagators
\cite{Matsubara:1995wd,Bernardeau:2008fa,Bernardeau:2010md}, which are
defined in equivalently various manners. Here we explain a
comprehensive definition according to Ref.~\cite{Matsubara:1995wd},
starting from a propagator in configuration space:
\begin{equation}
  \Gamma^{(n)}_{Xlm}\left(\bm{x}-\bm{x}_1,\ldots,\bm{x}-\bm{x}_n\right)
  \equiv
    (-i)^l
  \left\langle
    \frac{\delta^n {F}_{Xlm}(\bm{x})}{
      \delta\delta_\mathrm{L}(\bm{x}_1) \cdots \delta\delta_\mathrm{L}(\bm{x}_n)}
  \right\rangle.
  \label{eq:61}
\end{equation}
where $\delta_\mathrm{L}(\bm{x})$ is the linear density contrast,
$\delta/\delta\delta_\mathrm{L}(\bm{x})$ is the functional derivative
with respect to the linear density contrast, and
$\langle\cdots\rangle$ represents the ensemble average of the field
configurations. The phase factor $(-i)^l$ in front of
Eq.~(\ref{eq:61}) is our convention to define the propagators of the
tensor field, which we find convenient. Because of the translational
invariance of the Universe in a statistical sense, the rhs should be
invariant under the homogeneous translation,
$\bm{x} \rightarrow \bm{x} + \bm{x}_0$ and
$\bm{x}_i \rightarrow \bm{x}_i + \bm{x}_0$ for $i=1,\ldots,n$ with any
fixed displacement $\bm{x}_0$, and thus the arguments on the lhs of
Eq.~(\ref{eq:61}) should be given as those in the expression.

We apply the Fourier transform of the linear density contrast and the
tensor field,
\begin{align}
  \tilde{\delta}_\mathrm{L}(\bm{k})
  &=
  \int d^3\!x\, e^{-i\bm{k}\cdot\bm{x}} \delta_\mathrm{L}(\bm{x}),
  \label{eq:62}\\
  \tilde{F}_{Xlm}(\bm{k})
  &=
  \int d^3\!x\, e^{-i\bm{k}\cdot\bm{x}} F_{Xlm}(\bm{x}).
  \label{eq:63}
\end{align}
and the functional derivative in Fourier space is given by
\begin{equation}
  \frac{\delta}{\delta\tilde{\delta}_\mathrm{L}(\bm{k})}
  = \frac{1}{(2\pi)^3} \int d^3\!x\,e^{i\bm{k}\cdot\bm{x}}
  \frac{\delta}{\delta\delta_\mathrm{L}(\bm{x})}.
  \label{eq:64}
\end{equation}

Using the above equations, the Fourier counterpart of the rhs of
Eq.~(\ref{eq:61}) is calculated, and we have
\begin{multline}
  \left\langle
    \frac{\delta^n \tilde{F}_{Xlm}(\bm{k})}{
      \delta\tilde{\delta}_\mathrm{L}(\bm{k}_1) \cdots
      \delta\tilde{\delta}_\mathrm{L}(\bm{k}_n)}
  \right\rangle
  \\
  =
  i^l\,   
  (2\pi)^{3-3n}
  \delta_\mathrm{D}^3(\bm{k}_1+\cdots +\bm{k}_n - \bm{k})
  \tilde{\Gamma}^{(n)}_{Xlm}(\bm{k}_1,\ldots\bm{k}_n),
  \label{eq:65}
\end{multline}
where
\begin{multline}
  \tilde{\Gamma}^{(n)}_{Xlm}(\bm{k}_1,\ldots\bm{k}_n)
  = \int d^3\!x_1 \cdots  d^3\!x_n\,
  e^{-i(\bm{k}_1\cdot\bm{x}_1+\cdots +\bm{k}_n\cdot\bm{x}_n)}
  \\
  \times
  \Gamma^{(n)}_{Xlm}(\bm{x}_1,\ldots\bm{x}_n)
  \label{eq:66}
\end{multline}
is the Fourier transform of the propagator. The appearance of the
delta function on the rhs of Eq.~(\ref{eq:65}) is the consequence of
the statistical homogeneity of the Universe. Integrating
Eq.~(\ref{eq:65}), we have
\begin{equation}
  \tilde{\Gamma}^{(n)}_{Xlm}(\bm{k}_1,\ldots\bm{k}_n) =
    (-i)^l
  (2\pi)^{3n}
  \int \frac{d^3k}{(2\pi)^3}
  \left\langle
    \frac{\delta^n \tilde{F}_{Xlm}(\bm{k})}{
      \delta\tilde{\delta}_\mathrm{L}(\bm{k}_1) \cdots
      \delta\tilde{\delta}_\mathrm{L}(\bm{k}_n)}
  \right\rangle.
  \label{eq:67}
\end{equation}
The Fourier transform of Eq.~(\ref{eq:60}) is given by
\begin{equation}
  \tilde{F}_{Xlm}(\bm{k}) =
  \int d^3\!q\,e^{-i\bm{k}\cdot\bm{q} -i\bm{k}\cdot\bm{\Psi}(\bm{q})}
  F^\mathrm{L}_{Xlm}(\bm{q}).
  \label{eq:68}
\end{equation}

In the iPT, the displacement field $\bm{\Psi}(\bm{q})$ is expanded
according to the LPT,
\begin{multline}
  \bm{\Psi}(\bm{q}) =
  \sum_{n=1}^\infty \frac{i}{n!}
  \int \frac{d^3k_1}{(2\pi)^3} \cdots \frac{d^3k_n}{(2\pi)^3}
  e^{i(\bm{k}_1 + \cdots + \bm{k}_n)\cdot\bm{q}}
  \\ \times
  \bm{L}_n(\bm{k}_1,\ldots,\bm{k}_n)
  \tilde{\delta}_\mathrm{L}(\bm{k}_1) \cdots
  \tilde{\delta}_\mathrm{L}(\bm{k}_n).
  \label{eq:69}
\end{multline}
The perturbation kernels $\bm{L}_n$ are recursively obtained both for
the model of the Einstein-de Sitter Universe
\cite{Rampf:2012up,Zheligovsky:2013eca} and for general models
\cite{Matsubara:2015ipa}. The propagators of Eq.~(\ref{eq:67}) are
systematically evaluated by applying diagrammatic rules in the iPT.
The detailed derivation of the diagrammatic rules for the number
density field $\delta_X$ is given in Ref.~\cite{Matsubara:2011ck}.
Exactly the same derivation applies to the tensor field, simply
replacing $1+\delta_X \rightarrow F_{Xlm}$ and
$1+\delta^\mathrm{L}_X \rightarrow F^\mathrm{L}_{Xlm}$ in that
reference. In this method, the renormalized bias functions play
important roles. In the present case of the tensor field, the
renormalized bias functions $c^{(n)}_{Xlm}(\bm{k}_1,\ldots,\bm{k}_n)$
in Fourier space are defined by
\begin{multline}
  \left\langle
    \frac{\delta^n \tilde{F}^\mathrm{L}_{Xlm}(\bm{k})}{
      \delta\tilde{\delta}_\mathrm{L}(\bm{k}_1) \cdots
      \delta\tilde{\delta}_\mathrm{L}(\bm{k}_n)}
  \right\rangle
  \\
  =
    i^l\,
  (2\pi)^{3-3n}
  \delta_\mathrm{D}^3(\bm{k}_1+\cdots +\bm{k}_n - \bm{k})
  c^{(n)}_{Xlm}(\bm{k}_1,\ldots\bm{k}_n),
  \label{eq:70}
\end{multline}
or equivalently,
\begin{equation}
  c^{(n)}_{Xlm}(\bm{k}_1,\ldots\bm{k}_n) =
  (-i)^l
  (2\pi)^{3n}
  \int \frac{d^3k}{(2\pi)^3}
  \left\langle
    \frac{\delta^n \tilde{F}^\mathrm{L}_{Xlm}(\bm{k})}{
      \delta\tilde{\delta}_\mathrm{L}(\bm{k}_1) \cdots
      \delta\tilde{\delta}_\mathrm{L}(\bm{k}_n)}
  \right\rangle.
  \label{eq:71}
\end{equation}
As can be seen by comparing the above definition with
Eqs.~(\ref{eq:65}) and (\ref{eq:67}), the renormalized bias
functions are the counterparts of the propagator for the biasing in
Lagrangian space.

The diagrammatic rules of iPT are summarized in the Appendix of
Ref.~\cite{Matsubara:2013ofa}. The same rules apply, simply replacing
$\Gamma_X^{(n)} \rightarrow \Gamma_{Xlm}^{(n)}$,
$c_X^{(n)} \rightarrow c_{Xlm}^{(n)}$, etc. For the terms without
$c^{(n)}_X$ in that reference, an additional factor $c^{(0)}_{Xlm}$
should be multiplied, which is unity in the original formulation for
scalar fields. The propagators are given in a form,
\begin{equation}
  \tilde{\Gamma}^{(n)}_{Xlm}(\bm{k}_1,\ldots,\bm{k}_n) =
  \Pi(\bm{k})\,
  \hat{\Gamma}^{(n)}_{Xlm}(\bm{k}_1,\ldots,\bm{k}_n),
  \label{eq:72}
\end{equation}
where
\begin{equation}
  \Pi(\bm{k}) = \left\langle e^{-i\bm{k}\cdot\bm{\Psi}} \right\rangle
    =
    \exp\left[
      \sum_{n=2}^\infty \frac{(-i)^n}{n!}
      \left\langle (\bm{k}\cdot\bm{\Psi})^n \right\rangle_\mathrm{c}
    \right]
    \label{eq:73}
\end{equation}
is the vertex resummation factor in terms of the displacement field,
and $\langle\cdots\rangle_\mathrm{c}$ denotes the $n$-point connected
part of the random variables. The propagator
$\hat{\Gamma}^{(n)}_{Xlm}$ without the resummation factor
$\Pi(\bm{k})$ is called reduced propagators in the following.

Below, we give explicit forms of lower-order propagators, which can be
used to estimate power spectra, bispectra, etc. The vertex resummation
factor in the one-loop approximation of iPT is given by
\begin{equation}
  \Pi(\bm{k}) =
  \exp\left\{
    -\frac{1}{2} \int\frac{d^3p}{(2\pi)^3}
    \left[\bm{k}\cdot\bm{L}_1(\bm{p})\right]^2
    P_\mathrm{L}(p)
  \right\},
\label{eq:74}
\end{equation}
where
$P_\mathrm{L}(k)$ is the linear power
spectrum defined by
\begin{equation}
  \left\langle
    \tilde{\delta}_\mathrm{L}(\bm{k})
    \tilde{\delta}_\mathrm{L}(\bm{k})
  \right\rangle
  = (2\pi)^3 \delta_\mathrm{D}^3(\bm{k} + \bm{k}')
  P_\mathrm{L}(k).
  \label{eq:75}
\end{equation}
In the above equation, the delta function on the rhs appears due to
the statistical homogeneity of space in the initial density field, and
the linear power spectrum is a function of only the absolute magnitude
of wave vector $k=|\bm{k}|$ due to the statistical isotropy. As noted
above, reduced propagators of tensor fields are given by
straightforward generalizations of the corresponding propagators for
the scalar fields derived in Ref.~\cite{Matsubara:2013ofa}.

The reduced propagator of the first order, up to the one-loop
approximation, is given by
\begin{multline}
  \hat{\Gamma}_{Xlm}^{(1)}(\bm{k})
  = c_{Xlm}^{(1)}(\bm{k}) +
  \left[\bm{k}\cdot\bm{L}_1(\bm{k})\right] c_{Xlm}^{(0)}
\\
  + \int\frac{d^3p}{(2\pi)^3} P_\mathrm{L}(p)
  \biggl\{
    \left[\bm{k}\cdot\bm{L}_1(-\bm{p})\right]
    c_{Xlm}^{(2)}(\bm{k},\bm{p})
\\
      + \left[\bm{k}\cdot\bm{L}_1(-\bm{p})\right]
      \left[\bm{k}\cdot\bm{L}_1(\bm{k})\right]
      c_{Xlm}^{(1)}(\bm{p})
\\
      + \left[\bm{k}\cdot\bm{L}_2(\bm{k},-\bm{p})\right]
       c_{Xlm}^{(1)}(\bm{p})
\\
      + \frac{1}{2}
      \left[\bm{k}\cdot\bm{L}_3(\bm{k},\bm{p},-\bm{p})\right]
       c_{Xlm}^{(0)}
\\
      + \left[\bm{k}\cdot\bm{L}_1(\bm{p})\right]
      \left[\bm{k}\cdot\bm{L}_2(\bm{k},-\bm{p})\right]
       c_{Xlm}^{(0)}
  \biggr\}.
\label{eq:76}
\end{multline}
The first line of the above expression corresponds to the lowest, or
tree-level approximation. The reduced propagator of the second order in
the tree-level approximation is given by
\begin{multline}
  \hat{\Gamma}^{(2)}_{Xlm}(\bm{k}_1,\bm{k}_2) = 
  c^{(2)}_{Xlm}(\bm{k}_1,\bm{k}_2)
\\
  + \left[\bm{k}_{12}\cdot\bm{L}_1(\bm{k}_1)\right] c^{(1)}_{Xlm}(\bm{k}_2)
  + \left[\bm{k}_{12}\cdot\bm{L}_1(\bm{k}_2)\right] c^{(1)}_{Xlm}(\bm{k}_1)
\\
  +
  \left\{
    \left[\bm{k}_{12}\cdot\bm{L}_1(\bm{k}_1)\right]
    \left[\bm{k}_{12}\cdot\bm{L}_1(\bm{k}_2)\right]
    + \bm{k}_{12}\cdot\bm{L}_2(\bm{k}_1,\bm{k}_2)
  \right\} c^{(0)}_{Xlm}, 
\label{eq:77}
\end{multline}
where $\bm{k}_{12}=\bm{k}_1+\bm{k}_2$, and the reduced propagator of
the third order, again in the tree-level approximation, is given by
\begin{multline}
  \hat{\Gamma}^{(3)}_{Xlm}(\bm{k}_1,\bm{k}_2,\bm{k}_3) = 
  c^{(3)}_{Xlm}(\bm{k}_1,\bm{k}_2,\bm{k}_3)
  \\
  + \left[\bm{k}_{123}\cdot\bm{L}_1(\bm{k}_1)\right]
  c^{(2)}_{Xlm}(\bm{k}_2,\bm{k}_3) +\,\mathrm{cyc.}
  \\
  +
  \Bigl\{
    \left[\bm{k}_{123}\cdot\bm{L}_1(\bm{k}_1)\right]
    \left[\bm{k}_{123}\cdot\bm{L}_1(\bm{k}_2)\right]
    +
    \left[\bm{k}_{123}\cdot\bm{L}_2(\bm{k}_1,\bm{k}_2)\right]
    \Bigr\}
    \\
    \hspace{10pc} \times
    c^{(1)}_{Xlm}(\bm{k}_3) +\,\mathrm{cyc.}
  \\
  +
  \Bigl\{
    \left[\bm{k}_{123}\cdot\bm{L}_1(\bm{k}_1)\right]
    \left[\bm{k}_{123}\cdot\bm{L}_1(\bm{k}_2)\right]
    \left[\bm{k}_{123}\cdot\bm{L}_1(\bm{k}_3)\right]
    \\
    +
    \left[\bm{k}_{123}\cdot\bm{L}_1(\bm{k}_1)\right]
    \left[\bm{k}_{123}\cdot\bm{L}_2(\bm{k}_2,\bm{k}_3)\right]
    +\,\mathrm{cyc.}
    \\
    + \bm{k}_{123}\cdot\bm{L}_3(\bm{k}_1,\bm{k}_2,\bm{k}_3)
  \Bigr\} c^{(0)}_{Xlm}, 
\label{eq:78}
\end{multline}
where $\bm{k}_{123}=\bm{k}_1+\bm{k}_2+\bm{k}_3$, and
$+\,\mathrm{cyc.}$ represents two terms which are added with cyclic
permutations of each previous term.

In real space, the kernels of LPT in the standard theory of gravity
(in the Newtonian limit) are given by
\cite{Catelan:1994ze,Catelan:1996hw}
\begin{align}
& \bm{L}_1(\bm{k}) = \frac{\bm{k}}{k^2},
\label{eq:79}\\
& \bm{L}_2(\bm{k}_1,\bm{k}_2)
  =\frac37 \frac{\bm{k}_{12}}{{k_{12}}^2}
  \left[1 - \left(\frac{\bm{k}_1 \cdot \bm{k}_2}{k_1 k_2}\right)^2\right],
\label{eq:80}\\
&  \bm{L}_3(\bm{k}_1,\bm{k}_2,\bm{k}_3) =
  \frac13
  \left[\tilde{\bm{L}}_3(\bm{k}_1,\bm{k}_2,\bm{k}_3) + \mathrm{cyc.}\right],
\label{eq:81}\\
& \tilde{\bm{L}}_3(\bm{k}_1,\bm{k}_2,\bm{k}_3)
\nonumber\\
& \quad
  = \frac{\bm{k}_{123}}{{k_{123}}^2}
  \left\{
      \frac57
      \left[1 - \left(\frac{\bm{k}_1 \cdot \bm{k}_2}{k_1 k_2}\right)^2\right]
      \left[1 - \left(\frac{\bm{k}_{12} \cdot \bm{k}_3}
          {{k}_{12} k_3}\right)^2\right]
  \right.
\nonumber\\
& \qquad\quad
  \left.
  - \frac13
  \left[
      1 - 3\left(\frac{\bm{k}_1 \cdot \bm{k}_2}{k_1 k_2}\right)^2
      +\, 2 \frac{(\bm{k}_1 \cdot \bm{k}_2)(\bm{k}_2 \cdot \bm{k}_3)
        (\bm{k}_3 \cdot \bm{k}_1)}{{k_1}^2 {k_2}^2 {k_3}^2}
  \right]\right\}
\nonumber\\
  & \qquad
  + \frac{3}{7}\frac{\bm{k}_{123}}{{k_{123}}^2}
    \times
    \frac{(\bm{k}_1\times\bm{k}_{23})(\bm{k}_1\cdot\bm{k}_{23})}{{k_1}^2{k_{23}}^2}
    \left[1 - \left(\frac{\bm{k}_2 \cdot \bm{k}_3}{k_2 k_3}\right)^2\right].
\label{eq:82}
\end{align}
While the transverse part of the last line of Eq.~(\ref{eq:82}) does
not contribute to evaluating the power spectra up to one-loop order in
perturbation theory \cite{Matsubara:2013ofa}, this part does
contribute in more general occasions, such as two-loop corrections of
power spectrum, one-loop corrections of bispectrum,
etc.\footnote{These properties can be shown, e.g., by the diagrammatic
  method of iPT \cite{Matsubara:2011ck}} In the above, weak
dependencies on the time $t$ in the kernels are neglected
\cite{Bernardeau:2001qr,Matsubara:2015ipa}. Taking into account the
weak dependencies is also possible
\cite{Rampf:2015mza,Matsubara:2015ipa,Fasiello:2022lff,Rampf:2022tpg},
while the expressions are more complicated. In
Ref.~\cite{Matsubara:2015ipa}, for example, complete expressions of
the displacement kernels of LPT up to the seventh order are explicitly
given, together with a general way of recursively deriving the kernels
including weak dependencies on the time in general cosmology and
subleading growing modes. This method is also generalized to obtain
kernels of LPT in modified theories of gravity \cite{Aviles:2017aor}.

One of the benefits of the Lagrangian picture is that redshift-space
distortions are relatively easy to incorporate into the theory. The
displacement vector in redshift space $\bm{\Psi}^\mathrm{s}$ is
obtained from a linear mapping of that in real space by
$\bm{\Psi}^\mathrm{s} = \bm{\Psi} +
H^{-1}(\hat{\bm{z}}\cdot\dot{\bm{\Psi}})\hat{\bm{z}}$, where
$\dot{\bm{\Psi}} = \partial\bm{\Psi}/\partial t$ is the time
derivative of the displacement field, $\hat{\bm{z}}$ is a unit vector
directed to the line of sight, $H = \dot{a}/a$ is the time-dependent
Hubble parameter and $a(t)$ is the scale factor. A displacement kernel
in redshift space $\bm{L}^\mathrm{s}_n$ is simply related to the
kernel in real space in the same order by a linear mapping
\cite{Matsubara:2007wj}
\begin{equation}
  \bm{L}_n \rightarrow \bm{L}^\mathrm{s}_n = \bm{L}_n +
  nf\left(\hat{\bm{z}}\cdot\bm{L}_n\right)\hat{\bm{z}},
\label{eq:83}
\end{equation}
where $f=d\ln D/d\ln a = \dot{D}/HD$ is the linear growth rate, $D(t)$
is the linear growth factor. Throughout this paper, the
distant-observer approximation is assumed in redshift space and the
unit vector $\hat{\bm{z}}$ denotes the line-of-sight direction. The
expressions of Eqs.~(\ref{eq:74})--(\ref{eq:78}) apply as well in
redshift space when the displacement kernels $\bm{L}_n$ are replaced
by those in redshift space $\bm{L}^\mathrm{s}_n$.

\subsection{\label{subsec:PropSym}
  Symmetries
}

\subsubsection{\label{subsubsec:SymCC}
  Complex conjugate
}

We assume the Cartesian tensor field of Eq.~(\ref{eq:50}) is a real
tensor:
\begin{equation}
  F^*_{X\,i_1i_2\cdots i_l}(\bm{x}) = F_{X\,i_1i_2\cdots i_l}(\bm{x}),
  \label{eq:84}
\end{equation}
as it should be for physically observable quantities. For the spherical
decomposition of the traceless part, $F_{Xlm}(\bm{x})$,
Eqs.~(\ref{eq:29}), (\ref{eq:43}), (\ref{eq:48}) and the above
equation indicate
\begin{equation}
  F^*_{Xlm}(\bm{x}) = g_{(l)}^{mm'} F_{Xlm'}(\bm{x})
   = F_{Xl}^{\phantom{X}m}(\bm{x}),
  \label{eq:85}
\end{equation}
and thus the Fourier transform of Eq.~(\ref{eq:63}) satisfies
\begin{equation}
  \tilde{F}^*_{Xlm}(\bm{k})
  = g_{(l)}^{mm'}\tilde{F}_{Xlm'}(-\bm{k})
  = \tilde{F}_{Xl}^{\phantom{X}m}(-\bm{k}).
  \label{eq:86}
\end{equation}
That is, the complex conjugate of the irreducible tensor field in
Fourier space raises the azimuthal index, and inverts the sign of the
wave vector in the argument. The linear density contrast
$\delta_\mathrm{L}(\bm{x})$ is also a real field, and its Fourier
transform satisfies
${\tilde{\delta}_\mathrm{L}}^{\,*}(\bm{k}) =
\tilde{\delta}_\mathrm{L}(-\bm{k})$. Therefore, the renormalized bias
functions satisfy
\begin{align}
  c^{(n)*}_{Xlm}(\bm{k}_1,\ldots,\bm{k}_n)
  &= (-1)^l
    g_{(l)}^{mm'}c^{(n)}_{Xlm'}(-\bm{k}_1,\ldots,-\bm{k}_n)
    \nonumber\\
  &= (-1)^l
    c^{(n)m}_{Xl}(-\bm{k}_1,\ldots,-\bm{k}_n).
  \label{eq:87}
\end{align}
Similarly, the tensor propagator of Eq.~(\ref{eq:67}) satisfies
\begin{align}
  \tilde{\Gamma}^{(n)*}_{Xlm}(\bm{k}_1,\ldots,\bm{k}_n)
  &= (-1)^l
    g_{(l)}^{mm'}\tilde{\Gamma}^{(n)}_{Xlm'}(-\bm{k}_1,\ldots,-\bm{k}_n)
    \nonumber\\
  &= (-1)^l
    \tilde{\Gamma}^{(n)m}_{Xl}(-\bm{k}_1,\ldots,-\bm{k}_n).
  \label{eq:88}
\end{align}
In redshift space, the propagators also depend on the direction of
lines of sight, $\hat{\bm{z}}$. The above property of the complex
conjugate equally applies to those in redshift space, and we have
\begin{align}
  \tilde{\Gamma}^{(n)*}_{Xlm}(\bm{k}_1,\ldots,\bm{k}_n;\hat{\bm{z}})
  &= (-1)^l
    g_{(l)}^{mm'}
    \tilde{\Gamma}^{(n)}_{Xlm'}(-\bm{k}_1,\ldots,-\bm{k}_n;\hat{\bm{z}})
    \nonumber\\
  &= (-1)^l
    \tilde{\Gamma}^{(n)m}_{Xl}(-\bm{k}_1,\ldots,-\bm{k}_n;\hat{\bm{z}}).
  \label{eq:89}
\end{align}

\subsubsection{\label{subsubsec:SymRotation}
  Rotation
}

Under the passive rotation of Eq.~(\ref{eq:20}), vector
components $V_i$ generally transform as
\begin{equation}
  V_i \rightarrow V_i' = (R^{-1})_{ij}V_j = V_jR_{ji}.
  \label{eq:90}
\end{equation}
We denote the rotation of the position and wave vector components as,
e.g., $\bm{x} \rightarrow \bm{x}' = R^{-1}\bm{x}$ and
$\bm{k} \rightarrow \bm{k}' = R^{-1}\bm{k}$, respectively. This
notation does {\em not} mean the physical rotations of vectors and does
mean the rotation of components,
$x_i \rightarrow x_i' = x_jR_{ji}$ and
$k_i \rightarrow k_i' = k_jR_{ji}$, respectively, as a result of the
passive rotation of Cartesian basis, Eq.~(\ref{eq:20}). The
Cartesian tensor field of Eq.~(\ref{eq:50}) transforms as
\begin{equation}
  F_{X\,i_1i_2\cdots}(\bm{x}) \xrightarrow{\mathbb{R}}
  F_{X\,i_1i_2\cdots}'(\bm{x}') =
  F_{X\,j_1j_2\cdots}(\bm{x}) R_{j_1i_1} R_{j_2i_2}\cdots,
  \label{eq:91}
\end{equation}
and correspondingly the irreducible tensor of rank-$l$ transforms as
\begin{equation}
  F_{Xlm}(\bm{x}) \xrightarrow{\mathbb{R}}
  F_{Xlm}'(\bm{x}') = F_{Xlm'}(\bm{x}) D_{(l)m}^{m'}(R).
  \label{eq:92}
\end{equation}
The corresponding transformation in Fourier space,
Eq.~(\ref{eq:63}), is given by
\begin{equation}
  \tilde{F}_{Xlm}(\bm{k}) \xrightarrow{\mathbb{R}}
  \tilde{F}_{Xlm}'(\bm{k}') = \tilde{F}_{Xlm'}(\bm{k}) D_{(l)m}^{m'}(R),
  \label{eq:93}
\end{equation}
because the volume element $d^3\!x$ and the inner product
$\bm{k}\cdot\bm{x}$ are invariant under the rotation.

The same transformations apply to the tensor fields in
Lagrangian space, $F^\mathrm{L}_{Xlm}(\bm{q})$. Therefore, the
renormalized bias functions, Eq.~(\ref{eq:71}), transform as
\begin{multline}
  c^{(n)}_{Xlm}(\bm{k}_1,\ldots,\bm{k}_n) \xrightarrow{\mathbb{R}}
  c^{(n)\prime}_{Xlm}(\bm{k}_1',\ldots\bm{k}_n')
  \\
  = c^{(n)}_{Xlm'}(\bm{k}_1,\ldots,\bm{k}_n) D_{(l)m}^{m'}(R).
  \label{eq:94}
\end{multline}
Similarly, the propagators of Eq.~(\ref{eq:67}) transform as
\begin{multline}
  \tilde{\Gamma}^{(n)}_{Xlm}(\bm{k}_1,\ldots\bm{k}_n) \xrightarrow{\mathbb{R}}
  \tilde{\Gamma}^{(n)\prime}_{Xlm}(\bm{k}_1',\ldots\bm{k}_n')
  \\
  = \tilde{\Gamma}^{(n)}_{Xlm'}(\bm{k}_1,\ldots\bm{k}_n) D_{(l)m}^{m'}(R).
  \label{eq:95}
\end{multline}
In redshift space, the lines of sight also rotate,
$\hat{\bm{z}} \rightarrow \hat{\bm{z}}' = R^{-1}\hat{\bm{z}}$,
although simultaneous rotations of wave vectors and lines of sight
do not change the scalar products in the propagators, as in
Eqs.~(\ref{eq:74})--(\ref{eq:83}). Thus we have
\begin{multline}
  \tilde{\Gamma}^{(n)}_{Xlm}(\bm{k}_1,\ldots\bm{k}_n;\hat{\bm{z}})
  \xrightarrow{\mathbb{R}}
  \tilde{\Gamma}^{(n)\prime}_{Xlm}(\bm{k}_1',\ldots\bm{k}_n';\hat{\bm{z}}')
    \\
    =
  \tilde{\Gamma}^{(n)}_{Xlm'}(\bm{k}_1,\ldots\bm{k}_n;\hat{\bm{z}})
  D_{(l)m}^{m'}(R).
  \label{eq:96}
\end{multline}

\subsubsection{\label{subsubsec:SymParity}
  Parity
}

Next, we consider the property of parity symmetry. Keeping the
left-handed coordinates system, we consider an active parity
transformation of the physical system, instead of flipping the axes of
the coordinates system. With the active parity transformation, the
field values at a position $\bm{x}$ are mapped into those at
$-\bm{x}$, and the functional form of the tensor field is transformed
as
\begin{equation}
  F_{X\,i_1\cdots i_l}(\bm{x}) \xrightarrow{\mathbb{P}}
  F_{X\,i_1\cdots i_l}'(\bm{x}) =
  (-1)^{p_X+l} F_{X\,i_1\cdots i_l}(-\bm{x}),
  \label{eq:97}
\end{equation}
where $p_X=0$ for ordinary tensors and $p_X=1$ for pseudotensors. The
angular momentum is a typical example of a pseudotensor of rank-1. The
corresponding transformation for the irreducible tensor is given by
\begin{equation}
  F_{Xlm}(\bm{x}) \xrightarrow{\mathbb{P}}
  F_{Xlm}'(\bm{x}) = (-1)^{p_X+l} F_{Xlm}(-\bm{x})
  \label{eq:98}
\end{equation}
in configuration space, and 
\begin{equation}
  \tilde{F}_{Xlm}(\bm{k}) \xrightarrow{\mathbb{P}}
  \tilde{F}_{Xlm}'(\bm{k}) = (-1)^{p_X+l} \tilde{F}_{Xlm}(-\bm{k})
  \label{eq:99}
\end{equation}
in Fourier space.
The same applies to tensor fields in Lagrangian space.
Accordingly, the renormalized bias functions transform as
\begin{multline}
  c^{(n)}_{Xlm}(\bm{k}_1,\ldots,\bm{k}_n) \xrightarrow{\mathbb{P}}
  c^{(n)\prime}_{Xlm}(\bm{k}_1,\ldots,\bm{k}_n)
  \\ =
  (-1)^{p_X+l}\, c^{(n)}_{Xlm}(-\bm{k}_1,\ldots,-\bm{k}_n),
  \label{eq:100}
\end{multline}
for the linear density field transforms as
$\tilde{\delta}_\mathrm{L}(\bm{k}) \rightarrow
\tilde{\delta}_\mathrm{L}(-\bm{k})$.

Similarly, the propagators transform as
\begin{multline}
  \tilde{\Gamma}^{(n)}_{Xlm}(\bm{k}_1,\ldots,\bm{k}_n) \xrightarrow{\mathbb{P}}
  \tilde{\Gamma}^{(n)\prime}_{Xlm}(\bm{k}_1,\ldots,\bm{k}_n)
  \\
  = (-1)^{p_X+l}\,
  \tilde{\Gamma}^{(n)}_{Xlm}(-\bm{k}_1,\ldots,-\bm{k}_n).
  \label{eq:101}
\end{multline}
In redshift space, we have
\begin{multline}
  \tilde{\Gamma}^{(n)}_{Xlm}(\bm{k}_1,\ldots,\bm{k}_n;\hat{\bm{z}})
  \xrightarrow{\mathbb{P}}
  \tilde{\Gamma}^{(n)\prime}_{Xlm}(\bm{k}_1,\ldots,\bm{k}_n;\hat{\bm{z}})
  \\
  = (-1)^{p_X+l}\,
  \tilde{\Gamma}^{(n)}_{Xlm}(-\bm{k}_1,\ldots,-\bm{k}_n;-\hat{\bm{z}}).
  \label{eq:102}
\end{multline}

\subsubsection{\label{subsubsec:SymInterC}
  Interchange of arguments
}

An obvious symmetry of the renormalized bias functions and propagators
is that they are invariant under permutation of the wave vectors in the
argument. For the renormalized bias function, we have
\begin{equation}
  c^{(n)}_{Xlm}(\bm{k}_{\sigma(1)},\ldots,\bm{k}_{\sigma(n)}) =
  c^{(n)}_{Xlm}(\bm{k}_1,\ldots\bm{k}_n),
  \label{eq:103}
\end{equation}
where $\sigma \in \mathcal{S}_n$ is a permutation of the symmetric
group $\mathcal{S}_n$ of order $n$. Any permutation can be realized by
a series of interchange operations of adjacent arguments:
\begin{equation}
  c^{(n)}_{Xlm}(\bm{k}_1,\ldots,\bm{k}_i,\bm{k}_{i+1},\ldots\bm{k}_n)
  =
  c^{(n)}_{Xlm}(\bm{k}_1,\ldots,\bm{k}_{i+1},\bm{k}_i,\ldots\bm{k}_n),
  \label{eq:104}
\end{equation}
with arbitrary $i=1,\ldots,n-1$. Similarly, we have
\begin{equation}
  \tilde{\Gamma}^{(n)}_{Xlm}(\bm{k}_1,\ldots,\bm{k}_i,\bm{k}_{i+1},\ldots\bm{k}_n)
  =
  \tilde{\Gamma}^{(n)}_{Xlm}(\bm{k}_1,\ldots,\bm{k}_{i+1},\bm{k}_i,\ldots\bm{k}_n),
  \label{eq:105}
\end{equation}
for the propagators in real space, and
\begin{multline}
  \tilde{\Gamma}^{(n)}_{Xlm}(\bm{k}_1,\ldots,\bm{k}_i,\bm{k}_{i+1},\ldots\bm{k}_n;\hat{\bm{z}})
  \\
  =
  \tilde{\Gamma}^{(n)}_{Xlm}(\bm{k}_1,\ldots,\bm{k}_{i+1},\bm{k}_i,\ldots\bm{k}_n;\hat{\bm{z}}), 
  \label{eq:106}
\end{multline}
for the propagators in redshift space.

\section{\label{sec:InvariantFunctions}
  Rotationally invariant functions
}

Throughout this paper, we assume that the Universe is statistically
isotropic. In redshift space, the statistical quantities such as the
power spectrum and correlation functions are apparently anisotropic
and the line of sight is a special direction. However, when we rotate
the whole system including the direction to the line of sight,
$\hat{\bm{z}}$, rotations of the coordinates system should not alter
the physical degrees of freedom of the statistics. In order to
explicitly represent the property in our formalism, we introduce
invariant functions for the renormalized bias functions and
propagators so that the invariant functions contain only physical
degrees of freedom which do not depend on a particular coordinates
system of arbitrary choice.

\subsection{\label{subsec:RenBiasFn}
  Renormalized bias functions of tensor fields
}

In the following of this paper, we mostly work in Fourier space. For
notational simplicity, we omit the tildes above variables in Fourier
space, and denote $\delta_\mathrm{L}(\bm{k})$, $F_{Xlm}(\bm{k})$,
$\Gamma^{(n)}_{Xlm}(\bm{k}_1,\ldots)$ etc.~instead of
$\tilde{\delta}_\mathrm{L}(\bm{k})$, $\tilde{F}_{Xlm}(\bm{k})$,
$\tilde{\Gamma}^{(n)}_{Xlm}(\bm{k}_1,\ldots)$, respectively, as long
as when they are not confusing.

One of the essential parts of iPT is the introduction of the
renormalized bias function defined by Eq.~(\ref{eq:71}). They
characterize complicated fully nonlinear processes in the formation of
astronomical objects, such as various types of galaxies and clusters,
etc. If these complicated processes are described by some kind of
models and the function $F^\mathrm{L}_{Xlm}$ in Lagrangian space is
analytically given in terms of the linear density field
$\delta_\mathrm{L}$, then we can analytically calculate the
renormalized bias function $c^{(n)}_{Xlm}$ according to the model. In
the case of the number density of halos, the Press-Schechter model and
its extensions are applied to the concrete calculations along this
line \cite{Matsubara:2012nc,Matsubara:2013ofa,Matsubara:2016wth}. When
this kind of models is not known, the renormalized bias functions have
infinite degrees of freedom which is difficult to calculate from the
fundamental level. However, the rotational symmetry considered in the
previous subsection places some constraints on the functional form of
the renormalized bias functions as we show below.

We consider the renormalized bias functions given in the form of
Eq.~(\ref{eq:71}). In the simplified notation without tildes,
\begin{equation}
  c^{(n)}_{Xlm}(\bm{k}_1,\ldots\bm{k}_n) =
  (-i)^l
  (2\pi)^{3n}
  \int \frac{d^3k}{(2\pi)^3}
  \left\langle
    \frac{\delta^n F^\mathrm{L}_{Xlm}(\bm{k})}{
      \delta\delta_\mathrm{L}(\bm{k}_1) \cdots
      \delta\delta_\mathrm{L}(\bm{k}_n)}
  \right\rangle.
  \label{eq:107}
\end{equation}
The angular dependencies of the wave vectors in the arguments can be
decomposed in a series expansion with spherical harmonics with Racah's
normalization, Eq.~(\ref{eq:17}), as
\begin{multline}
  c^{(n)}_{Xlm}(\bm{k}_1,\ldots\bm{k}_n) =
  \sum_{l_1,\ldots,l_n}
  c^{(n)\,l_1\cdots l_n}_{Xlm;m_1\cdots m_n}(k_1,\ldots,k_n)
  \\ \times
  C_{l_1m_1}^*(\hat{\bm{k}}_1)\cdots C_{l_nm_n}^*(\hat{\bm{k}}_n),
  \label{eq:108}
\end{multline}
where $k_i = |\bm{k}_i|$ are absolute values of the wave vectors, and
$\hat{\bm{k}}_i = (\theta_i,\phi_i)$ represents the spherical
coordinates for the direction of wave vectors $\bm{k}_i$. The repeated
indices $m$, $m_1,\ldots,m_n$ in Eq.~(\ref{eq:108}) are summed over
according to the Einstein summation convention as in the previous
section. Because of the orthonormality relation for the spherical
harmonics, Eq.~(\ref{eq:397}), the expansion of Eq.~(\ref{eq:108}) can
be inverted and we have
\begin{multline}
  c^{(n)\,l_1\cdots l_n}_{Xlm;m_1\cdots m_n}(k_1,\ldots,k_n) =
  (2l_1+1)\cdots(2l_n+1)
  \\ \times
  \int \frac{d^2\hat{k}_1}{4\pi} \cdots  \frac{d^2\hat{k}_n}{4\pi}
  c^{(n)}_{Xlm}(\bm{k}_1,\ldots\bm{k}_n)
  \\ \times 
  C_{l_1m_1}(\hat{\bm{k}}_1)\cdots C_{l_nm_n}(\hat{\bm{k}}_n),
  \label{eq:109}
\end{multline}
where $d^2\hat{k}_i = \sin\theta_i\,d\theta_i\,d\phi_i$
represents the angular integration of the wave vector $\bm{k}_i$. Under
the passive rotation of Eq.~(\ref{eq:20}), the function of
Eq.~(\ref{eq:109}) covariantly transforms as
\begin{multline}
  c^{(n)\,l_1\cdots l_n}_{Xlm;m_1\cdots m_n}(k_1,\ldots,k_n)
  \xrightarrow{\mathbb{R}}
  c^{(n)\,l_1\cdots l_n}_{Xlm';m_1'\cdots m_n'}(k_1,\ldots,k_n)
  \\ \times
  D_{(l)m}^{m'}(R) D_{(l_1)m_1}^{m_1'}(R) \cdots  D_{(l_n)m_n}^{m_n'}(R),
  \label{eq:110}
\end{multline}
as obviously indicated by the lower positions of the azimuthal indices
$m, m_1,\ldots m_n$.

For the statistically isotropic Universe, the functions of
Eq.~(\ref{eq:109}) are rotationally invariant as these functions are
given by ensemble averages of the field variables as seen by
Eqs.~(\ref{eq:107}) and (\ref{eq:109}), and they should not depend
on the choice of coordinates system $\hat{\mathbf{e}}_i$ to describe
wave vectors $\bm{k}_1,\ldots,\bm{k}_n$ of the renormalized bias
functions $c^{(n)}_{Xlm}(\bm{k}_1,\ldots,\bm{k}_n)$. Therefore, they
are invariant under the transformation of Eq.~(\ref{eq:110}), and we
have
\begin{multline}
  c^{(n)\,l_1\cdots l_n}_{Xlm;m_1\cdots m_n}(k_1,\ldots,k_n)
  = c^{(n)\,l_1\cdots l_n}_{Xlm';m_1'\cdots m_n'}(k_1,\ldots,k_n)
  \\ \times
  \frac{1}{8\pi^2} \int [dR]
  D_{(l)m}^{m'}(R) D_{(l_1)m_1}^{m_1'}(R) \cdots  D_{(l_n)m_n}^{m_n'}(R).
  \label{eq:111}
\end{multline}
where
\begin{equation}
  \int [dR] \cdots
  = \int_0^{2\pi}d\alpha \int_0^\pi d\beta\,\sin\beta
  \int_0^{2\pi} d\gamma \cdots
\label{eq:112}
\end{equation}
 represents the integrals over Euler angles
 $(\alpha\beta\gamma)$ of the rotation $R$.

A product of two Wigner's rotation matrices of the same rotation
 reduces to a single matrix through vector-coupling coefficients as
 \cite{Edmonds:1955fi}
\begin{multline}
  D_{(l_1)m_1}^{m_1'}(R) D_{(l_2)m_2}^{m_2'}(R)
  \\
  =
  \sum_{l,m,m'} (2l+1)
  \begin{pmatrix}
    l_1 & l_2 & l \\
    m_1' & m_2' & m'
  \end{pmatrix}
  \begin{pmatrix}
    l_1 & l_2 & l \\
    m_1 & m_2 & m
  \end{pmatrix}
  D_{(l)m}^{m'\,*}(R),
  \label{eq:113}
\end{multline}
using the Wigner's $3j$-symbol. The complex conjugate of the rotation
matrix is given by
$D_{(l)m_1}^{m_2*}(R) = g_{(l)}^{m_1m_1'}
g^{(l)}_{m_2m_2'}D_{(l)m_1'}^{m_2'}(R)$ using our metric tensor for
spherical basis.

In the following, we use a simplified notation for the $3j$-symbols,
\begin{equation}
  \left(l_1\,l_2\,l_3\right)_{m_1m_2m_3} \equiv
    \begin{pmatrix}
      l_1 & l_2 & l_3 \\
      m_1 & m_2 & m_3
    \end{pmatrix},
  \label{eq:114}
\end{equation}
which is nonzero only when $m_1+m_2+m_3=0$. We consider that the
azimuthal indices can be raised or lowered by spherical metric, e.g.,
\begin{align}
  \left(l_1\,l_2\,l_3\right)_{m_1}^{\phantom{m_1}m_2m_3}
  &= g_{(l_2)}^{m_2m_2'} g_{(l_3)}^{m_3m_3'}
    \left(l_1\,l_2\,l_3\right)_{m_1m_2'm_3'}
    \nonumber \\
  &= (-1)^{m_2+m_3}
    \begin{pmatrix}
      l_1 & l_2 & l_3 \\
      m_1 & -m_2 & -m_3
    \end{pmatrix},
  \label{eq:115}
\end{align}
and so forth. The properties and formulas for Wigner's $3j$-symbol
with our notation, which are repeatedly used in this paper below and
in subsequent papers in the series, are summarized in
Appendix~\ref{app:3njSymbols}.
With our notation of the $3j$-symbol, Eq.~(\ref{eq:113}) is
equivalent to an expression,
\begin{multline}
  D_{(l_1)m_1}^{m_1'}(R) D_{(l_2)m_2}^{m_2'}(R)
  = (-1)^{l_1+l_2}
  \sum_{l=0}^\infty (-1)^l(2l+1)
  \\ \times
  \left(l_1\,l_2\,l\right)_{m_1m_2}^{\phantom{m_1m_2}m}
  \left(l_1\,l_2\,l\right)^{m_1'm_2'}_{\phantom{m_1'm_2'}m'}
  D_{(l)m}^{m'}(R),
  \label{eq:116}
\end{multline}
in which the rotational covariance is explicit. We use a property of
Eq.~(\ref{eq:421}) for the $3j$-symbol to derive the above equation.

Consecutive applications of Eq.~(\ref{eq:116}) to
Eq.~(\ref{eq:111}), the dependence on the rotation $R$ in the
integrand on the rhs can be represented by a linear combination of
products of less than or equal to three rotation matrices and the
coefficients of the expression are given by products of $3j$-symbols.
The averages of products of rotation matrices of three or fewer are
given by
\begin{align}
  &
    \frac{1}{8\pi^2} \int [dR]
  D_{(l)m}^{m'}(R)
    = \delta_{l0} \delta_{m0} \delta_{m'0},
  \label{eq:117}\\
  &
    \frac{1}{8\pi^2} \int [dR]
  D_{(l_1)m_1}^{m_1'}(R) D_{(l_2)m_2}^{m_2'}(R)
    = \frac{\delta_{l_1l_2}}{2l_1+1}
    g^{(l_1)}_{m_1m_2} g_{(l_1)}^{m_1'm_2'},
  \label{eq:118}\\
  &
    \frac{1}{8\pi^2} \int [dR]
    D_{(l_1)m_1}^{m_1'}(R) D_{(l_2)m_2}^{m_2'}(R) D_{(l_3)m_3}^{m_3'}(R)
    \nonumber\\
  & \hspace{5pc}
    = (-1)^{l_1+l_2+l_3} \left(l_1\,l_2\,l_3\right)_{m_1m_2m_3}
     \left(l_1\,l_2\,l_3\right)^{m_1'm_2'm_3'}.
  \label{eq:119}
\end{align}
Equations~(\ref{eq:117}) and (\ref{eq:118}) are also derived from
Eq.~(\ref{eq:119}) noting $D_{(0)0}^0(R)=1$ and Eqs.~(\ref{eq:427}).
Also, Eqs.~(\ref{eq:118}) and (\ref{eq:119}) are derived from
further applications of Eq.~(\ref{eq:116}) to reduce the number of
products and finally using Eq.~(\ref{eq:117}).

As a result of the above procedure, averaging over the rotation $R$,
the dependence of the indices of $m$'s in Eq.~(\ref{eq:111}) are all
represented by spherical metrics and $3j$-symbols. First, we
specifically derive the results for lower orders of $n$.
For $n=0$, using Eqs.~(\ref{eq:111}) and (\ref{eq:117}), we derive
\begin{equation}
  c^{(0)}_{Xlm} = c^{(0)}_{X00}\delta_{l0} \delta_{m0}.
  \label{eq:120}
\end{equation}
This result is trivial because the corresponding function of
Eq.~(\ref{eq:107}) for $n=0$ does not depend on the direction of the
wave vector. For notational simplicity, we simply denote
$c^{(0)}_X \equiv c^{(0)}_{X00}$, and we have
\begin{equation}
  c^{(0)}_{Xlm} = \delta_{l0} \delta_{m0} c^{(0)}_{X}.
  \label{eq:121}
\end{equation}
For a scalar function, $F_X$, the coefficient $c^{(0)}_X$ just
corresponds to the mean value,
$c^{(0)}_X = c^{(0)}_{X00} = \langle F_{X00} \rangle = \langle F_{X00}
\rangle \mathsf{Y}^{(0)}= \langle F_X \rangle $, as seen from
Eqs.~(\ref{eq:7}) and (\ref{eq:32}). Because of the assumption that
the field $F_X$ is a real field, the coefficient $c^{(0)}_X$ has a
real value:
\begin{equation}
  c^{(0)*}_X = c^{(0)}_X.
  \label{eq:122}
\end{equation}
The parity transformation is given by
\begin{equation}
  c^{(0)}_X \xrightarrow{\mathbb{P}} (-1)^{p_X} c^{(0)}_X.
  \label{eq:123}
\end{equation}
If the Universe is statistically invariant under the parity, we should
have $p_X=0$, because the mean value of a scalar field
$\langle F_X \rangle$ vanishes if the field is a pseudoscalar in the
Universe with parity symmetry.

For $n=1$, using Eqs.~(\ref{eq:111}) and (\ref{eq:118}), we derive
\begin{equation}
  c^{(1)\,l_1}_{Xlm;m_1}(k)
  = \frac{\delta_{l_1l}}{2l+1}
  g^{(l)}_{mm_1} g_{(l)}^{m'm_1'} c^{(1)\,l}_{Xlm';m_1'}(k),
  \label{eq:124}
\end{equation}
and therefore both indices $m,m_1$ appear only in the metric
$g^{(l)}_{mm_1}$. Defining a rotationally invariant function,
\begin{equation}
  c^{(1)}_{Xl}(k)
  \equiv
  \frac{(-1)^l}{\sqrt{2l+1}}
  g_{(l)}^{mm_1} c^{(1)\,l}_{Xlm;m_1}(k),
  \label{eq:125}
\end{equation}
one finds that Eq.~(\ref{eq:124}) reduces to a simple form,
\begin{equation}
  c^{(1)\,l_1}_{Xlm;m_1}(k)
  =  \frac{(-1)^l}{\sqrt{2l+1}}
    \delta_{ll_1} g^{(l)}_{mm_1} c^{(1)}_{Xl}(k).
  \label{eq:126}
\end{equation}
Substituting this expression into Eq.~(\ref{eq:108}) with $n=1$, we
derive
\begin{equation}
  c^{(1)}_{Xlm}(\bm{k}) =
  \frac{(-1)^l}{\sqrt{2l+1}}\,
  c^{(1)}_{Xl}(k) C_{lm}(\hat{\bm{k}}).
  \label{eq:127}
\end{equation}
The invariant coefficient $c^{(1)}_{Xl}(k)$ can also be read off
from an expression of the renormalized bias function
$c^{(1)}_{Xlm}(\bm{k})$ in which the angular dependence on
$\bm{k}$ is expanded by the spherical harmonics
$C_{lm}(\hat{\bm{k}})$. One can also explicitly invert the above
equation by using the orthonormality relation of spherical harmonics.

Combining the properties of complex conjugate for the spherical
harmonics, Eq.~(\ref{eq:394}), and for the renormalized bias function,
Eq.~(\ref{eq:87}) with $n=1$, we have
\begin{equation}
  c^{(1)*}_{Xl}(k) = c^{(1)}_{Xl}(k),
  \label{eq:128}
\end{equation}
and thus the reduced function is a real function. Combining the
properties of parity for the spherical harmonics, Eq.~(\ref{eq:393}),
and for the renormalized bias function, Eq.~(\ref{eq:100}) with
$n=1$, we have
\begin{equation}
  c^{(1)}_{Xl}(k) \xrightarrow{\mathbb{P}} (-1)^{p_X} c^{(1)}_{Xl}(k).
  \label{eq:129}
\end{equation}
If the Universe is statistically invariant under the parity, the
function is nonzero only when $p_X = 0$. This means that pseudotensors
do not have the first-order renormalized bias function in the presence
of parity symmetry. This property is partly because we consider the
case that only scalar perturbations $\delta_\mathrm{L}$ are
responsible for the properties of tensor fields. For a typical example
that is a manifestation of the last conclusion, angular momenta of
galaxies are not generated by the first-order effect in linear
perturbation theory \cite{Peebles1969}.

For $n=2$, using Eqs.~(\ref{eq:111}) and (\ref{eq:119}), we derive
\begin{equation}
  c^{(2)\,l_1l_2}_{Xlm;m_1m_2}(k_1,k_2)
  =
  \left(l\,l_1\,l_2\right)_{mm_1m_2}
    c^{(2)\,l}_{Xl_1l_2}(k_1,k_2),
  \label{eq:130}
\end{equation}
where we define a rotationally invariant function,
\begin{equation}
  c^{(2)\,l}_{Xl_1l_2}(k_1,k_2)
  \equiv
  (-1)^{l+l_1+l_2}
    \left(l\,l_1\,l_2\right)^{mm_1m_2}
    c^{(2)\,l_1l_2}_{Xlm;m_1m_2}(k_1,k_2).
  \label{eq:131}
\end{equation}
The above Eq.~(\ref{eq:130}) essentially corresponds to the
Wigner-Eckart theorem of representation theory and quantum mechanics
\cite{Sakurai:2011zz}, and the dependencies on azimuthal indices
$m,m_1,m_2$ appear only through $3j$-symbols. Substituting
Eq.~(\ref{eq:130}) into Eq.~(\ref{eq:108}) with $n=2$, we derive
\begin{multline}
  c^{(2)}_{Xlm}(\bm{k}_1,\bm{k}_2)
  =
  \sum_{l_1,l_2}
  c^{(2)\,l}_{Xl_1l_2}(k_1,k_2)
  \left(l\,l_1\,l_2\right)_{m}^{\phantom{m}m_1m_2}
  \\ \times
    C_{l_1m_1}(\hat{\bm{k}}_1) C_{l_2m_2}(\hat{\bm{k}}_2).
  \label{eq:132}
\end{multline}
The last expression contains the bipolar spherical harmonics
\cite{Khersonskii:1988krb} which is defined by Eq.~(\ref{eq:400}) in
Appendix~\ref{app:SphericalHarmonics} with our notation. Rather than
using the conventional notation for the bipolar spherical harmonics,
$\{Y_{l_1}(\hat{\bm{k}}_1)\otimes Y_{l_2}(\hat{\bm{k}}_2)\}_{lm}$, we
introduce a notation with convenient normalization of the bipolar
spherical harmonics,
\begin{equation}
  X^{l_1l_2}_{lm}(\hat{\bm{k}}_1,\hat{\bm{k}}_2) \equiv
  \frac{(-1)^l\,4\pi}{\sqrt{\{l\}\{l_1\}\{l_2\}}}
  \left\{
  Y_{l_1}(\hat{\bm{k}}_1)\otimes Y_{l_2}(\hat{\bm{k}}_2)
  \right\}_{lm},
  \label{eq:133}
\end{equation}
where we employ a notation,
\begin{equation}
  \{L\} \equiv 2L+1
  \label{eq:134}
\end{equation}
for a non-negative integer $L$ throughout the paper henceforth, as
this type of factor repeatedly appears later. Please do not confuse
with the curly brackets of the bipolar spherical harmonics, in which
the argument is not a non-negative integer. The bipolar spherical
harmonics satisfy the orthonormality relation, Eq.~(\ref{eq:405}), and
product formula, Eq.~(\ref{eq:407}), as shown in
Appendix~\ref{app:SphericalHarmonics}. With the above notation,
Eq.~(\ref{eq:132}) is concisely given by
\begin{equation}
  c^{(2)}_{Xlm}(\bm{k}_1,\bm{k}_2)
  = \sum_{l_1,l_2}
  c^{(2)\,l}_{Xl_1l_2}(k_1,k_2)
  X^{l_1l_2}_{lm}(\hat{\bm{k}}_1,\hat{\bm{k}}_2).
  \label{eq:135}
\end{equation}

Combining the properties of complex conjugate for the bipolar
spherical harmonics, Eq.~(\ref{eq:404}), and for the renormalized
bias function, Eq.~(\ref{eq:87}) with $n=2$, we have
\begin{equation}
  c^{(2)\,l*}_{Xl_1l_2}(k_1,k_2) = 
  c^{(2)\,l}_{Xl_1l_2}(k_1,k_2)
  \label{eq:136}
\end{equation}
and thus the reduced function is a real function. Combining the
properties of parity for the bipolar spherical harmonics,
Eq.~(\ref{eq:403}), and for the renormalized bias function,
Eq.~(\ref{eq:100}) with $n=2$, we have
\begin{equation}
  c^{(2)\,l}_{Xl_1l_2}(k_1,k_2)
  \xrightarrow{\mathbb{P}} (-1)^{p_X+l+l_1+l_2}
  c^{(2)\,l}_{Xl_1l_2}(k_1,k_2).
  \label{eq:137}
\end{equation}
If the Universe is statistically invariant under the parity, the
function is nonzero only when $p_X+l+l_1+l_2=\mathrm{even}$. The
interchange symmetry of Eq.~(\ref{eq:104}) in this case is given by
$c^{(2)}_{Xlm}(\bm{k}_1,\bm{k}_2) = c^{(2)}_{Xlm}(\bm{k}_2,\bm{k}_1)$.
Because of the symmetry of $3j$-symbols of Eq.~(\ref{eq:423}), the
bipolar spherical harmonics satisfy
\begin{equation}
  X^{l_1l_2}_{lm}(\hat{\bm{k}}_1,\hat{\bm{k}}_2)
  = (-1)^{l+l_1+l_2}
  X^{l_2l_1}_{lm}(\hat{\bm{k}}_2,\hat{\bm{k}}_1),
  \label{eq:138}
\end{equation}
and therefore the invariant function satisfies
\begin{equation}
  c^{(2)\,l}_{Xl_2l_1}(k_2,k_1)
  = (-1)^{l+l_1+l_2}
  c^{(2)\,l}_{Xl_1l_2}(k_1,k_2).
  \label{eq:139}
\end{equation}

For $n=3$, using Eqs.~(\ref{eq:111}), (\ref{eq:116}) and
(\ref{eq:119}), we derive
\begin{multline}
  c^{(3)\,l_1l_2l_3}_{Xlm;m_1m_2m_3}(k_1,k_2,k_3)
  =
  \sum_L (-1)^L\sqrt{\{L\}}
    \left(l\,l_1\,L\right)_{mm_1M}
  \\ \times
    \left(L\,l_2\,l_3\right)^{M}_{\phantom{M}m_2m_3}
    c^{(3)\,l;L}_{Xl_1l_2l_3}(k_1,k_2,k_3),
  \label{eq:140}
\end{multline}
where we define a rotationally invariant function,
\begin{multline}
  c^{(3)\,l;L}_{Xl_1l_2l_3}(k_1,k_2,k_3)
  \equiv (-1)^{l+l_1+l_2+l_3+L}
  \sqrt{\{L\}}
    \left(l\,l_1\,L\right)^{mm_1}_{\phantom{mm_1}M}
    \\ \times
    \left(L\,l_2\,l_3\right)^{Mm_2m_3}
    c^{(3)\,l_1l_2l_3}_{Xlm;m_1m_2m_3}(k_1,k_2,k_3).
  \label{eq:141}
\end{multline}
Substituting Eq.~(\ref{eq:140}) into Eq.~(\ref{eq:108}) with $n=3$,
we derive
\begin{multline}
  c^{(3)}_{Xlm}(\bm{k}_1,\bm{k}_2,\bm{k}_3)
  =
  \sum_{l_1,l_2,l_3,L} (-1)^L\sqrt{\{L\}}
  \\ \times
  c^{(3)\,l;L}_{Xl_1l_2l_3}(k_1,k_2,k_3)
  \left(l\,l_1\,L\right)_{m}^{\phantom{m}m_1M}
  \left(L\,l_2\,l_3\right)_{M}^{\phantom{M}m_2m_3}
  \\ \times
  C_{l_1m_1}(\hat{\bm{k}}_1) C_{l_2m_2}(\hat{\bm{k}}_2)
  C_{l_3m_3}(\hat{\bm{k}}_3).
  \label{eq:142}
\end{multline}
The last expression contains the tripolar spherical harmonics
\cite{Khersonskii:1988krb}.
Similar to Eq.~(\ref{eq:133}) of bipolar spherical harmonics, we
introduce a notation for the tripolar spherical harmonics,
\begin{multline}
  X^{l_1l_2l_3}_{L;lm}(\hat{\bm{k}}_1,\hat{\bm{k}}_2,\hat{\bm{k}}_3)
  \equiv
  \frac{(-1)^l\,(4\pi)^{3/2}}{\sqrt{\{l\}\{l_1\}\{l_2\}\{l_3\}}}
  \\ \times
  \left\{
    Y_{l_1}(\hat{\bm{k}}_1) \otimes
    \left\{
      Y_{l_2}(\hat{\bm{k}}_2) \otimes
      Y_{l_3}(\hat{\bm{k}}_3)
    \right\}_L
  \right\}_{lm}.
  \label{eq:143}
\end{multline}
The explicit form of the above harmonics is defined by
Eq.~(\ref{eq:408}), and their orthonormality relation,
Eq.~(\ref{eq:412}) and product formula, Eq.~(\ref{eq:413}) are
given in Appendix~\ref{app:SphericalHarmonics} with our notations for
convenience. Thereby Eq.~(\ref{eq:142}) is concisely represented as
\begin{equation}
  c^{(3)}_{Xlm}(\bm{k}_1,\bm{k}_2,\bm{k}_3)
  =
  \sum_{l_1,l_2,l_3,L}
  c^{(3)\,l;L}_{Xl_1l_2l_3}(k_1,k_2,k_3)
  X^{l_1l_2l_3}_{L;lm}(\hat{\bm{k}}_1,\hat{\bm{k}}_2,\hat{\bm{k}}_3).
  \label{eq:144}
\end{equation}

Combining the properties of complex conjugate for the tripolar
spherical harmonics, Eq.~(\ref{eq:411}), and for the renormalized
bias function, Eq.~(\ref{eq:87}) with $n=3$, we have
\begin{equation}
  c^{(3)\,l;L\,*}_{Xl_1l_2l_3}(k_1,k_2,k_3)
  = c^{(3)\,l;L}_{Xl_1l_2l_3}(k_1,k_2,k_3),
  \label{eq:145}
\end{equation}
and thus the reduced function is a real function. Combining the
properties of parity for the tripolar spherical harmonics,
Eq.~(\ref{eq:410}), and for the renormalized bias function,
Eq.~(\ref{eq:100}) with $n=3$, we have
\begin{equation}
  c^{(3)\,l;L}_{Xl_1l_2l_3}(k_1,k_2,k_3)
  \xrightarrow{\mathbb{P}} (-1)^{p_X+l+l_1+l_2+l_3}
  c^{(3)\,l;L}_{Xl_1l_2l_3}(k_1,k_2,k_3).
  \label{eq:146}
\end{equation}
If the Universe is statistically invariant under the parity, the
function is nonzero only when $p_X+l+l_1+l_2+l_3=\mathrm{even}$. The
interchange symmetries of Eq.~(\ref{eq:104}) in terms of invariant
functions in this case are derived from properties of interchanging
arguments in tripolar spherical harmonics. For an interchange of the
last two arguments, $2 \leftrightarrow 3$, we have
\begin{equation}
  X^{l_1l_2l_3}_{L;lm}(\hat{\bm{k}}_1,\hat{\bm{k}}_2,\hat{\bm{k}}_3)
  = (-1)^{l_2+l_3+L}
  X^{l_1l_3l_2}_{L;lm}(\hat{\bm{k}}_1,\hat{\bm{k}}_3,\hat{\bm{k}}_2),
  \label{eq:147}
\end{equation}
just similarly in the case of Eq.~(\ref{eq:138}). For an
interchange of the first two arguments, $1 \leftrightarrow 2$,
however, a recoupling of the $3j$-symbols appears, and we have
\begin{multline}
  X^{l_1l_2l_3}_{L;lm}(\hat{\bm{k}}_1,\hat{\bm{k}}_2,\hat{\bm{k}}_3)
  = (-1)^{l_1+l_2}
  \sum_{L'} \{L'\}
  \begin{Bmatrix}
    l_1 & l & L \\
    l_2 & l_3 & L'
  \end{Bmatrix}
  \\ \times
  X^{l_2l_1l_3}_{L';lm}(\hat{\bm{k}}_2,\hat{\bm{k}}_1,\hat{\bm{k}}_3),
  \label{eq:148}
\end{multline}
where the factor in front of the tripolar spherical harmonics on the
rhs is a $6j$-symbol defined by Eq.~(\ref{eq:434}) or equivalently by
Eq.~(\ref{eq:435}). Useful properties of $6j$-symbols are summarized
in Eqs.~(\ref{eq:436}), (\ref{eq:437}) and (\ref{eq:443}) of
Appendix~\ref{app:3njSymbols}. The above relation of
Eq.~(\ref{eq:148}) is derived by applying a sum rule of the
$3j$-symbols, Eq.~(\ref{eq:443}), to the definition of the tripolar
spherical harmonics, Eq.~(\ref{eq:408}). Correspondingly, the
interchange symmetries of invariant coefficients of
Eq.~(\ref{eq:144}) are given by
\begin{equation}
  c^{(3)l;L}_{Xl_ll_3l_2}(k_1,k_3,k_2)
  = (-1)^{l_2+l_3+L}
  c^{(3)l;L}_{Xl_ll_2l_3}(k_1,k_2,k_3),
  \label{eq:149}
\end{equation}
and
\begin{multline}
  c^{(3)l;L}_{Xl_2l_1l_3}(k_2,k_1,k_3)
  = (-1)^{l_1+l_2}
  \sum_{L'} \{L'\}
  \begin{Bmatrix}
    l_1 & l & L \\
    l_2 & l_3 & L'
  \end{Bmatrix}
  \\ \times
  c^{(3)l;L'}_{Xl_ll_2l_3}(k_1,k_2,k_3).
  \label{eq:150}
\end{multline}
Combining Eqs.~(\ref{eq:149}) and (\ref{eq:150}), all the other
symmetries concerning the permutation of $(1,2,3)$ in the subscripts
of the arguments are straightforwardly obtained. A similar kind of
considerations of the interchange symmetry in a formulation of angular
trispectrum of the cosmic microwave background is found in
Ref.~\cite{Hu:2001fa} to enforce permutation symmetry in the angular
trispectrum.

In the same way, one derives the results for general orders $n > 3$.
The expansion coefficient of order
$n$ is given by
\begin{multline}
  c^{(n)\,l_1\cdots l_n}_{Xlm;m_1\cdots m_n}(k_1,\ldots,k_n)
  \\
  = \sum_{L_2,\ldots,L_{n-1}}
  (-1)^{L_2+\cdots L_{n-1}}
  \sqrt{\{L_2\}\cdots \{L_{n-1}\}}
    \left(l\,l_1\,L_2\right)_{mm_1M_2}
    \\ \times
    \left(L_2\,l_2\,L_3\right)^{M_2}_{\phantom{M_2}m_2M_3}
    \cdots
    \left(L_{n-2}\,l_{n-2}\,L_{n-1}\right)^{M_{n-2}}_{\phantom{M_{n-2}}m_2M_{n-1}}
    \\ \times
    \left(L_{n-1}\,l_{n-1}\,l_n\right)^{M_{n-1}}_{\phantom{M_{n-1}}m_{n-1}m_n}
    c^{(n)\,l;L_2\cdots L_{n-1}}_{Xl_1\cdots l_n}(k_1,\ldots,k_n),
  \label{eq:151}
\end{multline}
where we define
\begin{multline}
  c^{(n)\,l;L_2\cdots L_{n-1}}_{Xl_1\cdots l_n}(k_1,\ldots,k_n)
  \equiv
  (-1)^{l+l_1+\cdots +l_n}
  \\ \times
  (-1)^{L_2+\cdots+L_{n-1}}
  \sqrt{\{L_2\}\cdots\{L_{n-1}\}}
    \left(l\,l_1\,L_2\right)^{mm_1}_{\phantom{mm_1}M_2}
    \\ \times
    \left(L_2\,l_2\,L_3\right)^{M_2m_2}_{\phantom{M_2m_2}M_3}
    \cdots
    \left(L_{n-2}\,l_{n-2}\,L_{n-1}\right)^{M_{n-2}m_{n-2}}_{\phantom{M_{n-2}m_{n-2}}M_{n-1}}
    \\ \times
    \left(L_{n-1}\,l_{n-1}\,l_n\right)^{M_{n-1}m_{n-1}m_n}
    c^{(n)\,l_1\cdots l_n}_{Xlm;m_1\cdots m_n}(k_1,\ldots,,k_n).
  \label{eq:152}
\end{multline}
Thus the azimuthal indices of the decomposed renormalized bias
functions, Eq.~(\ref{eq:109}) are all represented by combinations of
$3j$-symbols, and the physical contents of the renormalized bias
functions are given by the invariant functions,
$c^{(n)\,l;L_2\cdots L_{n-1}}_{Xl_1\cdots l_n}(k_1,\ldots,k_n)$. The
fully rotational symmetry is the reason why the azimuthal indices of
decomposed renormalized bias function,
$c^{(n)\,l;l_1\cdots l_n}_{Xm;m_1\cdots m_n}(\ldots)$ are represented
by the combinations of $3j$-symbols. Because of the symmetry of
$3j$-symbols, we have $m+m_1+\cdots +m_n=0$, which is a manifestation
of the axial symmetry of the rotation around the third axis
$\hat{\mathbf{e}}_3$.

The renormalized bias function of Eq.~(\ref{eq:108}) is given by
\begin{multline}
  c^{(n)}_{Xlm}(\bm{k}_1,\ldots,\bm{k}_n)
  =
  \sum_{\substack{l_1,\ldots,l_n\\L_2,\ldots,L_{n-1}}}
  (-1)^{L_2 + \cdots + L_{n-1}}
  \sqrt{\{L_2\}\cdots\{L_{n-1}\}}
  \\ \times
  c^{(n)\,l;L_2\cdots L_{n-1}}_{Xl_1\cdots l_n}(k_1,\ldots,k_n)
  \left(l\,l_1\,L_2\right)_{m}^{\phantom{m}m_1M_2}
  \\ \times
  \left(L_2\,l_2\,L_3\right)_{M_2}^{\phantom{M_2}m_2M_3} \cdots
  \left(L_{n-2}\,l_{n-2}\,L_{n-1}\right)_{M_{n-2}}^{\phantom{M_{n-2}}m_{n-2}M_{n-1}}
  \\ \times
  \left(L_{n-1}\,l_{n-1}\,l_n\right)_{M_{n-1}}^{\phantom{M_{n-1}}m_{n-1}m_n}
  C_{l_1m_1}(\hat{\bm{k}}_1) \cdots C_{l_nm_n}(\hat{\bm{k}}_n).
  \label{eq:153}
\end{multline}
The last expression contains the polypolar spherical harmonics defined
by Eq.~(\ref{eq:414}), and their orthonormality relation,
Eq.~(\ref{eq:418}) and product formula, Eq.~(\ref{eq:419}) are given
in Appendix~\ref{app:SphericalHarmonics} with our notations for
convenience. Similar to Eqs.~(\ref{eq:133}) and (\ref{eq:143}) of
bipolar and tripolar spherical harmonics, we introduce a notation for
the polypolar spherical harmonics,
\begin{multline}
  X^{l_1\cdots l_n}_{L_2\cdots L_{n-1};lm}
  (\hat{\bm{k}}_1,\ldots,\hat{\bm{k}}_n)
  \equiv
  \frac{(-1)^l\,(4\pi)^{n/2}}{\sqrt{\{l\}\{l_1\}\cdots \{l_n\}}}
  \\ \times
  \left\{
    Y_{l_1}(\hat{\bm{k}}_1) \otimes
    \left\{
      Y_{l_2}(\hat{\bm{k}}_2) \otimes
      \left\{
        \cdots \otimes
        Y_{l_n}(\hat{\bm{k}}_n)
      \right\}_{L_{n-1}} \cdots
    \right\}_{L_2}
  \right\}_{lm},
  \label{eq:154}
\end{multline}
and Eq.~(\ref{eq:153}) is concisely represented as
\begin{multline}
  c^{(n)}_{Xlm}(\bm{k}_1,\cdots,\bm{k}_n)
  =
  \sum_{\substack{l_1,\ldots,l_n\\L_2,\ldots,L_{n-1}}}
  c^{(n)\,l;L_2\cdots L_{n-1}}_{Xl_1\cdots l_n}(k_1,\ldots,k_n)
  \\ \times
  X^{l_1\cdots l_n}_{L_2\cdots L_{n-1};lm}
  (\hat{\bm{k}}_1,\ldots,\hat{\bm{k}}_n).
  \label{eq:155}
\end{multline}

Combining the properties of complex conjugate for the polypolar
spherical harmonics, Eq.~(\ref{eq:417}), and for the renormalized
bias function, Eq.~(\ref{eq:87}), we have
\begin{equation}
  c^{(n)\,l;L_2\cdots L_{n-1}*}_{Xl_1\cdots l_n}(k_1,\ldots,k_n)
  = c^{(n)\,l;L_2\cdots L_{n-1}}_{Xl_1\cdots l_n}(k_1,\ldots,k_n),
  \label{eq:156}
\end{equation}
and thus the reduced function is a real function. Combining the
properties of parity for the polypolar spherical harmonics,
Eq.~(\ref{eq:416}), and for the renormalized bias function,
Eq.~(\ref{eq:100}), we have
\begin{multline}
  c^{(n)\,l;L_2\cdots L_{n-1}}_{Xl_1\cdots l_n}(k_1,\ldots,k_3)
  \\
  \xrightarrow{\mathbb{P}} (-1)^{p_X+l+l_1+\cdots +l_n}
  c^{(n)\,l;L_2\cdots L_{n-1}}_{Xl_1\cdots l_n}(k_1,\ldots,k_3).
  \label{eq:157}
\end{multline}
If the Universe is statistically invariant under the parity, the
function is nonzero only when $p_X+l+l_1+\cdots +l_n=\mathrm{even}$.

The interchange symmetries of Eq.~(\ref{eq:104}) in terms of
invariant function in this case are also derived following a similar
way in the case of $n=3$. For the interchange of the last two
arguments, we have
\begin{multline}
  c^{(n)l;L_2\cdots L_{n-2}L_{n-1}}_{Xl_l\cdots l_{n-2}l_nl_{n-1}}
  (k_1,\ldots,k_{n-2},k_n,k_{n-1})
  \\
  = (-1)^{l_{n-1}+l_n+L_{n-1}}
  c^{(n)l;L_2\cdots L_{n-1}}_{Xl_l\cdots l_n}
  (k_1,\ldots,k_n),
  \label{eq:158}
\end{multline}
and for the interchange of the other adjacent arguments, $i
\leftrightarrow i+1$, we have
\begin{multline}
  c^{(n)\,l;L_2\cdots L_{i-1}L_{i+1}L_iL_{i+2}\cdots L_{n-1}}
  _{Xl_1\cdots l_{i-1}l_{i+1}l_il_{i+2}\cdots l_n}
  (k_1,\ldots,k_{i-1},k_{i+1},k_i,k_{i+2},\ldots,k_n)
  \\
  = (-1)^{l_i+l_{i+1}}
  \sum_{L'} \{L'\}
  \begin{Bmatrix}
    l_i & L_i & L_{i+1} \\
    l_{i+1} & L_{i+2} & L'
  \end{Bmatrix}
  \\ \times
  c^{(n)\,l;L_2\cdots L_iL'L_{i+2}\cdots L_{n-1}}_{Xl_1\cdots l_n}(k_1,\ldots,k_n),
  \label{eq:159}
\end{multline}
where $i=1,\ldots,n-2$. Combining Eqs.~(\ref{eq:158}) and
(\ref{eq:159}), all the other symmetries concerning the permutation
of $(1,\ldots,n)$ in the subscripts of the arguments are
straightforwardly obtained.

\subsection{\label{subsec:Propagators}
  Propagators of tensor fields
}

The renormalized bias functions describe the bias mechanisms in the
physical space, and thus there is not any corresponding concept of
redshift space. However, the apparent clustering in redshift space
introduces anisotropy in the propagators in Eulerian space. In the
following, we separately consider the propagators in real space and
redshift space.

\subsubsection{\label{subsec:PropReal}
  Real space
}

In real space, where the fully rotational symmetry is satisfied even
in Eulerian space, the reduced propagators $\hat{\Gamma}^{(n)}_{Xlm}$
are similarly decomposed into rotationally invariant coefficients as
in the case of renormalized bias functions explained in the previous
subsection. The derivation of the invariant propagators is mostly
parallel to the latter case. In real space, formally replacing the
renormalized bias functions $c^{(n)}_{Xlm}$ with normalized
propagators $\hat{\Gamma}^{(n)}_{Xlm}$ in Sec.~\ref{subsec:RenBiasFn}
is enough to derive the necessary equations of invariant propagators. We
summarize some important equations explicitly below for later use. The
derivations are exactly the same as those of renormalized bias
functions. 

The propagator of the first order, $n=1$, is given by
\begin{equation}
  \hat{\Gamma}^{(1)}_{Xlm}(\bm{k}) =
  \frac{(-1)^l}{\sqrt{2l+1}}\,
  \hat{\Gamma}^{(1)}_{Xl}(k) C_{lm}(\hat{\bm{k}}),
  \label{eq:160}
\end{equation}
with the invariant propagator $\hat{\Gamma}^{(1)}_{Xl}(k)$ which is a
real function. If the Universe is statistically invariant under the
parity, the first-order propagator is nonzero only when $p_X=0$ and
vanishes for pseudotensors. The parity transformation of the invariant
function is given by
\begin{equation}
  \hat{\Gamma}^{(1)}_{Xl}(k) \xrightarrow{\mathbb{P}}
  (-1)^{p_X} \hat{\Gamma}^{(1)}_{Xl}(k).
  \label{eq:161}
\end{equation}

The propagator of second order, $n=2$, is given by
\begin{equation}
  \hat{\Gamma}^{(2)}_{Xlm}(\bm{k}_1,\bm{k}_2)
  = \sum_{l_1,l_2}
  \hat{\Gamma}^{(2)\,l}_{Xl_1l_2}(k_1,k_2)
  X^{l_1l_2}_{lm}(\hat{\bm{k}}_1,\hat{\bm{k}}_2).
  \label{eq:162}
\end{equation}
with the invariant propagator $\hat{\Gamma}^{(2)}_{Xl_1l_2}(k_1,k_2)$
which is a real function. The parity transformation of the invariant
function is given by
\begin{equation}
  \hat{\Gamma}^{(2)\,l}_{Xl_1l_2}(k_1,k_2)
  \xrightarrow{\mathbb{P}} (-1)^{p_X+l+l_1+l_2}
  \hat{\Gamma}^{(2)\,l}_{Xl_1l_2}(k_1,k_2).
  \label{eq:163}
\end{equation}
If the Universe is statistically invariant under the parity, the
function is nonzero only when $p_X+l+l_1+l_2=\mathrm{even}$. The
interchange symmetry for the second-order propagator is given by
\begin{equation}
  \hat{\Gamma}^{(2)\,l}_{Xl_2l_1}(k_2,k_1)
  = (-1)^{l+l_1+l_2}
  \hat{\Gamma}^{(2)\,l}_{Xl_1l_2}(k_1,k_2).
  \label{eq:164}
\end{equation}

The propagator of third order, $n=3$, is given by
\begin{multline}
  \hat{\Gamma}^{(3)}_{Xlm}(\bm{k}_1,\bm{k}_2,\bm{k}_3)
  =
  \sum_{l_1,l_2,l_3,L}
  \hat{\Gamma}^{(3)\,l;L}_{Xl_1l_2l_3}(k_1,k_2,k_3)
  \\ \times
  X^{l_1l_2l_3}_{L;lm}(\hat{\bm{k}}_1,\hat{\bm{k}}_2,\hat{\bm{k}}_3),
  \label{eq:165}
\end{multline}
with the invariant propagator
$\hat{\Gamma}^{(3)\,l;L}_{Xl_1l_2l_3}(k_1,k_2,k_3)$ which is a real
function. The parity transformation of the invariant function is given
by
\begin{equation}
  \hat{\Gamma}^{(3)\,l;L}_{Xl_1l_2l_3}(k_1,k_2,k_3)
  \xrightarrow{\mathbb{P}} (-1)^{p_X+l+l_1+l_2+l_3}
  \hat{\Gamma}^{(3)\,l;L}_{Xl_1l_2l_3}(k_1,k_2,k_3).
  \label{eq:166}
\end{equation}
If the Universe is statistically invariant under the parity, the
function is nonzero only when $p_X+l+l_1+l_2+l_3=\mathrm{even}$. The
interchange symmetries for the third-order propagator are given by
\begin{equation}
  \hat{\Gamma}^{(3)l;L}_{Xl_ll_3l_2}(k_1,k_3,k_2)
  = (-1)^{l_2+l_3+L}
  \hat{\Gamma}^{(3)l;L}_{Xl_ll_2l_3}(k_1,k_2,k_3),
  \label{eq:167}
\end{equation}
and
\begin{multline}
  \hat{\Gamma}^{(3)l;L}_{Xl_2l_1l_3}(k_2,k_1,k_3)
  = (-1)^{l_1+l_2}
  \sum_{L'} \{L'\}
  \begin{Bmatrix}
    l_1 & l & L \\
    l_2 & l_3 & L'
  \end{Bmatrix}
  \\ \times
  \hat{\Gamma}^{(3)l;L'}_{Xl_ll_2l_3}(k_1,k_2,k_3).
  \label{eq:168}
\end{multline}

The propagators of order $n > 3$ are given by
\begin{multline}
  \hat{\Gamma}^{(n)}_{Xlm}(\bm{k}_1,\ldots,\bm{k}_n)
  =
  \sum_{\substack{l_1,\ldots,l_n\\L_2,\ldots,L_{n-1}}}
  \hat{\Gamma}^{(n)\,l;L_2\cdots L_{n-1}}_{Xl_1\cdots l_n}(k_1,\ldots,k_n)
  \\ \times
  X^{l_1\cdots l_n}_{L_2\cdots L_{n-1};lm}
  (\hat{\bm{k}}_1,\ldots,\hat{\bm{k}}_n),
  \label{eq:169}
\end{multline}
with the invariant propagators
$\hat{\Gamma}^{(n)\,l;L_2\cdots L_{n-1}}_{Xl_1\cdots
  l_n}(k_1,\ldots,k_n)$ which are real functions. The parity
transformation is given by
\begin{multline}
  \hat{\Gamma}^{(n)\,l;L_2\cdots L_{n-1}}_{Xl_1\cdots l_n}(k_1,\ldots,k_3)
  \\
  \xrightarrow{\mathbb{P}} (-1)^{p_X+l+l_1+\cdots +l_n}
  \hat{\Gamma}^{(n)\,l;L_2\cdots L_{n-1}}_{Xl_1\cdots l_n}(k_1,\ldots,k_3).
  \label{eq:170}
\end{multline}
The interchange symmetries for the higher-order propagators are given
by
\begin{multline}
  \hat{\Gamma}^{(n)l;L_2\cdots L_{n-2}L_{n-1}}_{Xl_l\cdots l_{n-2}l_nl_{n-1}}
  (k_1,\ldots,k_{n-2},k_n,k_{n-1})
  \\
  = (-1)^{l_{n-1}+l_n+L_{n-1}}
  \hat{\Gamma}^{(n)l;L_2\cdots L_{n-1}}_{Xl_l\cdots l_n}
  (k_1,\ldots,k_n),
  \label{eq:171}
\end{multline}
and
\begin{multline}
  \hat{\Gamma}^{(n)\,l;L_2\cdots L_{i-1}L_{i+1}L_iL_{i+2}\cdots L_{n-1}}
  _{Xl_1\cdots l_{i-1}l_{i+1}l_il_{i+2}\cdots l_n}
  (k_1,\ldots,k_{i-1},k_{i+1},k_i,k_{i+2},\ldots,k_n)
  \\
  = (-1)^{l_i+l_{i+1}}
  \sum_{L'} \{L'\}
  \begin{Bmatrix}
    l_i & L_i & L_{i+1} \\
    l_{i+1} & L_{i+2} & L'
  \end{Bmatrix}
  \\ \times
  \hat{\Gamma}^{(n)\,l;L_2\cdots L_iL'L_{i+2}\cdots
    L_{n-1}}_{Xl_1\cdots l_n}(k_1,\ldots,k_n). 
  \label{eq:172}
\end{multline}
Symmetric properties of all the other permutations are obtained by
combining the above two cases.

\subsubsection{\label{subsubsec:PropRed}
  Redshift space
}

In redshift space, the propagators also depend on the direction of the
line of sight, $\hat{\bm{z}}$. In this paper, we approximately
consider the direction of the line of sight to be fixed in the observed
space. This approximation is usually called the plane-parallel, or
distant-observer approximation, and is known as a good approximation
for most practical purposes. In order to keep the rotational
covariance in the theory, the line of sight, $\hat{\bm{z}}$, is
considered to be arbitrarily directed in the coordinates system.
In literature, it is a common practice to fix the direction of the line of
sight in the third axis $\hat{\bf{e}}_3$ in the distant-observer
approximation. In this paper, we do not fix the direction, and we have
$\hat{\bm{z}} \ne \hat{\mathbf{e}}_3$ in general.

In redshift space, the propagators depend on the direction of the line
of sight $\hat{\bm{z}}$ as well as the wave vectors. We decompose the
directional dependencies on directions of both the line of sight
$\hat{\bm{z}}$ and wave vectors $\hat{\bm{k}}_1,\ldots,\hat{\bm{k}}_n$
by spherical harmonics in a similar manner of Eq.~(\ref{eq:108}) in
the case of renormalized bias functions. However, it turns out to be
often useful {\em not} to expand all the angular dependencies, and
keeping the angular dependencies unexpanded for the sum of the wave
vectors, $\bm{k} =\bm{k}_1 + \cdots + \bm{k}_n$. The sum of wave
vectors corresponds to ``the external wave vector''
\cite{Matsubara:2011ck}, which is not integrated over in the nonlinear
spectra calculated by the perturbation theory. Therefore, the
amplitude $k=|\bm{k}|$ and directional cosine with respect to the line
of sight $\mu=\hat{\bm{z}}\cdot\hat{\bm{k}}$ can be fixed in the
calculation of propagators in redshift space. Some parts of the
directional dependencies of the propagators on the line of sight
$\hat{\bm{z}}$ and the wave vectors
$\hat{\bm{k}}_1,\hat{\bm{k}}_2,\ldots$, which is included in the
parameters $k$ and $\mu$ are optionally not need to be expanded in the
spherical harmonics. While it is sometimes required to expand every
angular dependency into spherical harmonics (e.g., Paper~IV), the
calculations are more complicated than keeping the dependencies
unexpanded in $k$ and $\mu$. In the following, we consider the general
case that the propagators conveniently have an explicit dependence on
$k$ and $\mu$.

The expansion is given by
\begin{multline}
  \hat{\Gamma}^{(n)}_{Xlm}(\bm{k}_1,\ldots\bm{k}_n;\hat{\bm{z}};k,\mu)
  \\
  = \sum_{l_z,l_1,\ldots,l_n}
  \hat{\Gamma}^{(n)\,l_z;l_1\cdots l_n}_{Xlm;m_z;m_1\cdots m_n}(k_1,\ldots,k_n;k,\mu)
  \\ \times
  C_{l_zm_z}^*(\hat{\bm{z}})
  C_{l_1m_1}^*(\hat{\bm{k}}_1) \cdots C_{l_nm_n}^*(\hat{\bm{k}}_n).
  \label{eq:173}
\end{multline}
As the parameters $k$ and $\mu$ are functions of $\hat{\bm{z}}$ and
$\bm{k}_1,\ldots,\bm{k}_n$, the above expansion is not unique in
general. Nevertheless, if which parts of the dependencies on the
parameters $k$ and $\mu$ are consistently fixed in the expression on
the lhs of Eq.~(\ref{eq:173}) throughout, the expansion is unique, at
least artificially. Using the expansion of Eq.~(\ref{eq:173}),
deriving the invariant propagators in redshift space is quite similar
to the previous cases of renormalized bias functions and propagators
in real space. Comparing Eq.~(\ref{eq:173}) with Eq.~(\ref{eq:109}),
the similarity is apparent, while there are additional dependencies on
the direction to the line of sight, $\hat{\bm{z}}$, and also possible
dependencies on $k$ and $\mu$. Bearing these differences in mind, we
apply the similar derivation in Sec.~\ref{subsec:RenBiasFn} to the
present case below.

The transformation under the passive
rotation of Eq.~(\ref{eq:20}) is given by
\begin{multline}
  \hat{\Gamma}^{(n)\,l_z;l_1\cdots l_n}_{Xlm;m_z;m_1\cdots m_n}
  (k_1,\ldots,k_n;k,\mu)
  \\
  \xrightarrow{\mathbb{R}}
  \hat{\Gamma}^{(n)\,l_z;l_1\cdots l_n}_{Xlm';m_z';m_1'\cdots m_n'}
  (k_1,\ldots,k_n;k,\mu)
  D_{(l)m}^{m'}(R)
  \\ \times
  D_{(l_z)m_z}^{m_z'}(R)
  D_{(l_1)m_1}^{m_1'}(R) \cdots
  D_{(l_n)m_n}^{m_n'}(R).
  \label{eq:174}
\end{multline}
When we rotate the direction of the line of sight together with
the coordinate system, the propagators should be invariant under the
rotation of the coordinate system in the Universe which satisfies the
cosmological principle. Thus we have
\begin{multline}
  \hat{\Gamma}^{(n)\,l_z;l_1\cdots l_n}_{Xlm;m_z;m_1\cdots m_n}
  (k_1,\ldots,k_n;k,\mu)
  \\
  = \hat{\Gamma}^{(n)\,l_z;l_1\cdots l_n}_{Xlm';m_z';m_1'\cdots m_n'}
  (k_1,\ldots,k_n;k,\mu)
  \frac{1}{8\pi^2} \int [dR]
  D_{(l)m}^{m'}(R)
  \\ \times
  D_{(l_z)m_z}^{m_z'}(R)
  D_{(l_1)m_1}^{m_1'}(R) \cdots D_{(l_n)m_n}^{m_n'}(R).
  \label{eq:175}
\end{multline}
Following the same procedure below Eq.~(\ref{eq:111}) in the case of
renormalized bias functions, one can represent the dependence of
propagators in redshift space by spherical metric, $3j$-symbols, and
invariant variables. The main difference is that an angular
dependence on the line of sight and dependencies on $k$ and $\mu$
additionally appear in the expansion.

For the propagator of first order, $n=1$, we have $k=k_1$ and one can
omit the dependence on $k_1$. We thus have
\begin{equation}
  \hat{\Gamma}^{(1)\,l_z;l_1}_{Xlm;m_z;m_1}(k,\mu)
  = \left(l\,l_z\,l_1\right)_{mm_zm_1}
  \hat{\Gamma}^{(1)\,l\,l_z}_{Xl_1}(k,\mu),
  \label{eq:176}
\end{equation}
with
\begin{equation}
  \hat{\Gamma}^{(1)l\,l_z}_{Xl_1}(k,\mu)
  \equiv
  (-1)^{l+l_z+l_1}
  \left(l\,l_z\,l_1\right)^{mm_zm_1}
  \hat{\Gamma}^{(1)\,l_z;l_1}_{Xlm;m_z;m_1}(k,\mu).
  \label{eq:177}
\end{equation}
The propagator is given by
\begin{equation}
  \hat{\Gamma}^{(1)}_{Xlm}(\bm{k};\hat{\bm{z}};k,\mu)
  = \sum_{l_z,l_1}
  \hat{\Gamma}^{(1)\,l\,l_z}_{Xl_1}(k,\mu)
  X^{l_zl_1}_{lm}(\hat{\bm{z}},\hat{\bm{k}}).
  \label{eq:178}
\end{equation}
Combining the properties of complex conjugate for the bipolar spherical
harmonics, Eq.~(\ref{eq:404}), and for the propagator in redshift space,
Eq.~(\ref{eq:89}) with $n=1$, we have
\begin{equation}
  \hat{\Gamma}^{(1)l\,l_z\,*}_{Xl_1}(k,\mu)
  = (-1)^{l_z} \hat{\Gamma}^{(1)l\,l_z}_{Xl_1}(k,-\mu).
  \label{eq:179}
\end{equation}
Combining the properties of parity for the bipolar spherical
harmonics, Eq.~(\ref{eq:403}), and for the propagator in redshift
space, Eq.~(\ref{eq:102}) with $n=1$, we have
\begin{equation}
  \hat{\Gamma}^{(1)l\,l_z}_{Xl_1}(k,\mu)
  \xrightarrow{\mathbb{P}}
  (-1)^{p_X+l+l_z+l_1} \hat{\Gamma}^{(1)l\,l_z}_{Xl_1}(k,\mu).
  \label{eq:180}
\end{equation}
If the Universe is statistically invariant under the parity, the
function is nonzero only when $p_X+l+l_z+l_1=\mathrm{even}$.

For the propagator of second order, $n=2$, we have
\begin{multline}
  \hat{\Gamma}^{(2)\,l_z;l_1l_2}_{Xlm;m_z;m_1m_2}(k_1,k_2;k,\mu)
  =
  \sum_L (-1)^L \sqrt{\{L\}}
    \left(l\,l_z\,L\right)_{mm_zM}
    \\ \times
    \left(L\,l_1\,l_2\right)^M_{\phantom{M}m_1m_2}
    \hat{\Gamma}^{(2)\,l\,l_z;L}_{Xl_1l_2}(k_1,k_2;k,\mu),
  \label{eq:181}
\end{multline}
with
\begin{multline}
  \hat{\Gamma}^{(2)\,l\,l_z;L}_{Xl_1l_2}(k_1,k_2;k,\mu)
  \equiv
  (-1)^{l+l_z+l_1+l_2+L} \sqrt{\{L\}}
    \left(l\,l_z\,L\right)^{mm_z}_{\phantom{mm_z}M}
    \\ \times
    \left(L\,l_1\,l_2\right)^{Mm_1m_2}
    \hat{\Gamma}^{(2)\,l\,l_z;l_1l_2}_{Xmm_z;m_1m_2}(k_1,k_2;k,\mu).
  \label{eq:182}
\end{multline}
The propagator is given by
\begin{multline}
  \hat{\Gamma}^{(2)}_{Xlm}(\bm{k}_1,\bm{k}_2;\hat{\bm{z}};k,\mu)
  \\
  =
  \sum_{l_z,l_1,l_2,L}
  \hat{\Gamma}^{(2)\,l\,l_z;L}_{Xl_1l_2}(k_1,k_2;k,\mu)
  X^{l_zl_1l_2}_{L;lm}(\hat{\bm{z}},\hat{\bm{k}}_1,\hat{\bm{k}}_2).
  \label{eq:183}
\end{multline}
Combining the properties of complex conjugate for the tripolar
spherical harmonics, Eq.~(\ref{eq:411}), and for the propagator in
redshift space, Eq.~(\ref{eq:89}) with $n=2$, we have
\begin{equation}
  \hat{\Gamma}^{(2)\,l\,l_z;L\,*}_{Xl_1l_2}(k_1,k_2;k,\mu)
  = (-1)^{l_z} \hat{\Gamma}^{(2)\,l\,l_z;L}_{Xl_1l_2}(k_1,k_2;k,-\mu),
  \label{eq:184}
\end{equation}
Combining the properties of parity for the tripolar spherical
harmonics, Eq.~(\ref{eq:410}), and for the propagator in redshift
space, Eq.~(\ref{eq:102}) with $n=2$, we have
\begin{equation}
  \hat{\Gamma}^{(2)\,l\,l_z;L}_{Xl_1l_2}(k_1,k_2;k,\mu)
  \xrightarrow{\mathbb{P}}
  (-1)^{p_X+l+l_z+l_1+l_2}
  \hat{\Gamma}^{(2)\,l\,l_z;L}_{Xl_1l_2}(k_1,k_2;k,\mu),
  \label{eq:185}
\end{equation}
If the Universe is statistically invariant under the parity, the
function is nonzero only when $p_X+l+l_z+l_1+l_2=\mathrm{even}$. The
interchange symmetry for the second-order propagator is given by
\begin{equation}
  \hat{\Gamma}^{(2)\,l\,l_z;L}_{Xl_2l_1}(k_2,k_1;k,\mu)
  = (-1)^{l_1+l_2+L}
  \hat{\Gamma}^{(2)\,l\,l_z;L}_{Xl_1l_2}(k_1,k_2;k,\mu).
  \label{eq:186}
\end{equation}

For the propagator of higher orders, $n \geq 3$, we have
\begin{multline}
  \hat{\Gamma}^{(n)\,l_z;l_1\cdots l_n}_{Xlm;m_z;m_1\cdots m_n}
  (k_1,\ldots,k_n;k,\mu)
  \\
  =
  \sum_{L_1,\ldots,L_{n-1}}
  (-1)^{L_1+\cdots +L_{n-1}}
    \sqrt{\{L_1\}\cdots\{L_{n-1}\}}
    \left(l\,l_z\,L_1\right)_{mm_zM_1}
    \\ \times
    \left(L_1\,l_1\,L_2\right)^{M_1}_{\phantom{M_1}m_1M_2}
    \cdots
    \left(L_{n-2}\,l_{n-2}\,L_{n-1}\right)^{M_{n-2}}_{\phantom{M_{n-2}}m_{n-2}M_{n-1}}
    \\ \times
    \left(L_{n-1}\,l_{n-1}\,l_n\right)^{M_{n-1}}_{\phantom{M_{n-1}}m_{n-1}m_n}
    \hat{\Gamma}^{(n)\,l\,l_z;L_1\cdots L_{n-1}}_{Xl_1\cdots l_n}
    (k_1,\ldots,k_n;k,\mu),
  \label{eq:187}
\end{multline}
with
\begin{multline}
  \hat{\Gamma}^{(n)\,l\,l_z;L_1\cdots L_{n-1}}_{Xl_1\cdots l_n}
  (k_1,\ldots,k_n;k,\mu)
  \equiv
  (-1)^{l+l_z+l_1+\cdots +l_n}
  \\ \times
  (-1)^{L_1+\cdots +L_{n-1}} \sqrt{\{L_1\}\cdots\{L_{n-1}\}}
    \left(l\,l_z\,L_1\right)^{mm_z}_{\phantom{mm_z}M_1}
    \\ \times
    \left(L_1\,l_1\,L_2\right)^{M_1m_1}_{\phantom{M_1m_1}M_2}
    \cdots
    \left(L_{n-2}\,l_{n-2}\,L_{n-1}\right)^{M_{n-2}m_{n-2}}_{\phantom{M_{n-2}m_{n-2}}M_{n-1}}
    \\ \times
    \left(L_{n-1}\,l_{n-1}\,l_n\right)^{M_{n-1}m_{n-1}m_n}
    \hat{\Gamma}^{(n)\,l_z;l_1\cdots l_n}_{Xlm;m_z;m_1\cdots m_n}
    (k_1,\ldots,k_n;k,\mu),
  \label{eq:188}
\end{multline}
The propagator is given by
\begin{multline}
  \hat{\Gamma}^{(n)}_{Xlm}(\bm{k}_1,\ldots,\bm{k}_n;\hat{\bm{z}};k,\mu)
  \\ =
  \sum_{\substack{l_z,l_1,\ldots,l_n\\L_1,\ldots,L_{n-1}}}
  \hat{\Gamma}^{(n)\,l\,l_z;L_1\cdots L_{n-1}}_{Xl_1\cdots l_n}
  (k_1,\ldots,k_n;k,\mu)
  \\ \times
  X^{l_zl_1\cdots l_n}_{L_1\cdots L_{n-1};lm}
  (\hat{\bm{z}},\hat{\bm{k}}_1,\ldots,\hat{\bm{k}}_n).
  \label{eq:189}
\end{multline}
Combining the properties of complex conjugate for the polypolar
spherical harmonics, Eq.~(\ref{eq:417}), and for the propagator in
redshift space, Eq.~(\ref{eq:89}), we have
\begin{multline}
  \hat{\Gamma}^{(n)\,l\,l_z;L_1\cdots L_{n-1}\,*}_{Xl_1\cdots l_n}
  (k_1,\ldots,k_n;k,\mu)
  \\
  = (-1)^{l_z}
  \hat{\Gamma}^{(n)\,l\,l_z;L_1\cdots L_{n-1}}_{Xl_1\cdots l_n}
  (k_1,\ldots,k_n;k,-\mu).
  \label{eq:190}
\end{multline}
Combining the properties of parity for the polypolar spherical
harmonics, Eq.~(\ref{eq:416}), and for the propagator in redshift
space, Eq.~(\ref{eq:102}), we have
\begin{multline}
  \hat{\Gamma}^{(n)\,l\,l_z;L_1\cdots L_{n-1}}_{Xl_1\cdots l_n}
  (k_1,\ldots,k_n;k,\mu)
  \\
  \xrightarrow{\mathbb{P}}
  (-1)^{p_X+l+l_1+\cdots +l_n}
  \hat{\Gamma}^{(n)\,l\,l_z;L_1\cdots L_{n-1}}_{Xl_1\cdots l_n}
  (k_1,\ldots,k_n;k,\mu).
  \label{eq:191}
\end{multline}
When the Universe is statistically invariant under the parity, the
function is nonzero only when
$p_X+l+l_z+l_1+\cdots +l_n=\mathrm{even}$. Corresponding to
Eqs.~(\ref{eq:158}) and (\ref{eq:159}) of the renormalized bias
function, the interchange symmetries for the higher-order propagators
are given by
\begin{multline}
  \hat{\Gamma}^{(n)l\,l_z;L_1\cdots L_{n-2}L_{n-1}}_{Xl_l\cdots l_{n-2}l_nl_{n-1}}
  (k_1,\ldots,k_{n-2},k_n,k_{n-1};k,\mu)
  \\
  = (-1)^{l_{n-1}+l_n+L_{n-1}}
  \hat{\Gamma}^{(n)l\,l_z;L_1\cdots L_{n-1}}_{Xl_l\cdots l_n}
  (k_1,\ldots,k_n;k,\mu),
  \label{eq:192}
\end{multline}
and
\begin{multline}
  \hat{\Gamma}^{(n)\,l\,l_z;L_1\cdots L_{i+1}L_i\cdots L_{n-1}}
  _{Xl_1\cdots l_{i-1}l_{i+1}l_il_{i+2}\cdots l_n}
  (k_1,\ldots,k_{i+1},k_i,\ldots,k_n;k,\mu)
  \\
  = (-1)^{l_i+l_{i+1}}
  \sum_{L'} (2L'+1)
  \begin{Bmatrix}
    l_i & L_i & L_{i+1} \\
    l_{i+1} & L_{i+2} & L'
  \end{Bmatrix}
  \\ \times
  \hat{\Gamma}^{(n)\,l\,l_z;L_1\cdots L_iL'L_{i+2}\cdots L_{n-1}}_{Xl_1\cdots l_n}
  (k_1,\ldots,k_n;k,\mu),
  \label{eq:193}
\end{multline}
where $i\geq 2$. In the invariant propagator on the lhs of
Eq.~(\ref{eq:193}), the order of only two arguments $k_i$ and
$k_{i+1}$ is reversed.

\subsection{Sample calculations of decomposed propagators
  \label{subsec:LowProp}
}

One can straightforwardly derive explicit forms of reduced propagators
with the formal decomposition as explained above. A simple way to
derive the invariant coefficients is that we represent the angular
dependence of the propagators in terms of the spherical harmonics and
polypolar spherical harmonics, and read off the coefficients from the
resulting expressions.

\subsubsection{Tree-level first-order propagator in real space}

The simplest example is the tree-level approximation of first-order
propagator of Eq.~(\ref{eq:76}):
\begin{equation}
  \hat{\Gamma}_{Xlm}^{(1)}(\bm{k})
  = c_{Xlm}^{(1)}(\bm{k}) +
  \left[\bm{k}\cdot\bm{L}_1(\bm{k})\right] c_{Xlm}^{(0)}.
  \label{eq:194}
\end{equation}
The first-order Lagrangian kernel $\bm{L}_1(\bm{k})$ is given by
Eq.~(\ref{eq:79}), and we have $\bm{k}\cdot\bm{L}_1(\bm{k}) = 1$. The
relevant renormalized bias functions are represented by invariant
quantities through Eqs.~(\ref{eq:121}) and (\ref{eq:127}). Therefore
the propagator of Eq.~(\ref{eq:194}) is represented in a form of
expansion with spherical harmonics as
\begin{equation}
  \hat{\Gamma}_{Xlm}^{(1)}(\bm{k}) =
  \frac{(-1)^l}{\sqrt{2l+1}}
  \left[ c^{(1)}_{Xl}(k)
      + \delta_{l0} c^{(0)}_X 
    \right]
    C_{lm}(\hat{\bm{k}}).
  \label{eq:195}
\end{equation}
Comparing this expression with Eq.~(\ref{eq:160}), we have
\begin{equation}
  \hat{\Gamma}^{(1)}_{Xl}(k) =
  c^{(1)}_{Xl}(k) + \delta_{l0} c^{(0)}_X.
  \label{eq:196}
\end{equation}
When the Universe is statistically symmetric under the parity, the rhs
of Eq.~(\ref{eq:196}) is nonzero only for normal tensors, $p_X=0$, as
discussed in Sec.~\ref{subsec:RenBiasFn} and also the lhs consistently
vanishes as discussed in Sec.~\ref{subsec:Propagators}.

\subsubsection{Tree-level first-order propagator in redshift space
 with the complete expansion}

In redshift space, the tree-level approximation of the first-order
propagator is given by
\begin{equation}
  \hat{\Gamma}_{Xlm}^{(1)}(\bm{k};\hat{\bm{z}})
  = c_{Xlm}^{(1)}(\bm{k}) +
  \left[\bm{k}\cdot\bm{L}^\mathrm{s}_1(\bm{k};\hat{\bm{z}})\right]
  c_{Xlm}^{(0)},
  \label{eq:197}
\end{equation}
where the displacement kernel is replaced by the one in redshift space
as in Eq.~(\ref{eq:83}), and the relevant factor in the above
equation is given by
\begin{equation}
  \bm{k}\cdot\bm{L}^\mathrm{s}_1(\bm{k};\hat{\bm{z}})
  = 1 + f (\hat{\bm{z}}\cdot\hat{\bm{k}})^2
  = 1 + \frac{f}{3}
  + \frac{2f}{3} \mathit{P}_2(\hat{\bm{z}}\cdot\hat{\bm{k}}),
  \label{eq:198}
\end{equation}
where $\mathit{P}_2(x)=(3x^2-1)/2$ is the second Legendre polynomial,
$\mathit{P}_l(x)$ with $l=2$.

First, we explain step-by-step calculations to obtain the invariant
propagator in the case that the dependence of $\mu$ is completely
expanded into spherical harmonics. The following calculations can be
regarded as a prototype for more involved calculations in later
sections and in a subsequent Paper~II. The Legendre polynomials
$\mathit{P}_l(\hat{\bm{z}}\cdot\hat{\bm{k}})$ can be decomposed into a
superposition of products of spherical harmonics by the addition
theorem of Eq.~(\ref{eq:395}), and represented by a special case of
bipolar spherical harmonics as
\begin{equation}
  \mathit{P}_l(\hat{\bm{z}}\cdot\hat{\bm{k}})
    = (-1)^l \sqrt{2l+1}\, X^{ll}_{00}(\hat{\bm{z}},\hat{\bm{k}}),
  \label{eq:199}
\end{equation}
where the last function on the rhs is a special case of normalized
bipolar spherical harmonics given by Eq.~(\ref{eq:133}). 

Combining Eqs.~(\ref{eq:197})--(\ref{eq:199}) above, we derive
\begin{multline}
  \hat{\Gamma}^{(1)}_{Xlm}(\bm{k};\hat{\bm{z}})
  = \frac{(-1)^l}{\sqrt{2l+1}}
  c^{(1)}_{Xl}(k) C_{lm}(\hat{\bm{k}})
  \\
  +
  \delta_{l0} \delta_{m0} c^{(0)}_X
  \left[
    1 + \frac{f}{3} + \frac{2f}{3} \sqrt{5}\,
    X^{22}_{00}(\hat{\bm{z}},\hat{\bm{k}})
  \right].
  \label{eq:200}
\end{multline}
Each term of the above equation can be represented by special cases of
bipolar spherical harmonics, because we have
\begin{align}
  \delta_{l0} \delta_{m0}
  &= \delta_{l0}\,X^{00}_{lm}(\hat{\bm{z}},\hat{\bm{k}}),
    \label{eq:201}\\
  C_{lm}(\hat{\bm{k}})
  &= (-1)^l\sqrt{2l+1}\,
    X^{0l}_{lm}(\hat{\bm{z}},\hat{\bm{k}}),
    \label{eq:202}
\end{align}
as straightforwardly shown from the definitions. One can also rewrite
the bipolar spherical harmonics appearing in Eq.~(\ref{eq:200}) as
\begin{equation}
  \delta_{l0} \delta_{m0}
  X^{22}_{00}(\hat{\bm{z}},\hat{\bm{k}})
  =  \delta_{l0}
  X^{22}_{lm}(\hat{\bm{z}},\hat{\bm{k}}).
  \label{eq:203}
\end{equation}
Substituting the above identities into Eq.~(\ref{eq:200}), one can
readily read off the coefficient of the bipolar spherical harmonics
$X^{l\,l_z}_{lm}(\hat{\bm{z}},\hat{\bm{k}})$ of
Eq.~(\ref{eq:178}) as
\begin{multline}
  \hat{\Gamma}^{(1)l\,l_z}_{Xl_1}(k) =
  \delta_{l_z0}\delta_{l_1l}\, c^{(1)}_{Xl}(k)
  \\
  + \delta_{l0} c^{(0)}_X
  \left[
    \left(1 + \frac{f}{3}\right)
    \delta_{l_z0}\delta_{l_10} +
    \frac{2\sqrt{5}f}{3} \delta_{l_z2}\delta_{l_12}
  \right].
  \label{eq:204}
\end{multline}
Putting $f=0$, $l_z=0$ and $l_1=l$ in the above equation consistently
recovers the result in real space of Eq.~(\ref{eq:196}). When the
Universe is statistically invariant under the parity, the rhs of
Eq.~(\ref{eq:204}) only nonzero when $p_X=0$ as discussed in
Sec.~\ref{subsec:RenBiasFn}, and apparently the rhs is nonzero only
for $l+l_z+l_1=\mathrm{even}$. Thus, the lhs is nonzero only when
$p_X+l+l_z+l_1=\mathrm{even}$, which is consistent with the discussion
in Sec.~\ref{subsec:Propagators}. The same result of
Eq.~(\ref{eq:204}) can also be derived more straightforwardly by using
an orthonormality relation of the bipolar spherical harmonics given in
Eq.~(\ref{eq:405}). According to this orthonormality relation, the
Eq.~(\ref{eq:178}) is inverted as
\begin{equation}
  \hat{\Gamma}^{(1)l\,l_z}_{Xl_1}(k)
  = (-1)^{l+l_z+l_1}\{l\}\{l_z\}\{l_1\}
  \int \frac{d^2\hat{z}}{4\pi}\frac{d^2\hat{k}}{4\pi}\,
  \hat{\Gamma}^{(1)}_{Xl0}(\bm{k};\hat{\bm{z}})
  X^{l_zl_1}_{l0}(\hat{\bm{z}},\hat{\bm{k}}).
  \label{eq:205}
\end{equation}
Substituting Eq.~(\ref{eq:200}) into the above equation, using
Eqs.~(\ref{eq:201})--(\ref{eq:203}) with the orthonormal relation
of Eq.~(\ref{eq:405}), we derive the same result of
Eq.~(\ref{eq:204}) as well.

\subsubsection{Tree-level first-order propagator in redshift space
 with a partial expansion}

The above result of Eq.~(\ref{eq:204}) for the first-order
invariant propagator in redshift space is obtained by completely
expanding the angular dependencies in the propagator. As explained in
Sec.~\ref{subsubsec:PropRed}, the complete expansion is required in
some situations. In some other simple situations, angular dependencies
that are included in the external wave vector can optionally be kept
unexpanded to make the calculations simpler. In the case of the
tree-level first-order propagator, the argument of the propagator in
Eq.~(\ref{eq:197}) is itself the external wave vector.

Combining Eqs.~(\ref{eq:197}) and (\ref{eq:198}), we trivially have
\begin{equation}
  \hat{\Gamma}_{Xlm}^{(1)}(\bm{k})
  = c_{Xlm}^{(1)}(\bm{k}) + \left(1  + f\mu^2\right) c_{Xlm}^{(0)}.
  \label{eq:206}
\end{equation}
Substituting Eqs.~(\ref{eq:121}), (\ref{eq:127}), (\ref{eq:201}), and
(\ref{eq:202}) into the above equation, the invariant propagator can
be read off from Eq.~(\ref{eq:178}), resulting in
\begin{equation}
  \hat{\Gamma}^{(1)l\,l_z}_{Xl_1}(k,\mu) =
  \delta_{l_z0} \delta_{l_1l}
  \left[
    c^{(1)}_{Xl}(k) + \delta_{l0} (1+f\mu^2) c^{(0)}_X
  \right].
  \label{eq:207}
\end{equation}
The same result is also derived by substituting Eq.~(\ref{eq:206})
into Eq.~(\ref{eq:205}), formally regarding as if the variable $\mu$
does not explicitly depend on $\hat{\bm{k}}$ and $\hat{\bm{z}}$.

\subsubsection{Tree-level second-order propagator in real space}

For another example of calculating the invariant propagators, we
consider the second-order propagator $\hat{\Gamma}^{(2)}_{Xlm}$ in
real space, Eq.~(\ref{eq:162}). The calculation in this case is
quite similar to the derivation of Eq.~(\ref{eq:204}). With the
tree-level approximation, the second-order propagator is given by
Eq.~(\ref{eq:77}). Applying a similar procedure as in the above
derivation using Eqs.~(\ref{eq:199}) and
(\ref{eq:201})--(\ref{eq:203}), the rhs of Eq.~(\ref{eq:77}) is
represented in a form of the rhs of Eq.~(\ref{eq:162}), and we can
read off the invariant coefficient from the resulting expression.

In the course of calculation, one needs to represent a product of
spherical harmonics of the same argument through the formula given by
Eq.~(\ref{eq:399}). The coefficient of the spherical harmonics in
this formula is known as the Gaunt integral. Similarly to the
simplified notation for the $3j$-symbols we introduced above, it is
useful to use a simplified symbol for the Gaunt integral
$[l_1\,l_2\,l_3]_{m_1m_2m_3}$ defined by Eq.~(\ref{eq:428}) in our
notation. We consider the azimuthal indices in the Gaunt integral can
be raised and lowered by the spherical metric, such as
\begin{equation}
  \left[l_1\,l_2\,l_3\right]^{m_1m_2}_{\phantom{m_1m_2}m_3}
  = g_{(l_1)}^{m_1m_1'} g_{(l_2)}^{m_2m_2'}
  \left[l_1\,l_2\,l_3\right]_{m_1'm_2'm_3},
  \label{eq:208}
\end{equation}
and so forth. The Gaunt integral of Eq.~(\ref{eq:428}) is nonzero only
when $l_1+l_2+l_3$ is an even number. With this notation,
Eq.~(\ref{eq:399}) is simply represented by
\begin{equation}
  C_{l_1m_1}(\hat{\bm{k}})\,C_{l_2m_2}(\hat{\bm{k}})
  = \sum_{l} (2l+1) \left[l_1\,l_2\,l\right]_{m_1m_2}^{\phantom{m_1m_2}m}
  C_{lm}(\hat{\bm{k}}).
  \label{eq:209}
\end{equation}
Properties of the Gaunt integral, Eqs.~(\ref{eq:428})--(\ref{eq:433})
are useful in analytical deductions of invariant propagators.

Following the procedures above to derive the invariant propagator of
first order, and additionally using the formula of
Eq.~(\ref{eq:209}), the invariant propagator of the second order can be
straightforwardly derived. The result in real space is given by
\begin{multline}
  \hat{\Gamma}^{(2)l}_{Xl_1l_2}(k_1,k_2) =
  c^{(2)l}_{Xl_1l_2}(k_1,k_2)
  \\ +
  c^{(1)}_{Xl}(k_1)
  \left[
    \delta_{l_1l}\delta_{l_20}
    + \frac{k_1}{k_2} \frac{(-1)^l}{\sqrt{\{l\}}}\,\{l_1\}
    \begin{pmatrix}
      l & l_1 & 1 \\
      0 & 0 & 0
    \end{pmatrix}
    \delta_{l_21}
  \right]
  \\ +
  c^{(1)}_{Xl}(k_2)
  \left[
    \delta_{l_10}\delta_{l_2l}
    + \frac{k_2}{k_1} \frac{(-1)^l}{\sqrt{\{l\}}}\,\{l_2\}
    \begin{pmatrix}
      l & l_2 & 1 \\
      0 & 0 & 0
    \end{pmatrix}
    \delta_{l_11}
  \right]
  \\
  + \delta_{l0} c^{(0)}_X
  \left[
    \frac{34}{21}\delta_{l_10}\delta_{l_20}
    - \sqrt{3}
    \left( \frac{k_1}{k_2} + \frac{k_2}{k_1} \right)
    \delta_{l_11} \delta_{l_21}
    \right.
      \\
    \left.
    + \frac{8\sqrt{5}}{21} \delta_{l_12} \delta_{l_22}
    \right].
  \label{eq:210}
\end{multline}
If the Universe is statistically invariant under the parity, the terms
on the rhs of Eq.~(\ref{eq:210}) except the first term is nonzero only
for normal tensors, $p_X=0$, as discussed in
Sec.~\ref{subsec:RenBiasFn}, and apparently nonzero only for
$l+l_1+l_2=\mathrm{even}$. The first term on the rhs is nonzero only
for $p_X+l+l_1+l_2=\mathrm{even}$. The lhs is nonzero only for
$p_X+l+l_1+l_2=\mathrm{even}$, and therefore the properties are
consistent with each other. The corresponding result in redshift space
is explicitly given in Paper~II \cite{PaperII}.

Other propagators to arbitrary orders, both in real space and redshift
space, can be calculated in a similar way as illustrated above,
although the calculations are more or less demanding in the case of
calculation including higher-order perturbations.

\section{The spectra of tensor fields in perturbation theory
  \label{sec:TensorSpectrum}
}

In this section, we consider how one can calculate the power spectrum
and higher-order spectra such as the bispectrum and so forth, for
tensor fields in general.

\subsection{The power spectrum
  \label{subsec:PowerSpec}
}

\subsubsection{\label{subsubsec:InvPS}
  The power spectrum in real space
}

The power spectrum $P^{(l_1l_2)}_{X_1X_2\,m_1m_2}(\bm{k})$ of
irreducible components of tensor fields $F_{Xlm}(\bm{k})$ can be
defined by
\begin{multline}
  \left\langle
    F_{X_1l_1m_1}(\bm{k}_1) F_{X_2l_2m_2}(\bm{k}_2)
  \right\rangle_\mathrm{c}
  \\
  =
  (2\pi)^3\delta_\mathrm{D}^3(\bm{k}_1+\bm{k}_2)
  P^{(l_1l_2)}_{X_1X_2m_1m_2}(\bm{k}_1),
  \label{eq:211}
\end{multline}
where $\langle\cdots\rangle_\mathrm{c}$ indicates the two-point
cumulant, or the connected part, and the appearance of the delta
function is due to translational symmetry. We consider generally the
cross power spectra between two fields, $X_1$ and $X_2$ with
irreducible components $l_1$ and $l_2$, respectively. The autopower
spectrum is straightforwardly obtained by putting $X_1 = X_2 = X$.

The power spectrum $P^{(l_1l_2)}_{X_1X_2m_1m_2}(\bm{k})$ defined above
depends on the coordinates system, as it has azimuthal indices and
depends on the direction of wave vector $\bm{k}$ in the arguments. One
can also construct the reduced power spectrum which is invariant under
the rotation, in a similar manner to introducing the invariant
propagators in Sec.~\ref{subsec:Propagators}.

First, we consider the power spectrum in real space. The directional
dependence on the wave vector can be expanded by spherical harmonics
as
\begin{equation}
  P^{(l_1l_2)}_{X_1X_2m_1m_2}(\bm{k}) =
  \sum_l
  P^{l_1l_2;l}_{X_1X_2m_1m_2;m}(k)\,C^*_{lm}(\hat{\bm{k}}).
  \label{eq:212}
\end{equation}
Because of the rotational transformations of Eqs.~(\ref{eq:93}) and
(\ref{eq:23}), and the rotational invariance of the delta function in
the definition of the power spectrum, Eq.~(\ref{eq:211}), the
rotational transformation of the expansion coefficient is given by
\begin{equation}
  P^{l_1l_2;l}_{X_1X_2m_1m_2;m}(k) \xrightarrow{\mathbb{R}}
  P^{l_1l_2;l}_{X_1X_2m_1'm_2';m'}(k)
  D^{m_1'}_{(l_1)m_1}(R) D^{m_2'}_{(l_2)m_2}(R) D^{m'}_{(l)m}(R).
  \label{eq:213}
\end{equation}
When the Universe is statistically isotropic, the functional form of
the power spectrum should not depend on the choice of coordinates
system. Following the same procedure as in
Sec.~\ref{subsec:RenBiasFn}, the power spectrum in a statistically
isotropic Universe is shown to have a form,
\begin{equation}
  P^{(l_1l_2)}_{X_1X_2m_1m_2}(\bm{k}) =
  i^{l_1+l_2}
  \sum_l
  \left(l_1\,l_2\,l\right)_{m_1m_2}^{\phantom{m_1m_2}m}
  C_{lm}(\hat{\bm{k}}) P^{l_1l_2;l}_{X_1X_2}(k),
    \label{eq:214}
\end{equation}
where
\begin{equation}
  P^{l_1l_2;l}_{X_1X_2}(k) =
  i^{l_1+l_2}(-1)^l
  \left(l_1\,l_2\,l\right)^{m_1m_2m}
  P^{l_1l_2;l}_{X_1X_2m_1m_2;m}(k)
  \label{eq:215}
\end{equation}
is the invariant power spectrum. The physical degrees of freedom of
the power spectrum are represented by the last invariant spectrum,
$P^{l_1l_2;l}_{X_1X_2}(k)$. There are a finite number of independent
functions of power spectra which depend only on the modulus of the
wave vector $k=|\bm{k}|$ of Fourier modes, due to the triangle
inequality of the $3j$-symbol, $|l_1-l_2| \leq l \leq l_1+l_2$ for a
fixed set of integers $(l_1,l_2)$. For example, when $l_1=l_2=2$, only
$l=0,1,\ldots, 4$ are allowed and there are five independent
components in the power spectrum.\footnote{In Ref.~\cite{Vlah:2019byq},
  the power spectrum $P_{22}^{(q)}(k)$ of intrinsic alignment is
  defined in a coordinate-dependent way, where $q=0,\pm 1,\pm2$. The
  degree of freedom of their power spectrum is 5 and is consistent
  with ours.}

\subsubsection{Symmetries in real space}

According to Eqs.~(\ref{eq:86}) and (\ref{eq:211}), the complex
conjugate of the power spectrum is given by
\begin{equation}
  P^{(l_1l_2)*}_{X_1X_2m_1m_2}(\bm{k}) =
  P^{(l_1l_2)m_1m_2}_{X_1X_2}(-\bm{k}).
  \label{eq:216}
\end{equation}
Using the property of the $3j$-symbol, Eq.~(\ref{eq:421}), the
symmetry of the complex conjugate is shown to be given by
\begin{equation}
  P^{l_1l_2;l\,*}_{X_1X_2}(k) = P^{l_1l_2;l}_{X_1X_2}(k),
  \label{eq:217}
\end{equation}
and thus the invariant power spectrum is a real function.
Additionally, the parity transformation of the power spectrum is given
by
\begin{equation}
  P^{(l_1l_2)}_{X_1X_2m_1m_2}(\bm{k}) \xrightarrow{\mathbb{P}}
  (-1)^{p_{X_1}+p_{X_2}+l_1+l_2} P^{(l_1l_2)}_{X_1X_2m_1m_2}(-\bm{k}),
  \label{eq:218}
\end{equation}
because of Eqs.~(\ref{eq:99}) and (\ref{eq:211}). Accordingly, the
parity transformation of the invariant power spectrum is given by
\begin{equation}
  P^{l_1l_2;l}_{X_1X_2}(k) \xrightarrow{\mathbb{P}}
  (-1)^{p_{X_1}+p_{X_2}+l_1+l_2+l} P^{l_1l_2;l}_{X_1X_2}(k).
  \label{eq:219}
\end{equation}
If the Universe is statistically invariant under the parity, the power
spectrum is invariant under the parity transformation, we thus have
\begin{equation}
  p_{X_1} + p_{X_2} + l_1 + l_2 + l = \mathrm{even}.
  \label{eq:220}
\end{equation}
Together with the triangle inequality,
$|l_1-l_2| \leq l \leq l_1+l_2$, mentioned above, the numbers of
independent components in the invariant power spectrum
$P^{l_1l_2;l}_{X_1X_2}(k)$ are further reduced in the presence of
parity symmetry. For example, when $l_1=l_2=2$ and
$p_{X_1}=p_{X_2}=0$, only $l=0,2,4$ are allowed and there are three
independent components in the power spectrum.\footnote{In
  Ref.~\cite{Vlah:2019byq}, their power spectrum of intrinsic
  alignment satisfies $P_{22}^{(q)}(k)=P_{22}^{(-q)}(k)$ in the
  presence of parity symmetry, and thus only three components,
  $q=0,1,2$, are independent. Again, the degree of freedom of theirs
  is consistent with ours.} The power spectrum in real space has an
interchange symmetry,
\begin{equation}
  P^{(l_2l_1)}_{X_2X_1m_2m_1}(\bm{k})
  = P^{(l_1l_2)}_{X_1X_2m_1m_2}(-\bm{k}),
  \label{eq:221}
\end{equation}
and the corresponding symmetry for the invariant spectrum is given by
\begin{equation}
  P^{l_2l_1;l}_{X_2X_1}(k) =
  (-1)^{l_1+l_2} P^{l_1l_2;l}_{X_1X_2}(k).
  \label{eq:222}
\end{equation}

\subsubsection{The power spectrum in redshift space}

Next, we consider the power spectrum in redshift space. The directional
dependence of the wave vector and the line of sight can be expanded by
spherical harmonics as
\begin{equation}
  P^{(l_1l_2)}_{X_1X_2m_1m_2}(\bm{k};\hat{\bm{z}}) =
  \sum_{l,l_z}
  P^{l_1l_2;l\,l_z}_{X_1X_2m_1m_2;mm_z}(k,\mu)
  C^*_{l_zm_z}(\hat{\bm{z}})
  C^*_{lm}(\hat{\bm{k}}).
  \label{eq:223}
\end{equation}
Corresponding to the discussion about the propagator in
Sec.~\ref{subsubsec:PropRed}, some part of the dependence on
directional cosine $\mu = \hat{\bm{z}}\cdot\hat{\bm{k}}$ can be kept
unexpanded, and we generally include the argument $\mu$ in the
coefficients in the above expansion. When the dependence on $\mu$ is
completely expanded into the spherical harmonics, the argument $\mu$
is absent, and the expansion of Eq.~(\ref{eq:223}) is unique.
Otherwise, unless which dependence on $\mu$ is specified and fixed,
the expansion is not unique.

Following similar considerations of rotational symmetry in the rest of
this paper, e.g., in deriving Eq.~(\ref{eq:178}), the power spectrum
in the statistically isotropic Universe is shown to have a form,
\begin{multline}
  P^{(l_1l_2)}_{X_1X_2m_1m_2}(\bm{k};\hat{\bm{z}}) =
    i^{l_1+l_2}
  \sum_L
  (-1)^L \sqrt{\{L\}}
  \left(l_1\,l_2\,L\right)_{m_1m_2}^{\phantom{m_1m_2}M}
  \\ \times
  \sum_{l,l_z}
  X^{l_zl}_{LM}(\hat{\bm{z}},\hat{\bm{k}})
  P^{l_1l_2;l\,l_z;L}_{X_1X_2}(k,\mu),
  \label{eq:224}
\end{multline}
where
\begin{multline}
  P^{l_1l_2;l\,l_z;L}_{X_1X_2}(k,\mu) =
  i^{l_1+l_2}(-1)^{l+l_z+L}
  \sqrt{\{L\}}
  \\ \times
  \left(l_1\,l_2\,L\right)^{m_1m_2}_{\phantom{m_1m_2}M}
  \left(L\,l\,l_z\right)^{Mmm_z}
  P^{l_1l_2;l\,l_z}_{X_1X_2m_1m_2;mm_z}(k,\mu)
  \label{eq:225}
\end{multline}
is the invariant power spectrum in redshift space.

Practically, it should be sometimes convenient to choose a coordinates
system in which the direction to the line of sight is chosen to be the
$z$ axis, $\hat{\bm{z}} = (0,0,1)$. In this case, we have
$X^{l_zl}_{LM}(\hat{\bm{z}},\hat{\bm{k}}) =
(L\,l_z\,l)_M^{\phantom{M}0m} C_{lm}(\hat{\bm{k}})$, and
Eq.~(\ref{eq:224}) reduces to
\begin{multline}
  P^{(l_1l_2)}_{X_1X_2m_1m_2}(\bm{k}) =
  i^{l_1+l_2}
  \sum_L
  (-1)^L \sqrt{\{L\}}
  \begin{pmatrix}
    l_1 & l_2 & L \\ m_1 & m_2 & -m_{12}
  \end{pmatrix}
  \\ \times
  \sum_{l,l_z}
  \begin{pmatrix}
    L & l_z & l \\ m_{12} & 0 & -m_{12}
  \end{pmatrix}
  C_{l\,m_{12}}(\hat{\bm{k}})
  P^{l_1l_2;l\,l_z;L}_{X_1X_2}(k,\mu),
  \label{eq:226}
\end{multline}
where $m_{12} \equiv m_1+m_2$ and $\mu = k_z/k$ in this coordinates
system. In a simple case that the invariant spectrum does not
explicitly depend on the direction cosine $\mu$, the above equation
can be inverted by using orthogonality relations of the spherical
harmonics and $3j$-symbols, Eqs.~(\ref{eq:397}) and (\ref{eq:424}).
The result is given by
\begin{multline}
  P^{l_1l_2;l\,l_z;L}_{X_1X_2}(k) =
  i^{l_1+l_2} (-1)^l \{l\}\{l_z\}\sqrt{\{L\}}
  \sum_{m_1,m_2}
  \begin{pmatrix}
    l_1 & l_2 & L \\ m_1 & m_2 & -m_{12}
  \end{pmatrix}
  \\ \times
  \begin{pmatrix}
    L & l_z & l \\ m_{12} & 0 & -m_{12}
  \end{pmatrix}
  \int \frac{d^2\hat{k}}{4\pi} C^*_{l\,m_{12}}(\hat{\bm{k}})
  P^{(l_1l_2)}_{X_1X_2m_1m_2}(\bm{k}).
  \label{eq:227}
\end{multline}
As the dependence of azimuthal angle $\phi_k$ of the wave vector
$\bm{k}$ in the power spectrum of Eq.~(\ref{eq:226}) is given by
$e^{im_{12}\phi_k}$ in the spherical harmonics of Eq.~(\ref{eq:17}),
the above equation further reduces to
\begin{multline}
  P^{l_1l_2;l\,l_z;L}_{X_1X_2}(k) =
  i^{l_1+l_2} (-1)^l \{l\}\{l_z\}\sqrt{\{L\}}
  \sum_{m_1,m_2}
  \sqrt{\frac{(l-m_{12})!}{(l+m_{12})!}}
  \\ \times
  \begin{pmatrix}
    l_1 & l_2 & L \\ m_1 & m_2 & -m_{12}
  \end{pmatrix}
  \begin{pmatrix}
    L & l_z & l \\ m_{12} & 0 & -m_{12}
  \end{pmatrix}
  \\ \times
  \frac{1}{2}
  \int_{-1}^1 d\mu\,P_l^{m_{12}}(\mu)\,
  P^{(l_1l_2)}_{X_1X_2m_1m_2}(k,\mu),
  \label{eq:228}
\end{multline}
where $P_l^m(\mu)$ are associate Legendre polynomials, and
$P^{(l_1l_2)}_{X_1X_2m_1m_2}(k,\mu)$ is a simplified notation for the
power spectrum $P^{(l_1l_2)}_{X_1X_2m_1m_2}(\bm{k})$ with a constraint
$\phi_k=0$ (i.e., $\bm{k}$ is on the $x$-$z$ plane).

\subsubsection{Symmetries in redshift space}

As we work in the distant-observer approximation, it is reasonable to
assume that the power spectrum is unchanged when the direction of the
line of sight is flipped, $\hat{\bm{z}} \rightarrow -\hat{\bm{z}}$,
keeping the physical system unchanged:
\begin{equation}
  P^{(l_1l_2)}_{X_1X_2m_1m_2}(\bm{k};\hat{\bm{z}}) =
  P^{(l_1l_2)}_{X_1X_2m_1m_2}(\bm{k};-\hat{\bm{z}}).
  \label{eq:229}
\end{equation}
On one hand, when the directional dependence of
$\mu=\hat{\bm{z}}\cdot\hat{\bm{k}}$ is completely expanded in the
expansion of Eq.~(\ref{eq:224}), the invariant spectrum satisfies
$P^{l_1l_2;l\,l_z;L}_{X_1X_2}(k) =
(-1)^{l_z}P^{l_1l_2;l\,l_z;L}_{X_1X_2}(k)$, and thus the integers
$l_z$ in Eq.~(\ref{eq:224}) should be even numbers to have nonzero
values of the power spectrum. On the other hand, when the dependence
of $\mu$ is included in the invariant spectrum as
$P^{l_1l_2;l\,l_z;L}_{X_1X_2}(k,\mu)$, the integers $l_z$ are not
necessarily even numbers, and the invariant spectrum instead satisfies
\begin{equation}
  P^{l_1l_2;l\,l_z;L}_{X_1X_2}(k,\mu) =
  (-1)^{l_z}P^{l_1l_2;l\,l_z;L}_{X_1X_2}(k,-\mu).
  \label{eq:230}
\end{equation}

According to Eqs.~(\ref{eq:86}) and (\ref{eq:211}) and the symmetry
of the $3j$-symbol, Eq.~(\ref{eq:421}), the complex conjugate of the
invariant function is shown to be given by
\begin{equation}
  P^{l_1l_2;l\,l_z;L\,*}_{X_1X_2}(k,\mu) =
  (-1)^{l_z}P^{l_1l_2;l\,l_z;L}_{X_1X_2}(k,-\mu).
\label{eq:231}
\end{equation}
Combining the above equation with the flip symmetry of
Eq.~(\ref{eq:230}), we derive
\begin{equation}
  P^{l_1l_2;l\,l_z;L\,*}_{X_1X_2}(k,\mu)
  = P^{l_1l_2;l\,l_z;L}_{X_1X_2}(k,\mu),
  \label{eq:232}
\end{equation}
i.e., the invariant spectra are real functions. This conclusion holds
regardless of whether the dependence of $\mu$ is present or not in the
invariant spectrum.

The parity transformation of the power spectrum is given by
\begin{equation}
  P^{(l_1l_2)}_{X_1X_2m_1m_2}(\bm{k};\hat{\bm{z}}) \xrightarrow{\mathbb{P}}
  (-1)^{p_{X_1}+p_{X_2}+l_1+l_2}
  P^{(l_1l_2)}_{X_1X_2m_1m_2}(-\bm{k};-\hat{\bm{z}}),
  \label{eq:233}
\end{equation}
because of Eqs.~(\ref{eq:99}) and (\ref{eq:211}). Accordingly, the 
transformation of the invariant power spectrum is given by
\begin{equation}
  P^{l_1l_2;l\,l_z;L}_{X_1X_2}(k,\mu) \xrightarrow{\mathbb{P}}
  (-1)^{p_{X_1}+p_{X_2}+l_1+l_2+l+l_z} P^{l_1l_2;l\,l_z;L}_{X_1X_2}(k,\mu).
  \label{eq:234}
\end{equation}
When the Universe is statistically invariant under the parity, the
power spectrum in redshift space is invariant under the parity
transformation, we thus have
\begin{equation}
  p_{X_1} + p_{X_2} + l_1 + l_2 + l + l_z = \mathrm{even}.
  \label{eq:235}
\end{equation}
The power spectrum in redshift space has an interchange symmetry,
\begin{equation}
  P^{(l_2l_1)}_{X_2X_1m_2m_1}(\bm{k};\hat{\bm{z}})
  = P^{(l_1l_2)}_{X_1X_2m_1m_2}(-\bm{k};\hat{\bm{z}}),
  \label{eq:236}
\end{equation}
and the corresponding symmetry for the invariant spectrum is given by
\begin{equation}
  P^{l_2l_1;l\,l_z;L}_{X_2X_1}(k,\mu) =
  (-1)^{l_1+l_2+l+L} P^{l_1l_2;l\,l_z;L}_{X_1X_2}(k,-\mu).
  \label{eq:237}
\end{equation}

\subsubsection{The linear power spectrum in real space
  \label{subsubsec:LinearPSreal}
}

We first consider the power spectrum with the lowest-order
approximation, or the linear power spectrum, which is the simplest
application of our formalism. The formal expression for the power
spectrum of scalar fields in the propagator formalism with Gaussian
initial conditions
\cite{Matsubara:1995wd,Bernardeau:2008fa,Matsubara:2011ck} is
straightforwardly generalized to the case of tensor field. In the
lowest order, we simply have
\begin{equation}
  P^{(l_1l_2)}_{X_1X_2m_1m_2}(\bm{k})
  =
  i^{l_1+l_2}  
  \Pi^2(\bm{k})
  \hat{\Gamma}^{(1)}_{X_1l_1m_1}(\bm{k})
  \hat{\Gamma}^{(1)}_{X_2l_2m_2}(-\bm{k})
  P_\mathrm{L}(k),
  \label{eq:238}
\end{equation}
where the parity symmetry for the vertex resummation factor,
$\Pi(-\bm{k})=\Pi(\bm{k})$ is taken into account.

The vertex resummation factor $\Pi(\bm{k})$ is given by
Eq.~(\ref{eq:74}). In the lowest order of the perturbation theory,
this factor can be approximated by $\Pi(\bm{k})=1$. This factor
strongly damps the power spectrum on small scales, where the linear
theory does not apply. However, sometimes it is useful to keep this
exponent even in the linear theory, especially in redshift space where
the peculiar velocities damp the power on small scales. When the
resummation factor is retained, the linear approximation for this
factor is derived from Eqs.~(\ref{eq:74}), (\ref{eq:79}), and
(\ref{eq:83}), and we have
\cite{Matsubara:2007wj,Matsubara:2011ck,Matsubara:2013ofa}
\begin{equation}
  \Pi(k) =
  \exp\left[
    -\frac{k^2}{12\pi^2}
    \int dp P_\mathrm{L}(p)
  \right],
  \label{eq:239}
\end{equation}
in real space, and 
\begin{equation}
  \Pi(k,\mu) =
  \exp\left\{
    -\frac{k^2}{12\pi^2}
    \left[ 1 + f(f+2)\mu^2 \right]
    \int dp P_\mathrm{L}(p)
  \right\},
  \label{eq:240}
\end{equation}
in redshift space, where $\mu \equiv \hat{\bm{z}}\cdot\hat{\bm{k}}$ is
the direction cosine between the wave vector and the line of sight, as
usual. Substituting $f=0$ in the expression in redshift space reduces
to the expression in real space as a matter of course.

In real space, the first-order propagator in terms of the invariant
propagator is given by Eq.~(\ref{eq:160}). Thus Eq.~(\ref{eq:238})
reduces to
\begin{multline}
  P^{(l_1l_2)}_{X_1X_2m_1m_2}(\bm{k})
  =
  \frac{(-i)^{l_1-l_2}}{\sqrt{\{l_1\}\{l_2\}}}\,
  \hat{\Gamma}^{(1)}_{X_1l_1}(k)
  \hat{\Gamma}^{(1)}_{X_2l_2}(k)
  \Pi^2(k)\, P_\mathrm{L}(k)
  \\ \times
  \sum_l \{l\}\left[l_1\,l_2\,l\right]_{m_1m_2}^{\phantom{m_1m_2}m}
  C_{lm}(\hat{\bm{k}}).
  \label{eq:241}
\end{multline}
The invariant power spectrum is easily read off from the above
expression and Eq.~(\ref{eq:214}), and we have
\begin{multline}
  P^{l_1l_2;l}_{X_1X_2}(k)
  = \frac{(-1)^{l_1}\{l\}}{\sqrt{{\{l_1\}\{l_2\}}}}
  \begin{pmatrix}
    l_1 & l_2 & l \\
    0 & 0 & 0
  \end{pmatrix}
  \\ \times
  \hat{\Gamma}^{(1)}_{X_1l_1}(k)
  \hat{\Gamma}^{(1)}_{X_2l_2}(k)
  \Pi^2(k) P_\mathrm{L}(k).
  \label{eq:242}
\end{multline}
Further substituting the lowest-order approximation of the first-order
propagator, Eq.~(\ref{eq:196}), the above equation is explicitly
given by
\begin{multline}
    P^{l_1l_2;l}_{X_1X_2}(k) =
    \Pi^2(k) P_\mathrm{L}(k)
    \\ \times
    \Biggl\{
  \frac{(-1)^{l_1}\,\{l\}}{\sqrt{{\{l_1\}\{l_2\}}}}
    \begin{pmatrix}
      l_1 & l_2 & l \\ 0 & 0 & 0
    \end{pmatrix}
    c^{(1)}_{X_1l_1}(k) c^{(1)}_{X_2l_2}(k)
    \\
    +
    \left[
      \delta_{l_1l} \delta_{l_20}
      c^{(1)}_{X_1l_1}(k) c^{(0)}_{X_2}
      + (-1)^l
      \delta_{l_10} \delta_{l_2l}
      c^{(0)}_{X_1} c^{(1)}_{X_2l_2}(k)
    \right]
    \\
    + \delta_{l0} \delta_{l_10} \delta_{l_20}
    c^{(0)}_{X_1} c^{(0)}_{X_2}
  \Biggr\}.
  \label{eq:243}
\end{multline}
This is one of the generic predictions of our theory. 

As a consistency check, we consider the scalar case, $l_1=l_2=0$ of
autopower spectrum, $X_1=X_2=X$. In this case, Eq.~(\ref{eq:243})
simply reduces to
\begin{equation}
  P^{00;0}_{XX}(k) =
  \left[b_X(k)\right]^2
  \Pi^2(k) P_\mathrm{L}(k),
  \label{eq:244}
\end{equation}
where
\begin{equation}
  b_X(k) \equiv c^{(1)}_{X0}(k) + c^{(0)}_X
  \label{eq:245}
\end{equation}
corresponds to the linear bias factor. Equation~(\ref{eq:244}) is
consistent with the result of the usual linear power spectrum of
biased density field.

The Kronecker's symbols in Eq.~(\ref{eq:243}) indicate that the
resulting expressions are quite different between scalar fields
($l_1=0$ or $l_2=0$) and nonscalar fields ($l_1\ge 1$ or $l_2\ge 1$).
When the fields $X_1$ and $X_2$ are both nonscalar fields with
$l_1\ne 0$ and $l_2 \ne 0$, Eq.~(\ref{eq:243}) simply reduces to
\begin{multline}
    P^{l_1l_2;l}_{X_1X_2}(k) =
    \frac{(-1)^{l_1}\,\{l\}}{\sqrt{{\{l_1\}\{l_2\}}}}
    \begin{pmatrix}
      l_1 & l_2 & l \\ 0 & 0 & 0
    \end{pmatrix}
    c^{(1)}_{X_1l_1}(k)
    c^{(1)}_{X_2l_2}(k)
    \Pi^2(k)
    P_\mathrm{L}(k),
    \\
    (l_1\ne 0 \mathrm{\ \ and\ \ } l_2 \ne 0).
  \label{eq:246}
\end{multline}
When $X_1$ is a nonscalar field and $X_2$ is a scalar field with
$l_1\ne 0$ and $l_2=0$, Eq.~(\ref{eq:243}) reduces to
\begin{equation}
  P^{l_10;l}_{X_1X_2}(k) =
  \delta_{l_1l}
  c^{(1)}_{X_1l_1}(k) b_{X_2}(k) \Pi^2(k)  P_\mathrm{L}(k),
  \quad
  (l_1\ne 0),
  \label{eq:247}
\end{equation}
where $b_X(k)$ is given by Eq.~(\ref{eq:245}). The above equation
survives only when $l=l_1$. When $X_1$ and $X_2$ are both scalar fields
with $l_1=l_2=0$, Eq.~(\ref{eq:243}) reduces to
\begin{equation}
    P^{00;l}_{X_1X_2}(k) =
    \delta_{l0}
    b_{X_1}(k) b_{X_2}(k) \Pi^2(k) P_\mathrm{L}(k),
    \label{eq:248}
\end{equation}
which survives only when $l=0$. The last equation coincides with
Eq.~(\ref{eq:244}) when $X_1=X_2=X$ and $l=0$.

\subsubsection{The linear power spectrum in redshift space
  \label{subsubsec:LinearPSred}
}

In redshift space, the lowest-order propagator is given by
Eq.~(\ref{eq:178}), which can be substituted into
Eq.~(\ref{eq:238}). As a result, there appears a product of bipolar
spherical harmonics, which can be represented by a superposition of a
single bipolar spherical harmonics using a formula with $9j$-symbols
\cite{Khersonskii:1988krb}, as given in Eq.~(\ref{eq:407}) of
Appendix~\ref{app:SphericalHarmonics} in our notation. The
definition of the $9j$-symbol is given by Eq.~(\ref{eq:438}), and
useful properties are summarized in
Eqs.~(\ref{eq:439})--(\ref{eq:442}). After some calculations along
the line described above, we have
\begin{multline}
  P^{(l_1l_2)}_{X_1X_2m_1m_2}(\bm{k};\hat{\bm{z}})
  =
    i^{l_1+l_2}
  \Pi^2(\bm{k}) P_\mathrm{L}(k)
  \sum_{l_{z1},l_{z2},l_1',l_2'} (-1)^{l_2'}
  \hat{\Gamma}^{(1)l_1l_{z1}}_{X_1l_1'}(k,\mu)
  \\ \times
  \hat{\Gamma}^{(1)l_2l_{z2}}_{X_2l_2'}(k,-\mu)
  \sum_{l,l_z,L} (-1)^{l+l_z+L}\{l\}\{l_z\}\{L\}
  \left(l_1\,l_2\,L\right)_{m_1m_2}^{\phantom{m_1m_2}M}
  \\ \times
  \begin{pmatrix}
    l_{z1} & l_{z2} & l_z \\
    0 & 0 & 0
  \end{pmatrix}
  \begin{pmatrix}
    l_1' & l_2' & l \\
    0 & 0 & 0
  \end{pmatrix}
  \begin{Bmatrix}
    l_{z1} & l_{z2} & l_z \\
    l_1' & l_2' & l\\
    l_1 & l_2 & L
  \end{Bmatrix}
  X^{l_zl}_{LM}(\hat{\bm{z}},\hat{\bm{k}}).
  \label{eq:249}
\end{multline}
Substitution of Eq.~(\ref{eq:204}) into the above gives an explicit
result. The invariant power spectrum is easily read off from the
expression of Eq.~(\ref{eq:224}), including the case that the
dependence of the directional cosine $\mu$ in invariant propagators is
present. We derive
\begin{multline}
  P^{l_1l_2;l\,l_z;L}_{X_1X_2}(k,\mu) =
  \Pi^2(k,\mu) P_\mathrm{L}(k)\,
  (-1)^{l+l_z} \{l\} \{l_z\} \sqrt{\{L\}}
  \\ \times
  \sum_{l_{z1},l_{z2},l_1',l_2'}
  \begin{pmatrix}
    l_{z1} & l_{z2} & l_z \\
    0 & 0 & 0
  \end{pmatrix}
  \begin{pmatrix}
    l_1' & l_2' & l \\
    0 & 0 & 0
  \end{pmatrix}
  \begin{Bmatrix}
    l_{z1} & l_{z2} & l_z \\
    l_1' & l_2' & l \\
    l_1 & l_2 & L
  \end{Bmatrix}
  \\ \times
  (-1)^{l_2'}
  \hat{\Gamma}^{(1)l_1l_{z1}}_{X_1l_1'}(k,\mu)
  \hat{\Gamma}^{(1)l_2l_{z2}}_{X_2l_2'}(k,-\mu).
  \label{eq:250}
\end{multline}
This is one of the generic predictions of our theory.

As a consistency check, one can recover the result of the power
spectrum in real space, Eqs.~(\ref{eq:241}) and (\ref{eq:242}). In
fact, substituting $l_{z1}=l_{z2}=0$ into Eqs.~(\ref{eq:249}) and
(\ref{eq:250}), using identities
$\hat{\Gamma}^{(1)l0}_{Xl'}(k,\mu) =
\delta_{l'l}\hat{\Gamma}^{(1)}_{Xl}(k)$ when $f=0$,
$(00l)_{000}=\delta_{l0}$, and a special case of $9j$-symbols
\cite{Khersonskii:1988krb},
\begin{equation}
  \begin{Bmatrix}
    0 & 0 & 0 \\
    l_1' & l_2' & l_3' \\
    l_1 & l_2 & l_3
  \end{Bmatrix}
  = \frac{\delta_{l_1l_1'}\delta_{l_2l_2'}\delta_{l_3l_3'}}
  {\sqrt{\{l_1\}\{l_2\}\{l_3\}}},
  \label{eq:251}
\end{equation}
the corresponding equations in real space are recovered.

As another consistency check, one can also see if Eq.~(\ref{eq:249})
reproduces the well-known result for the scalar case, i.e., the
Kaiser's formula \cite{Kaiser:1987qv}. In the case that the
directional dependence is completely expanded, substituting
Eq.~(\ref{eq:204}) into Eq.~(\ref{eq:249}), putting
$l_1=l_2=m_1=m_2=0$, $X_1=X_2=X$, and using Eqs.~(\ref{eq:427}) and
(\ref{eq:251}), we derive
\begin{multline}
  P^{(00)}_{XX00}(\bm{k};\hat{\bm{z}}) =
  \Pi^2(k,\mu) \left[b_X(k)\right]^2P_\mathrm{L}(k)
  \\ \times
  \left[
    \left(1 + \frac{2}{3} \beta + \frac{1}{5}\beta^2\right)
    \mathit{P}_0(\mu)
    + \left(\frac{4}{3}\beta + \frac{4}{7}\beta^2\right) 
    \mathit{P}_2(\mu)
    \right.
    \\
    \left.
    + \frac{8}{35}\beta^2 \mathit{P}_4(\mu)
  \right],
  \label{eq:252}
\end{multline}
where $\beta \equiv c^{(0)}_Xf/b_X(k)$ corresponds to the
redshift-space distortion parameter. When the scalar field is given by
density field of biased objects, this expression is equivalent to a
known result of the linear power spectrum of a biased (scalar) field
in redshift space \cite{Hamilton:1992zz}.

In the lowest-order approximation of the first-order propagator in
redshift space, Eq.~(\ref{eq:204}), either $l$ or $l_z$ is zero in
each term. Therefore, at least two of the indices $l_1$, $l_2$,
$l_{z1}$, $l_{z2}$ are zero in the $9j$-symbol of Eq.~(\ref{eq:250}).
When two of the indices in the $9j$-symbol are zero, there are formulas
\cite{Khersonskii:1988krb},
\begin{align}
  \begin{Bmatrix}
    l_1 & l_2 & l_3 \\
    l_4 & l_5 & l_6 \\
    l_7 & 0 & 0
  \end{Bmatrix}
  &= \delta_{l_70}
  \frac{\delta_{l_1l_4}\delta_{l_2l_5}\delta_{l_3l_6}}
  {\sqrt{\{l_1\}\{l_2\}\{l_3\}}},
  \label{eq:253}\\
  \begin{Bmatrix}
    l_1 & l_2 & l_3 \\
    l_4 & 0 & l_6 \\
    l_7 & l_8 & 0
  \end{Bmatrix}
  &= (-1)^{l_1-l_2-l_3}
  \frac{\delta_{l_4l_6}\delta_{l_2l_8}\delta_{l_3l_6}\delta_{l_7l_8}}
  {\sqrt{\{l_2\}\{l_3\}}},
  \label{eq:254}
\end{align}
A $9j$-symbol, in which two of the indices are zero, can always be
represented in a form of the lhs of the above equations by means of
the symmetric properties of the $9j$-symbol,
Eqs.~(\ref{eq:439})--(\ref{eq:441}). Thus, with the lowest-order
approximation of the first-order propagator given by
Eq.~(\ref{eq:204}), the summation of Eq.~(\ref{eq:250}) is
explicitly expanded to give
\begin{widetext}
\begin{multline}
    P^{l_1l_2;l\,l_z;L}_{X_1X_2}(k,\mu) =
    \Pi^2(k,\mu) P_\mathrm{L}(k)
    \\ \times
    \Biggl\{
    \delta_{l_z0} \delta_{lL}
    \frac{ (-1)^{l_1}\,\{l\}}{\sqrt{\{l_1\}\{l_2\}}}
    \begin{pmatrix}
      l_1 & l_2 & l \\ 0 & 0 & 0
    \end{pmatrix}
    c^{(1)}_{X_1l_1}(k) c^{(1)}_{X_2l_2}(k)
    + \left(1 + \frac{f}{3}\right)
    \delta_{l_z0} \delta_{lL}
    \Biggl[
    \delta_{l_1l} \delta_{l_20}
    c^{(1)}_{X_1l_1}(k) c^{(0)}_{X_2}
    + (-1)^l \delta_{l_10} \delta_{l_2l}
    c^{(0)}_{X_1} c^{(1)}_{X_2l_2}(k)
    \Biggr]
    \\
    + \frac{2f}{3} \delta_{l_z2}\{l\}
    \Biggl[
    (-1)^l \delta_{l_1L} \delta_{l_20}
    \frac{1}{\sqrt{\{l_1\}}}
    \begin{pmatrix}
      2 & l_1 & l \\ 0 & 0 & 0
    \end{pmatrix}
    c^{(1)}_{X_1l_1}(k) c^{(0)}_{X_2}
    + \delta_{l_10} \delta_{l_2L}
    \frac{1}{\sqrt{\{l_2\}}}
    \begin{pmatrix}
      2 & l_2 & l \\ 0 & 0 & 0
    \end{pmatrix}
    c^{(0)}_{X_1} c^{(1)}_{X_2l_2}(k)
    \Biggl]
    \\
    + \delta_{L0} \delta_{l_10} \delta_{l_20} \delta_{l\,l_z}
    \Biggl[
    \delta_{l_z0} \left( 1 + \frac{2f}{3} + \frac{f^2}{5} \right)
    + \sqrt{5} \delta_{l_z2} \left(\frac{4f}{3} + \frac{4f^2}{7} \right)
    + 3\,\delta_{l_z4} \frac{8f^2}{35}
    \Biggr]
    c^{(0)}_{X_1} c^{(0)}_{X_2}
  \Biggr\},
  \label{eq:255}
\end{multline}
\end{widetext}
where an identity 
\begin{equation}
  \begin{pmatrix}
    2 & 2 & l \\ 0 & 0 & 0
  \end{pmatrix}^2
  = \frac{1}{5} \delta_{l0}
  + \frac{2}{35} \delta_{l2}
  + \frac{2}{35} \delta_{l4}
  \label{eq:256}
\end{equation}
is used in deriving the above result. Equation~(\ref{eq:255}) is one
of the generic predictions of our theory.

The above Eq.~(\ref{eq:255}) is nonzero only when $l_z=0,2,4$, and the
redshift-space distortions in linear theory contain only monopole
($l_z=0$), quadrupole ($l_z=2$) and hexadecapole ($l_z=4$) components,
as is well known in the scalar perturbation theory,
Eq.~(\ref{eq:252}), while higher-order corrections contain
higher-order multipoles for the redshift-space distortions.

In linear theory besides the resummation factor $\Pi(k,\mu)$,
higher-rank tensors are not affected by the redshift-space distortions
as the first-order propagator of Eq.~(\ref{eq:204}) does not
neither. Thus, if neither $l_1$ nor $l_2$ is zero, Eq.~(\ref{eq:255})
is simply given by
\begin{multline}
    P^{l_1l_2;l\,l_z;L}_{X_1X_2}(k,\mu) =
    \delta_{l_z0}\delta_{lL}
    \Pi^2(k,\mu) P_\mathrm{L}(k)
    \\ \times
    \frac{(-1)^{l_1}\,\{l\}}{\sqrt{\{l_1\}\{l_2\}}}
    \begin{pmatrix}
      l_1 & l_2 & l \\ 0 & 0 & 0
    \end{pmatrix}
    c^{(1)}_{X_1l_1}(k) c^{(1)}_{X_2l_2}(k),
    \\
    (l_1 \ne 0 \mathrm{\ \ and\ \ } l_2\ne 0),
  \label{eq:257}
\end{multline}
which survives only when $l_z=0$ and $l=L$. The above equation without
the resummation factor exactly coincides with the result in real
space, Eq.~(\ref{eq:246}). Namely, redshift-space distortions on
higher-rank tensors are nonlinear effects. In fact, nonlinear loop
corrections in the power spectrum do introduce the redshift-space
distortion effects, as explicitly shown in Paper~II \cite{PaperII}.

As in the case of real space, Kronecker's symbols in
Eq.~(\ref{eq:255}) indicate that the resulting expression is different
between scalar fields ($l_1=0$ or $l_2=0$) and nonscalar fields
($l_1\ge 1$ or $l_2\ge 1$). When the fields $X_1$ and $X_2$ are both
nonscalar fields with $l_1\ne 0$ and $l_2 \ne 0$, the expression
reduces to Eq.~(\ref{eq:257}) above. When $X_1$ is a nonscalar field
and $X_2$ is a scalar field with $l_1\ne 0$ and $l_2=0$,
Eq.~(\ref{eq:255}) reduces to
\begin{multline}
  P^{l_10;l\,l_z;L}_{X_1X_2}(k,\mu) =
  \delta_{Ll_1}
  \Pi^2(k,\mu) P_\mathrm{L}(k) b_{X_2}(k) c^{(1)}_{X_1l_1}(k)
    \\ \times
    \left[
    \delta_{l_z0} \delta_{ll_1}
    \left(
      1 + \frac{\beta_2}{3}
    \right)
    \right.
    \\
    \left.
      + (-1)^l \delta_{l_z2}\frac{2\beta_2}{3}
     \frac{\{l\}}{\sqrt{\{l_1\}}}
  \begin{pmatrix}
    2 & l_1 & l \\ 0 & 0 & 0
  \end{pmatrix}
  \right],
  \label{eq:258}
\end{multline}
where $\beta_2 \equiv c^{(0)}_{X_2}f/b_{X_2}(k)$ and $b_X(k)$ is the
linear bias factor of Eq.~(\ref{eq:245}).
When $X_1$ and $X_2$ are both scalar
fields with $l_1=l_2=0$, Eq.~(\ref{eq:255}) reduces to
\begin{multline}
    P^{00;l\,l_z;L}_{X_1X_2}(k,\mu) =
    \delta_{L0}\delta_{ll_z}
    \Pi^2(k,\mu)
    P_\mathrm{L}(k)
    b_{X_1}(k) b_{X_2}(k)
    \\
    \times
    \Biggl\{
    \delta_{l_z0}
    \left[
      1 + \frac{1}{3}\left(\beta_1 + \beta_2\right)
      + \frac{1}{5}\beta_1\beta_2
    \right]
    \\
    + \sqrt{5} \delta_{l_z2}
    \left[
      \frac{2}{3}\left(\beta_1 + \beta_2\right)
      + \frac{4}{7} \beta_1 \beta_2
    \right]
    + 3\,\delta_{l_z4} \frac{8}{35} \beta_1 \beta_2
  \Biggr\},
  \label{eq:259}
\end{multline}
where $\beta_1 \equiv c^{(0)}_{X_1}f/b_{X_1}(k)$. The last form is
consistent with Eq.~(\ref{eq:252}).

\subsubsection{Alternative representation of the linear power spectrum
  in redshift space}

In the above examples, we consider the case that the dependence on the
directional cosine $\mu$ of the first-order normalized propagator in
redshift space is completely expanded into spherical harmonics, and
the propagator is given by Eq.~(\ref{eq:204}). We now consider the
other case that the dependence is not completely expanded and the
propagator is given by Eq.~(\ref{eq:207}). In this case, we
substitute Eq.~(\ref{eq:207}) into Eq.~(\ref{eq:250}). Because of
the simple form of Eq.~(\ref{eq:207}), the summations over $l_{z1}$,
$l_{z2}$, $l_1'$, $l_2'$ are trivially evaluated, resulting in,
\begin{multline}
    P^{l_1l_2;l\,l_z;L}_{X_1X_2}(k,\mu) =
    \delta_{l_z0} \delta_{Ll}
    \Pi^2(k,\mu) P_\mathrm{L}(k)
    \\ \times
    \Biggl\{
    \frac{(-1)^{l_1}\,\{l\}}{\sqrt{\{l_1\}\{l_2\}}}
    \begin{pmatrix}
      l_1 & l_2 & l \\ 0 & 0 & 0
    \end{pmatrix}
    c^{(1)}_{X_1l_1}(k) c^{(1)}_{X_2l_2}(k)
    \\
    + \left(1+f\mu^2\right)
    \left[
      \delta_{l_1l} \delta_{l_20}
      c^{(1)}_{X_1l_1}(k) c^{(0)}_{X_2}
      + (-1)^l
      \delta_{l_10} \delta_{l_2l}
      c^{(0)}_{X_1} c^{(1)}_{X_2l_2}(k)
    \right]
    \\
    + \left(1+f\mu^2\right)^2
    \delta_{l0} \delta_{l_10} \delta_{l_20}
    c^{(0)}_{X_1} c^{(0)}_{X_2}
  \Biggr\}.
  \label{eq:260}
\end{multline}
As a consistency check, one immediately sees that this equation reduces
to Eq.~(\ref{eq:243}) when $f=0$. 

When the fields $X_1$ and $X_2$ are both nonscalar fields with
$l_1\ne 0$ and $l_2 \ne 0$, Eq.~(\ref{eq:260}) simply reduces to
\begin{multline}
  P^{l_1l_2;l\,l_z;L}_{X_1X_2}(k,\mu) =
  \delta_{l_z0} \delta_{Ll}
    \Pi^2(k,\mu) P_\mathrm{L}(k)
    \\ \times
    \frac{(-1)^{l_1}\,\{l\}}{\sqrt{\{l_1\}\{l_2\}}}
    \begin{pmatrix}
      l_1 & l_2 & l \\ 0 & 0 & 0
    \end{pmatrix}
    c^{(1)}_{X_1l_1}(k)
    c^{(1)}_{X_2l_2}(k),
    \\
    (l_1\ne 0 \mathrm{\ \ and\ \ } l_2 \ne 0).
  \label{eq:261}
\end{multline}
which survives only when $l_z=0$ and $L=l$. This equation is exactly
the same as Eq.~(\ref{eq:257}). When $X_1$ is a nonscalar field and
$X_2$ is a scalar field with $l_1\ne 0$ and $l_2=0$,
Eq.~(\ref{eq:260}) reduces to
\begin{multline}
  P^{l_10;l\,l_z;L}_{X_1X_2}(k,\mu) =
  \delta_{l_z0}\delta_{lL} \delta_{l_1l}
  \Pi^2(k,\mu)  P_\mathrm{L}(k)
  \\ \times
  \left(1 + f\mu^2\right)
  c^{(1)}_{X_1l_1}(k) b_{X_2}(k),
  \quad (l_1\ne 0),
  \label{eq:262}
\end{multline}
which survives only when $l_z=0$ and $L=l=l_1$. When $X_1$ and $X_2$
are both scalar fields with $l_1=l_2=0$, Eq.~(\ref{eq:260}) reduces
to
\begin{equation}
    P^{00;l\,l_z;L}_{X_1X_2}(k,\mu) =
    \delta_{l0} \delta_{l_z0}\delta_{L0}
    \Pi^2(k) P_\mathrm{L}(k)
    \left(1 + f\mu^2\right)^2
    b_{X_1}(k) b_{X_2}(k),
    \label{eq:263}
\end{equation}
which survives only when $l=l_z=L=0$. The last equation corresponds to
the Kaiser's formula for the density power spectrum in redshift space
\cite{Kaiser:1987qv}.

\subsection{\label{subsec:CorrFunc}%
  The correlation function
}

While the power spectrum is defined in Fourier space, the counterpart
in configuration space is the correlation function,
$\xi^{(l_1l_2)}_{X_1X_2m_1m_2}(\bm{x})$, which is defined by
\begin{equation}
  \left\langle
    F_{X_1l_1m_1}(\bm{x}_1)
    F_{X_2l_2m_2}(\bm{x}_2)
  \right\rangle_\mathrm{c} =
  \xi^{(l_1l_2)}_{X_1X_2m_1m_2}(\bm{x}_1-\bm{x}_2),
  \label{eq:264}
\end{equation}
where the tensor field $F_{Xlm}(\bm{x})$ on the lhs now corresponds to
the one in the configuration space. On the rhs, the correlation function
is a function of the relative vector between the two positions,
$\bm{x}_1-\bm{x}_2$, because of the statistical homogeneity of the
Universe. It is standard that the correlation function and the power
spectrum are related by a three-dimensional Fourier transform as
\begin{equation}
  \xi^{(l_1l_2)}_{X_1X_2m_1m_2}(\bm{x}) =
  \int \frac{d^3k}{(2\pi)^3} e^{i\bm{k}\cdot\bm{x}}
  P^{(l_1l_2)}_{X_1X_2m_1m_2}(\bm{k}).
  \label{eq:265}
\end{equation}

First, we consider the correlation function in real space. In exactly
the same way of deriving Eq.~(\ref{eq:214}), the rotational symmetry
requires that the correlation function should have a form,
\begin{equation}
  \xi^{(l_1l_2)}_{X_1X_2m_1m_2}(\bm{x}) =
    i^{l_1+l_2}
  \sum_l i^l
  \left(l_1\,l_2\,l\right)_{m_1m_2}^{\phantom{m_1m_2}m}
  C_{lm}(\hat{\bm{x}})\, \xi^{l_1l_2;l}_{X_1X_2}(x),
  \label{eq:266}
\end{equation}
where the last factor $\xi^{l_1l_2;l}_{X_1X_2}(x)$ corresponds to the
invariant correlation function. Compared with the definition of the
invariant power spectrum of Eq.~(\ref{eq:214}), we should note that
the phase factor $i^l$ is additionally present. One can show that the
invariant correlation function of $\xi^{l_1l_2;l}_{X_1X_2}(x)$ defined
above with the phase factor is a real function.

Substituting Eq.~(\ref{eq:214}) into Eq.~(\ref{eq:265}), and
using a formula of Eq.~(\ref{eq:398}) in
Appendix~\ref{app:SphericalHarmonics}, we have
\begin{equation}
  \xi^{l_1l_2;l}_{X_1X_2}(x) =
  \int \frac{k^2dk}{2\pi^2} j_l(kx) 
  P^{l_1l_2;l}_{X_1X_2}(k),
  \label{eq:267}
\end{equation}
i.e., the invariant correlation function in real space is given by a
Hankel transform of the invariant power spectrum. Using the
completeness relation of the spherical Bessel functions, the inverse
relation of the above transform is given by
\begin{equation}
  P^{l_1l_2;l}_{X_1X_2}(k) =
  4\pi \int x^2 dx j_l(kx) 
  \xi^{l_1l_2;l}_{X_1X_2}(x). 
  \label{eq:268}
\end{equation}
The correlation function in redshift space is also represented by the
invariant spectrum in redshift space. The relation between the
correlation function and the power spectrum is just given by
Eq.~(\ref{eq:265}) as well in redshift space, provided that both
explicitly depend on the direction of the line of sight,
$\hat{\bm{z}}$,
\begin{equation}
  \xi^{(l_1l_2)}_{X_1X_2m_1m_2}(\bm{x};\hat{\bm{z}}) =
  \int \frac{d^3k}{(2\pi)^3} e^{i\bm{k}\cdot\bm{x}}
  P^{(l_1l_2)}_{X_1X_2m_1m_2}(\bm{k};\hat{\bm{z}}).
  \label{eq:269}
\end{equation}
Because of the rotational symmetry, we have
\begin{multline}
  \xi^{(l_1l_2)}_{X_1X_2m_1m_2}(\bm{x};\hat{\bm{z}}) =
    i^{l_1+l_2}
  \sum_{L} (-1)^L \sqrt{\{L\}}
  \left(l_1\,l_2\,L\right)_{m_1m_2}^{\phantom{m_1m_2}M}
  \\ \times
  \sum_{l,l_z}  i^l X^{l_zl}_{LM}(\hat{\bm{z}},\hat{\bm{x}})
  \xi^{l_1l_2;l\,l_z;L}_{X_1X_2}(x),
  \label{eq:270}
\end{multline}
just as in the case of the power spectrum of Eq.~(\ref{eq:224}), and
the last factor $\xi^{l_1l_2;l\,l_z;L}_{X_1X_2}(x)$ corresponds to the
invariant correlation function. When the invariant power spectrum in
redshift space $P^{l_1l_2;l\,l_z;L}_{X_1X_2}(k)$ does not contain any
dependence on the directional cosine of the wave vector,
$\mu = \cos\theta_k$, the invariant correlation function
$\xi^{l_1l_2;l\,l_z;L}_{X_1X_2}(x)$ is straightforwardly related to
the invariant spectrum via the Hankel transform just in the similar
way in the case of the real space above,
\begin{equation}
  \xi^{l_1l_2;l\,l_z;L}_{X_1X_2}(x) =
  \int \frac{k^2dk}{2\pi^2} j_l(kx) 
  P^{l_1l_2;l\,l_z;L}_{X_1X_2}(k),
  \label{eq:271}
\end{equation}
and
\begin{equation}
  P^{l_1l_2;l\,l_z;L}_{X_1X_2}(k) =
  4\pi \int r^2 dr j_l(kx) 
  \xi^{l_1l_2;l\,l_z;L}_{X_1X_2}(x). 
  \label{eq:272}
\end{equation}

  As in Eqs.~(\ref{eq:226})--(\ref{eq:228}) for the power spectrum, we
  consider expressions of the correlation function in a coordinates
  system in which the line of sight is directed to the $z$ axis,
  $\hat{\bm{z}}=(0,0,1)$. In this case, Eq.~(\ref{eq:270}) reduces to
\begin{multline}
  \xi^{(l_1l_2)}_{X_1X_2m_1m_2}(\bm{x}) =
  i^{l_1+l_2}
  \sum_L
  (-1)^L \sqrt{\{L\}}
  \begin{pmatrix}
    l_1 & l_2 & L \\ m_1 & m_2 & -m_{12}
  \end{pmatrix}
  \\ \times
  \sum_{l,l_z} i^l
  \begin{pmatrix}
    L & l_z & l \\ m_{12} & 0 & -m_{12}
  \end{pmatrix}
  C_{l\,m_{12}}(\hat{\bm{x}})
  \xi^{l_1l_2;l\,l_z;L}_{X_1X_2}(x),
  \label{eq:273}
\end{multline}
where $m_{12} \equiv m_1+m_2$. Similar to the case of the power spectrum,
Eq.~(\ref{eq:227}), the above equation can be inverted to give
\begin{multline}
  \xi^{l_1l_2;l\,l_z;L}_{X_1X_2}(x) =
  i^{l_1+l_2+l} \{l\}\{l_z\}\sqrt{\{L\}}
  \sum_{m_1,m_2}
  \begin{pmatrix}
    l_1 & l_2 & L \\ m_1 & m_2 & -m_{12}
  \end{pmatrix}
  \\ \times
  \begin{pmatrix}
    L & l_z & l \\ m_{12} & 0 & -m_{12}
  \end{pmatrix}
  \int \frac{d^2\hat{x}}{4\pi} C^*_{l\,m_{12}}(\hat{\bm{x}})
  \xi^{(l_1l_2)}_{X_1X_2m_1m_2}(\bm{x}),
  \label{eq:274}
\end{multline}
and also
\begin{multline}
  \xi^{l_1l_2;l\,l_z;L}_{X_1X_2}(x) =
  i^{l_1+l_2+l} \{l\}\{l_z\}\sqrt{\{L\}}
  \sum_{m_1,m_2}
  \sqrt{\frac{(l-m_{12})!}{(l+m_{12})!}}
  \\ \times
  \begin{pmatrix}
    l_1 & l_2 & L \\ m_1 & m_2 & -m_{12}
  \end{pmatrix}
  \begin{pmatrix}
    L & l_z & l \\ m_{12} & 0 & -m_{12}
  \end{pmatrix}
  \\ \times
  \frac{1}{2}
  \int_{-1}^1 d\mu_x\,P_l^{m_{12}}(\mu_x)\,
  \xi^{(l_1l_2)}_{X_1X_2m_1m_2}(x,\mu_x),
  \label{eq:275}
\end{multline}
where $\xi^{(l_1l_2)}_{X_1X_2m_1m_2}(x,\mu_x)$ is a simplified
notation for the correlation function
$\xi^{(l_1l_2)}_{X_1X_2m_1m_2}(\bm{x})$ with a constraint $\phi_x=0$
(i.e., $\bm{x}$ is on the $x$-$z$ plane), and $\cos\theta_x = \mu_x$.

However, when the invariant spectrum in redshift space
$P^{l_1l_2;l\,l_z;L}_{X_1X_2}(k,\mu)$ does depend on the directional
cosine, the above equations no longer hold. In this case, the
invariant correlation function can still be represented by the
invariant power spectrum. In order to derive the relation, we first
note that Eq.~(\ref{eq:270}) can be inverted by applying orthogonality
relations of the $3j$-symbols, Eq.~(\ref{eq:424}), and those of the
bipolar spherical harmonics, Eq.~(\ref{eq:405}). As a result, we have
\begin{multline}
  \xi^{l_1l_2;l\,l_z;L}_{X_1X_2}(x) =
  i^{l+l_1+l_2} (-1)^{l_z+L}
  \{l\}\{l_z\}\sqrt{\{L\}}
  \\ \times
  \int \frac{d^2\hat{z}}{4\pi} \frac{d^2\hat{x}}{4\pi}
  X^{l_zl}_{LM}(\hat{\bm{z}},\hat{\bm{x}})
  \xi^{(l_1l_2)}_{X_1X_2m_1m_2}(\bm{x};\hat{\bm{z}}).
  \label{eq:276}
\end{multline}
We substitute the expression of the power spectrum,
Eq.~(\ref{eq:224}), into Eq.~(\ref{eq:269}), and then the result is
substituted into Eq.~(\ref{eq:276}). The resulting equation is
calculated by using orthogonality relations of the $3j$-symbols,
Eq.~(\ref{eq:424}), the Rayleigh expansion formula,
Eq.~(\ref{eq:396}), sum rules of spherical harmonics and $6j$-symbols,
Eqs.~(\ref{eq:395}) and (\ref{eq:443}), and the formulas of
Eqs.~(\ref{eq:399}) and (\ref{eq:400}). After these manipulations, the
expression of the invariant correlation function in redshift space
finally reduces to
\begin{multline}
  \xi^{l_1l_2;l\,l_z;L}_{X_1X_2}(x)
  = (-1)^L \{l\}\{l_z\}
  \sum_{l',l_z',L'} (-1)^{L'} \{L'\}
  \\ \times
  \begin{pmatrix}
    l_z & l_z' & L' \\ 0 & 0 & 0
  \end{pmatrix}
  \begin{pmatrix}
    l & l' & L' \\ 0 & 0 & 0
  \end{pmatrix}
  \begin{Bmatrix}
    l_z & l_z' & L' \\
    l' & l & L
  \end{Bmatrix}
  \\ \times
  \int \frac{d\mu}{2} P_{L'}(\mu)
  \int \frac{k^2dk}{2\pi^2} j_l(kx)
  P^{l_1l_2;l'l_z';L}_{X_1X_2}(k,\mu),
  \label{eq:277}
\end{multline}
where $P_{L'}(\mu)$ is the Legendre polynomial of order $L'$. 
When the invariant spectrum does not depend on $\mu$, it is
straightforwardly shown that the above equation reduces to the
previous result of Eq.~(\ref{eq:271}), noting a special case of
the $6j$-symbol \cite{Khersonskii:1988krb},
\begin{equation}
  \begin{Bmatrix}
    l_z & l_z' & 0 \\
    l' & l & L
  \end{Bmatrix}
  = (-1)^{l+l_z+L}
  \frac{\delta_{l_z'l_z}\delta_{l'l}}{\sqrt{\{l_z\}\{l\}}}.
  \label{eq:278}
\end{equation}

\subsection{\label{subsec:PowerSpecNG} Scale-dependent bias in the
  power spectrum with non-Gaussian initial conditions }

In the presence of bias, the local-type non-Gaussianity is known to
produce the scale-dependent bias in the power spectrum on very large
scales \cite{Dalal:2007cu}. The same effect appears even in the shape
correlations \cite{Schmidt:2015xka,Akitsu:2020jvx}, which is
considered to be a second-rank ($l=2$) tensor field in our context.
Higher moments of the shape correlations are also investigated
\cite{Kogai:2020vzz}.

In our formalism, these previous findings are elegantly reproduced as
shown below. The derivation of scale-dependent bias from the
primordial non-Gaussianity in the iPT formalism of scalar fields is
already given in Ref.~\cite{Matsubara:2012nc}. We can simply
generalize the last method to the case of higher-rank tensor fields.
We consider the clustering in real space below, while the
generalization to that in redshift space is straightforward.

The primordial bispectrum $B_\mathrm{L}(\bm{k}_1,\bm{k}_2,\bm{k}_3)$
is defined by three-point correlations of linear density contrast in
Fourier space,
\begin{equation}
  \left\langle
    \delta_\mathrm{L}(\bm{k}_1)
    \delta_\mathrm{L}(\bm{k}_2)
    \delta_\mathrm{L}(\bm{k}_3)
  \right\rangle_\mathrm{c} =
  (2\pi)^3 \delta_\mathrm{D}^3(\bm{k}_1+\bm{k}_2+\bm{k}_3)
  B_\mathrm{L}(\bm{k}_1,\bm{k}_2,\bm{k}_3),
  \label{eq:279}
\end{equation}
where $\langle\cdots\rangle_\mathrm{c}$ denotes the three-point
cumulant, or the connected part. Applying a straightforward
generalization of the corresponding result of iPT
\cite{Matsubara:2012nc} to tensor fields, the lowest-order
contribution of the primordial bispectrum to the power spectrum in the
formalism of iPT is given by
\begin{multline}
  P^{\mathrm{NG}(l_1l_2)}_{X_1X_2m_1m_2}(\bm{k})
  = \frac{i^{l_1+l_2}}{2} \Gamma^{(1)}_{X_1l_1m_1}(-\bm{k})
  \\ \times
  \int \frac{d^3p}{(2\pi)^3}
  \Gamma^{(2)}_{X_2l_2m_2}(\bm{p},\bm{k}-\bm{p})
  B_\mathrm{L}(\bm{k},-\bm{p},\bm{p}-\bm{k})
  \\ + [(X_1l_1m_1) \leftrightarrow (X_2l_2m_2)],
  \label{eq:280}
\end{multline}
where we implicitly assume the parity symmetry of the Universe. The
linear power spectrum with the Gaussian initial condition,
Eq.~(\ref{eq:238}), is added to the above in order to obtain the total
power spectrum up to the linear order of the initial power spectrum
and bispectrum. However, the Gaussian contribution does not generate
scale-dependent bias in the limit of large scales, $k\rightarrow 0$,
and thus the non-Gaussian contribution of Eq.~(\ref{eq:280})
dominates in that limit.

The prefactor of the integral in Eq.~(\ref{eq:280}) is given by
Eq.~(\ref{eq:195}):
\begin{equation}
  \Gamma^{(1)}_{X_1l_1m_1}(\bm{k})
  =
  \frac{(-1)^{l_1}}{\sqrt{2l_1+1}}
  \left[
    c^{(1)}_{X_1l_1}(k)
    + \delta_{l_10} c^{(0)}_{X_1}
  \right]
  C_{l_1m_1}(\hat{\bm{k}}),
  \label{eq:281}
\end{equation}
where we consider the lowest-order approximation and the resummation
factor $\Pi(k)$ is just replaced by unity. In order to evaluate the
integral in Eq.~(\ref{eq:280}) with the spherical basis, one can
insert a unity
$\int d^3p' \delta_\mathrm{D}^3(\bm{p}+\bm{p}'-\bm{k}) = 1$ in the
integral, and reexpress the delta function by a Fourier integral, and
primordial bispectrum can be also expanded into spherical harmonics by
means of the plane-wave expansion of Eq.~(\ref{eq:396}). In this way,
the angular integrations are analytically performed, leaving only the
radial integrals over $p$ and $p'$.

We are interested in the scale-dependent bias from the primordial
non-Gaussianity, and instead of deriving a general expression
according to the above procedure, we directly consider a limiting case
of $k\rightarrow 0$ in Eq.~(\ref{eq:280}). In this case, the integral
is approximately given by
\begin{equation}
  \int \frac{d^3p}{(2\pi)^3}
  \Gamma^{(2)}_{X_2l_2m_2}(\bm{p},-\bm{p})
  B_\mathrm{L}(\bm{k},-\bm{p},\bm{p}).
  \label{eq:282}
\end{equation}
The second-order propagator in the integrand is given by
Eq.~(\ref{eq:77}) with $\bm{k}_{12}=\bm{0}$, $\bm{k}_1=\bm{p}$ and
$\bm{k}_2=-\bm{p}$, and thus we have
\begin{align}
  \Gamma^{(2)}_{X_2l_2m_2}(\bm{p},-\bm{p})
  &= c^{(2)}_{X_2l_2m_2}(\bm{p},-\bm{p})
    \nonumber\\
  &= \sum_{l_2',l_2''} (-1)^{l_2''} c^{(2)l_2}_{X_2l_2'l_2''}(p,p)
  X^{l_2'l_2''}_{l_2m_2}(\hat{\bm{p}},\hat{\bm{p}}),
\label{eq:283}
\end{align}
where Eq.~(\ref{eq:135}) is substituted. According to
Eqs.~(\ref{eq:400}), (\ref{eq:209}) and (\ref{eq:424}), we have
\begin{equation}
  X^{l_2'l_2''}_{l_2m_2}(\hat{\bm{p}},\hat{\bm{p}}) =
    \begin{pmatrix}
    l_2' & l_2'' & l_2 \\
    0 & 0 & 0
  \end{pmatrix}
  C_{l_2m_2}(\hat{\bm{p}}).
  \label{eq:284}
\end{equation}

We decompose the primordial bispectrum with the spherical harmonics as
\begin{multline}
  B_\mathrm{L}(\bm{k}_1,\bm{k}_2,\bm{k}_3)
  =
  \sum_{l_1,l_2,l_3}
  B_{\mathrm{L}}^{l_1l_2l_3}(k_1,k_2,k_3)
  \left(l_1\,l_2\,l_3\right)^{m_1m_2m_3}
  \\ \times
  C_{l_1m_1}(\hat{\bm{k}}_1)
  C_{l_2m_2}(\hat{\bm{k}}_2)
  C_{l_3m_3}(\hat{\bm{k}}_3).
  \label{eq:285}
\end{multline}
One can always decompose the primordial bispectrum in the above form,
which is shown in a similar way of deriving Eq.~(\ref{eq:142})
when $l=m=0$. The appearance of the $3j$-symbols is due to the
rotational symmetry. The inverse relation of the above is given by
\begin{multline}
  B_{\mathrm{L}}^{l_1l_2l_3}(k_1,k_2,k_3)
  =
    (-1)^{l_1+l_2+l_3}
  \{l_1\}\{l_2\}\{l_3\}
  \left(l_1\,l_2\,l_3\right)^{m_1m_2m_3}
  \\ \times
  \int \frac{d^2\hat{k}_1}{4\pi} \frac{d^2\hat{k}_2}{4\pi}
  \frac{d^2\hat{k}_3}{4\pi}
  B_\mathrm{L}(\bm{k}_1,\bm{k}_2,\bm{k}_3).
  \\ \times
  C_{l_1m_1}(\hat{\bm{k}}_1)
  C_{l_2m_2}(\hat{\bm{k}}_2)
  C_{l_3m_3}(\hat{\bm{k}}_3).
  \label{eq:286}
\end{multline}
The particular bispectrum in the integrand of Eq.~(\ref{eq:282}) is
given by
\begin{multline}
  B_\mathrm{L}(\bm{k},-\bm{p},\bm{p})
  = \sum_{l}
  C_{lm}(\hat{\bm{k}})
  C_{lm}^*(\hat{\bm{p}})
  \sum_{l',l''}
  (-1)^{l'}
  \\ \times
  \begin{pmatrix}
    l' & l'' & l \\
    0 & 0 & 0
  \end{pmatrix}
  B_{\mathrm{L}}^{l\,l'l''}(k,p,p).
  \label{eq:287}
\end{multline}

Using the above equations together for the integral of
Eq.~(\ref{eq:282}), the angular integration over $\hat{\bm{p}}$ can
be analytically performed by using the orthonormality relation of the
spherical harmonics, Eq.~(\ref{eq:397}). Comparing the resulting
equation with Eq.~(\ref{eq:214}), the corresponding invariant power
spectrum is derived, and the result is given by
\begin{multline}
  P^{\mathrm{NG}\,l_1l_2;l}_{X_1X_2}(k) =
    \frac{(-1)^{l_2}}{2}
  \frac{\{l\}}{\sqrt{\{l_1\}}\,\{l_2\}}
  \begin{pmatrix}
    l_1 & l_2 & l \\
    0 & 0 & 0
  \end{pmatrix}
  \left[c^{(1)}_{X_1l_1}(k) + \delta_{l_10} c^{(0)}_{X_1}\right]
  \\ \times
  \sum_{l_1',l_1'',l_2',l_2''}
  (-1)^{l_1''+l_2''}
  \begin{pmatrix}
    l_1' & l_1'' & l_2 \\
    0 & 0 & 0
  \end{pmatrix}
  \begin{pmatrix}
    l_2' & l_2'' & l_2 \\
    0 & 0 & 0
  \end{pmatrix}
  \\ \times
  \int \frac{p^2dp}{2\pi^2}
  c^{(2)l_2}_{X_2l_2'l_2''}(p,p)
  B_\mathrm{L}^{l_2l_1'l_1''}(k,p,p)
  \\
  + [(X_1l_1) \leftrightarrow (X_2l_2)].
  \label{eq:288}
\end{multline}
This is one of the generic predictions of our theory. 

The scale-dependent bias from the primordial non-Gaussianity is
conveniently derived from the cross power spectrum between the linear
density field and biased objects. In particular, we define the
lowest-order scale-dependent bias factor by
\begin{equation}
  \Delta b_{Xlm}(\bm{k}) \equiv
  \frac{P^{\mathrm{NG}(0l)}_{\delta X\,0m}(\bm{k})}{P_\mathrm{L}(k)},
  \label{eq:289}
\end{equation}
where $X_1=\delta$ indicates that the biased field
$F_{X_1l_1m_1}(\bm{k})$ is replaced by the mass density field
$\delta(\bm{k})$ with $l_1=m_1=0$, and thus $c^{(0)}_{X_1} = 1$ and
$c^{(n)}_{X_1l_1m_1} = 0$ for $n\geq 1$. Because of the rotational
symmetry, and as explicitly seen from Eq.~(\ref{eq:214}), the
directional dependence on the wave vector on the rhs of
Eq.~(\ref{eq:289}) should be proportional to $C_{lm}(\hat{\bm{k}})$,
thus we naturally define an invariant bias factor $\Delta b_{Xl}(k)$
by
\begin{equation}
  \Delta b_{Xlm}(\bm{k}) =
  \frac{(-i)^l}{\sqrt{\{l\}}}
  \Delta b_{Xl}(k)
  C_{lm}(\hat{\bm{k}}).
  \label{eq:290}
\end{equation}
Substituting Eq.~(\ref{eq:214}) into Eq.~(\ref{eq:289}), the above
definition is equivalent to
\begin{equation}
  \Delta b_{Xl}(k) =
  \frac{P^{\mathrm{NG}\,0l;l}_{\delta X}(k)}{P_\mathrm{L}(k)}.
  \label{eq:291}
\end{equation}
Substituting $l_1=0$, $l_2=l$, $c^{(0)}_{X_1}=1$, and
$c^{(1)}_{X_1l_1} = c^{(2)l_1}_{X_1l_1'l_1''} = 0$ into
Eq.~(\ref{eq:288}), the invariant scale-dependent bias of
Eq.~(\ref{eq:291}) reduces to
\begin{multline}
  \Delta b_{Xl}(k) =
  \frac{1}{2 P_\mathrm{L}(k)}
  \sum_{l_1',l_1'',l_2',l_2''}
    \frac{(-1)^{l_1''+l_2''}}{\sqrt{\{l\}}}
  \begin{pmatrix}
    l_1' & l_1'' & l \\
    0 & 0 & 0
  \end{pmatrix}
  \begin{pmatrix}
    l_2' & l_2'' & l \\
    0 & 0 & 0
  \end{pmatrix}
  \\ \times
  \int \frac{p^2dp}{2\pi^2}
  c^{(2)l}_{Xl_2'l_2''}(p,p)
  B_\mathrm{L}^{l\,l_1'l_1''}(k,p,p).
  \label{eq:292}
\end{multline}
This is one of the generic predictions of our theory.

The above formula is applicable for arbitrary models of primordial
non-Gaussianity with a given bispectrum at the lowest order. We
consider a particular model of non-Gaussianity, which is a
generalization of the so-called local-type model
\cite{Shiraishi:2013vja,Schmidt:2015xka}:
\begin{multline}
  B_\mathrm{L}(\bm{k}_1,\bm{k}_2,\bm{k}_3) =
  \frac{2\mathcal{M}(k_3)}{\mathcal{M}(k_1)\mathcal{M}(k_2)}
  \\ \times
  \sum_l f_\mathrm{NL}^{(l)}
  \mathit{P}_l(\hat{\bm{k}}_1\cdot\hat{\bm{k}}_2)
  P_\mathrm{L}(k_1) P_\mathrm{L}(k_2) + \mathrm{cyc.}
  \label{eq:293}
\end{multline}
The function $\mathcal{M}(k)$ in the prefactor is given by
\begin{equation}
  \mathcal{M}(k) =
  \frac{2}{3}
  \frac{D(z)}{(1+z_*)D(z_*)}
  \frac{k^2 T(k)}{{H_0}^2 \Omega_\mathrm{m0}},
  \label{eq:294}
\end{equation}
where $D(z)$ is the linear growth factor with an arbitrary
normalization, $z_*$ is an arbitrary redshift at the matter-dominated
epoch, $T(k)$ is the transfer function, $H_0$ is the Hubble constant,
and $\Omega_\mathrm{m0}$ is the density parameter of total matter. The
factor $(1+z_*)D(z_*)$ does not depend on the choice of $z_*$ as long
as $z_*$ is deep in the matter-dominated epoch. Some authors choose a
normalization of the linear growth factor by $(1+z_*)D(z_*)=1$ in which
case the expression of Eq.~(\ref{eq:294}) has the simplest form.
Substituting Eq.~(\ref{eq:293}) into Eq.~(\ref{eq:286}), we have
\begin{multline}
  B_{\mathrm{L}}^{l_1l_2l_3}(k_1,k_2,k_3)
  = 
  \frac{2\mathcal{M}(k_3)}{\mathcal{M}(k_1)\mathcal{M}(k_2)}
  P_\mathrm{L}(k_1)P_\mathrm{L}(k_2)
  \\ \times
  \delta_{l_1l_2} \delta_{l_30}
  (-1)^{l_1}\sqrt{\{l_1\}}\, f_\mathrm{NL}^{(l_1)}
  + \mathrm{cyc.}
  \label{eq:295}
\end{multline}
We need a special form of the bispectrum to further evaluate
Eqs.~(\ref{eq:288}) and (\ref{eq:292}). With the above model of
Eq.~(\ref{eq:293}), we have
\begin{multline}
    B_\mathrm{L}^{l\,l'l''}(k,p,p)
  \simeq 
  \frac{2 P_\mathrm{L}(k)}{\mathcal{M}(k)}
  P_\mathrm{L}(p)
    \left( \delta_{l'l} \delta_{l''0} + \delta_{l'0}
      \delta_{l''l} \right)
  \\ \times
  (-1)^l \sqrt{\{l\}}\, f_\mathrm{NL}^{(l)},
  \label{eq:296}
\end{multline}
where only two terms in which $\mathcal{M}(k)$ appeared in the
denominator are retained because the other term is negligible in the
limit of $k\rightarrow 0$.

Substituting Eq.~(\ref{eq:296}) into Eq.~(\ref{eq:288}), and using
Eq.~(\ref{eq:427}), we have
\begin{multline}
  P^{\mathrm{NG}\,l_1l_2;l}_{X_1X_2}(k) =
  \left[1 + (-1)^{l_2}\right]
  f_\mathrm{NL}^{(l_2)}
  \frac{P_\mathrm{L}(k)}{\mathcal{M}(k)}
    \frac{\{l\}}{\sqrt{\{l_1\}}\,\{l_2\}}
  \begin{pmatrix}
    l_1 & l_2 & l \\
    0 & 0 & 0
  \end{pmatrix}
  \\ \times
  \left[
    c^{(1)}_{X_1l_1}(k) + \delta_{l_10} c^{(0)}_{X_1} 
  \right]
  \sum_{l',l''} (-1)^{l''}
  \begin{pmatrix}
    l' & l'' & l_2 \\
    0 & 0 & 0
  \end{pmatrix}
  \\ \times
  \int \frac{p^2dp}{2\pi^2}
  c^{(2)l_2}_{X_2l'l''}(p,p) P_\mathrm{L}(p)
  \\ + [(X_1l_1) \leftrightarrow (X_2l_2)].
  \label{eq:297}
\end{multline}
Substituting $l_1=0$, $l_2=l$, $c^{(0)}_{X_1}=1$, and
$c^{(1)}_{X_1l_1} = c^{(2)l_1}_{X_1l_1'l_1''} = 0$ into the above
equation, and using Eq.~(\ref{eq:291}), we derive
\begin{multline}
  \Delta b_{Xl}(k) =
 \left[1 + (-1)^l\right]
  \frac{f_\mathrm{NL}^{(l)}}{\mathcal{M}(k)} 
    \frac{1}{\sqrt{\{l\}}}
  \sum_{l',l''} (-1)^{l''}
  \begin{pmatrix}
    l' & l'' & l \\
    0 & 0 & 0
  \end{pmatrix}
  \\ \times
  \int \frac{p^2dp}{2\pi^2}
  c^{(2)l}_{Xl'l''}(p,p) P_\mathrm{L}(p).
  \label{eq:298}
\end{multline}
Apparently, the rhs survives only in the case of $l=\mathrm{even}$.
This proves that the scale-dependent bias of the rank-$l$ tensor field
is sensitive to the multipole moment $l$ of the Legendre polynomials
$P_l(\hat{\bm{k}}_1\cdot\hat{\bm{k}}_2)$ in the model of
Eq.~(\ref{eq:293}) for the primordial non-Gaussianity. That is
previously shown in Ref.~\cite{Kogai:2020vzz} by quite different
considerations and a method.

As a consistency check, we can see if the known results of scalar
perturbations are derived from the general result above. In the scalar
case of $l=0$, Eq.~(\ref{eq:298}) reduces to
\begin{equation}
  \Delta b_{X0}(k) =
  \frac{2f_\mathrm{NL}}{\mathcal{M}(k)}
  \sum_{l} \frac{1}{\sqrt{\{l\}}}
  \int \frac{p^2dp}{2\pi^2}
  c^{(2)0}_{Xll}(p,p) P_\mathrm{L}(p),
  \label{eq:299}
\end{equation}
where $f_\mathrm{NL} \equiv f_\mathrm{NL}^{(0)}$ is the original
parameter of local-type non-Gaussianity. One can confirm that the last
equation is consistent with the previously known results for the
scale-dependent bias in the halo models. In fact, the second-order
renormalized bias function of a simple halo model is given by
\cite{Matsubara:2012nc}
\begin{equation}
  c^{(2)}_\mathrm{h}(\bm{k}_1,\bm{k}_2) =
  \frac{\delta_\mathrm{c}b^\mathrm{L}_1}{\sigma^2}
  W(k_1R) W(k_2R),
  \label{eq:300}
\end{equation}
where
\begin{equation}
  \sigma^2 = \int \frac{k^2dp}{2\pi^2}
   W^2(kR) P_\mathrm{L}(k),
  \label{eq:301}
\end{equation}
and $b^\mathrm{L}_1$ is a Lagrangian bias parameter of linear order.
Equation~(\ref{eq:300}) is a result for the simplest case with a
high-peaks (or high-mass) limit of halos in the Press-Schechter mass
function (see Ref.~\cite{Matsubara:2012nc} for results of other
extended halo models). The Lagrangian bias parameter $b^\mathrm{L}_1$
is related to the Eulerian bias parameter $b_1$ by
$b^\mathrm{L}_1 = b_1 - 1$. Comparing the above function with
Eq.~(\ref{eq:132}) or Eq.~(\ref{eq:135}), and noting
$c^{(2)0}_{\mathrm{h}00}(k_1,k_2) = c^{(2)}_{\mathrm{h}}(k_1,k_2)$,
the invariant coefficient in this case is given by
\begin{equation}
  c^{(2)0}_{\mathrm{h}\,l_1l_2}(k_1,k_2) =
  \delta_{l_10}\delta_{l_20}
  \frac{\delta_\mathrm{c}b^\mathrm{L}_1}{\sigma^2}
  W(k_1R) W(k_2R).
  \label{eq:302}
\end{equation}
Substituting the above expression into Eq.~(\ref{eq:299}), we have
\begin{equation}
  \Delta b_{\mathrm{h}0}(k) =
  \frac{2f_\mathrm{NL}\delta_\mathrm{c} b^\mathrm{L}_1}{\mathcal{M}(k)}.
  \label{eq:303}
\end{equation}
Because of the normalization of the spherical basis for the scalar
component, Eqs.~(\ref{eq:7}) and (\ref{eq:32}), the scale-dependent
bias of the halo number density is given by
$\Delta b_\mathrm{h}(k) = \Delta b_{\mathrm{h}00}(k) \mathsf{Y}^{(0)}
= \Delta b_{\mathrm{h}0}$, and thus we have
\begin{equation}
  \Delta b_\mathrm{h}(k)
  = \frac{2(b_1-1)f_\mathrm{NL}\delta_\mathrm{c}}{\mathcal{M}(k)}.
  \label{eq:304}
\end{equation}
This exactly reproduces a well-known result of the scale-dependent
bias in the simplest halo model, which was first derived from a
quite different method \cite{Dalal:2007cu}.

\subsection{\label{subsec:Bispectrum}
  The bispectrum
}

\subsubsection{\label{subsubsec:InvBispec}
  The invariant bispectrum
}

For yet another example of applications of our formalism, we consider
the bispectrum of the tensor field. For an illustrative purpose, we only
consider the bispectrum in real space below, while generalizing the
results to those in redshift space is straightforward with
similar methods employed for the power spectrum in
Sec.~\ref{subsec:PowerSpec}.

The bispectrum
$B^{(l_1l_2l_3)}_{X_1X_2X_3\,m_1m_2m_3}(\bm{k}_1,\bm{k}_2,\bm{k}_3)$
of irreducible components of tensor field $F_{Xlm}(\bm{k})$ can be
defined by
\begin{multline}
  \left\langle
    F_{X_1l_1m_1}(\bm{k}_1) F_{X_2l_2m_2}(\bm{k}_2)
    F_{X_3l_3m_3}(\bm{k}_3)
  \right\rangle_\mathrm{c}
  \\
  =
  (2\pi)^3\delta_\mathrm{D}^3(\bm{k}_1+\bm{k}_2+\bm{k}_3)
  B^{(l_1l_2l_3)}_{X_1X_2X_3\,m_1m_2m_3}(\bm{k}_1,\bm{k}_2,\bm{k}_3),
  \label{eq:305}
\end{multline}
as a simple generalization of the definition of the power spectrum of
Eq.~(\ref{eq:211}). We consider generally the cross bispectrum among
three fields, $X_1$, $X_2$ and $X_3$ with irreducible components
$l_1$, $l_2$ and $l_3$, respectively. The autobispectrum is
straightforwardly obtained by putting $X_1=X_2=X_3$.

The bispectrum
$B^{(l_1l_2l_3)}_{X_1X_2X_3\,m_1m_2m_3}(\bm{k}_1,\bm{k}_2,\bm{k}_3)$
defined above depends on the coordinates system. Generalizing the
method of constructing the invariant power spectrum, one can also
construct the invariant bispectrum as explicitly shown below. Because
of the presence of the delta function on the rhs of
Eq.~(\ref{eq:305}), the third argument $\bm{k}_3$ in the bispectrum
can be replaced by $-\bm{k}_1-\bm{k}_2$, and regarded as a function of
only $\bm{k}_1$ and $\bm{k}_2$. However, the resulting expression
loses apparent symmetry of permutating the subscripts $(1,2,3)$ of
$\bm{k}_i$, $l_i$ and $m_i$. As is usually the case in the analytic
predictions in perturbation theory, we can decompose the symmetric
bispectrum into a sum of asymmetric components as
\begin{equation}
  B^{(l_1l_2l_3)}_{X_1X_2X_3\,m_1m_2m_3}(\bm{k}_1,\bm{k}_2,\bm{k}_3)
  =
    \tilde{B}^{(l_1l_2l_3)}_{X_1X_2X_3\,m_1m_2m_3}(\bm{k}_1,\bm{k}_2)
    + \mathrm{cyc.}
  \label{eq:306}
\end{equation}
Although the decomposition into asymmetric components is not unique,
it is always possible to specify the originally symmetric bispectrum by
giving the asymmetric bispectra, and the predictions of perturbation
theory are usually given in the above form.

The directional dependence of the asymmetric bispectrum on the
wave vectors can be expanded by spherical harmonics as
\begin{multline}
  \tilde{B}^{(l_1l_2l_3)}_{X_1X_2X_3m_1m_2m_3}(\bm{k}_1,\bm{k}_2)
  =
  \sum_{l_1',l_2'}
  \tilde{B}^{l_1l_2l_3;l_1'l_2'}_{X_1X_2X_3m_1m_2m_3;m_1'm_2'}(k_1,k_2)
  \\ \times
  C^*_{l_1'm_1'}(\hat{\bm{k}}_1) C^*_{l_2'm_2'}(\hat{\bm{k}}_2).
  \label{eq:307}
\end{multline}
Because of the rotational transformations of Eqs.~(\ref{eq:93}) and
(\ref{eq:23}), and rotational invariance of the delta function in the
definition of the bispectrum, Eq.~(\ref{eq:305}), the rotational
transformation of the expansion coefficients is given by
\begin{multline}
  \tilde{B}^{l_1l_2l_3;l_1'l_2'}_{X_1X_2X_3m_1m_2m_3;m_1'm_2'}(k_1,k_2)
  \\
  \xrightarrow{\mathbb{R}}
  \tilde{B}^{l_1l_2l_3;l_1'l_2'}_{X_1X_2X_3m_1''m_2''m_3'';m_1'''m_2'''}(k_1,k_2)
  D^{m_1''}_{(l_1)m_1}(R)
  \\ \times
  D^{m_2''}_{(l_2)m_2}(R)
  D^{m_3''}_{(l_2)m_3}(R) D^{m_1'''}_{(l)m_1'}(R) D^{m_2'''}_{(l)m_2'}(R).
  \label{eq:308}
\end{multline}
For the statistically isotropic Universe, the functional form of the
bispectrum should not depend on the choice of coordinates system.
Following exactly the same procedure as in
Sec.~\ref{subsec:RenBiasFn}, the bispectrum in the statistically
isotropic Universe is shown to have a form,
\begin{multline}
  \tilde{B}^{(l_1l_2l_3)}_{X_1X_2X_3m_1m_2m_3}(\bm{k}_1,\bm{k}_2) =
  i^{l_1+l_2+l_3}
  \sum_{L,L'}
   (-1)^{L+L'}
   \sqrt{\{L\}\{L'\}}
   \\ \times
  \left(l_1\,l_2\,L\right)_{m_1m_2}^{\phantom{m_1m_2}M}
  \left(L\,l_3\,L'\right)_{Mm_3}^{\phantom{Mm_3}M'}
  \sum_{l_1',l_2'}
   \sqrt{\{l_1'\}\{l_2'\}}
  \\ \times
  X^{l_1'l_2'}_{L'M'}(\hat{\bm{k}}_1,\hat{\bm{k}}_2)
  \tilde{B}^{l_1l_2l_3;l_1'l_2';LL'}_{X_1X_2X_3}(k_1,k_2),
    \label{eq:309}
\end{multline}
where
\begin{multline}
  \tilde{B}^{l_1l_2l_3;l_1'l_2';LL'}_{X_1X_2X_3}(k_1,k_2) =
  i^{l_1+l_2+l_3}(-1)^{l_1'+l_2'+L+L'}
  \sqrt{\frac{\{L\}\{L'\}}{\{l_1'\}\{l_2'\}}}
  \\ \times
    \left(l_1\,l_2\,L\right)^{m_1m_2}_{\phantom{m_1m_2}M}
  \left(L\,l_3\,L'\right)^{Mm_3}_{\phantom{Mm_3}M'}
  \left(L'\,l_1'\,l_2'\right)^{M'm_1'm_2'}
  \\ \times
  \tilde{B}^{l_1l_2l_3;l_1'l_2'}_{X_1X_2X_3m_1m_2m_3;m_1'm_2'}(k_1,k_2)
  \label{eq:310}
\end{multline}
is the invariant bispectrum.

According to Eqs.~(\ref{eq:86}), (\ref{eq:305}) and the symmetry of
the $3j$-symbol, Eq.~(\ref{eq:421}), the symmetry of the complex
conjugate is shown to be given by
\begin{equation}
  B^{l_1l_2l_3;l_1'l_2';LL'\,*}_{X_1X_2X_3}(k) =
  B^{l_1l_2l_3;l_1'l_2';LL'}_{X_1X_2X_3}(k),
  \label{eq:311}
\end{equation}
and thus the invariant bispectrum is a real function. Additionally,
the parity transformation of the bispectrum is given by
\begin{multline}
  \tilde{B}^{(l_1l_2l_3)}_{X_1X_2X_3m_1m_2m_3}(\bm{k}_1,\bm{k}_2)
  \xrightarrow{\mathbb{P}}
  (-1)^{p_{X_1}+p_{X_2}+p_{X_3}+l_1+l_2+l_3}
  \\ \times
  \tilde{B}^{(l_1l_2l_3)}_{X_1X_2X_3m_1m_2m_3}(-\bm{k}_1,-\bm{k}_2),
  \label{eq:312}
\end{multline}
because of Eqs.~(\ref{eq:99}) and (\ref{eq:305}). Accordingly, the
parity transformation of the invariant bispectrum is given by
\begin{multline}
  \tilde{B}^{l_1l_2l_3;l_1'l_2'}_{X_1X_2X_3}(k_1,k_2) \xrightarrow{\mathbb{P}}
  (-1)^{p_{X_1}+p_{X_2}+p_{X_3}+l_1+l_2+l_3+l_1'+l_2'}
  \\ \times
  \tilde{B}^{l_1l_2l_3;l_1'l_2'}_{X_1X_2X_3}(k_1,k_2).
  \label{eq:313}
\end{multline}
In a Universe with the parity symmetry, the bispectrum is invariant
under the parity transformation, we thus have
\begin{equation}
  p_{X_1} + p_{X_2} + p_{X_3} + l_1 + l_2 + l_3 + l_1' + l_2'
  = \mathrm{even}.
  \label{eq:314}
\end{equation}

\subsubsection{\label{subsubsec:TreeBispec}
  The lowest-order bispectrum
}

We consider the bispectrum of the lowest-order approximation or the
tree-level bispectrum. The formal expression of the lowest-order
bispectrum in terms of propagators is given by
\begin{multline}
  \tilde{B}^{(l_1l_2l_3)}_{X_1X_2X_3\,m_1m_2m_3}(\bm{k}_1,\bm{k}_2)
  =
    i^{l_1+l_2+l_3}
  \Gamma^{(1)}_{X_1l_1m_1}(\bm{k}_1) \Gamma^{(1)}_{X_2l_2m_2}(\bm{k}_2)
  \\ \times
  \Gamma^{(2)}_{X_3l_3m_3}(-\bm{k}_1,-\bm{k}_2)
  P_\mathrm{L}(k_1) P_\mathrm{L}(k_2).
  \label{eq:315}
\end{multline}
We do not use the higher-order resummation factor and substitute
$\Pi(k) = 1$. Substituting Eqs.~(\ref{eq:160}) and (\ref{eq:162}),
and applying Eq.~(\ref{eq:209}), the above equation is represented
in terms of invariant functions as
\begin{multline}
  \tilde{B}^{(l_1l_2l_3)}_{X_1X_2X_3\,m_1m_2m_3}(\bm{k}_1,\bm{k}_2)
  =
  \frac{i^{l_1+l_2-l_3}}{\sqrt{\{l_1\}\{l_2\}\{l_3\}}}
  P_\mathrm{L}(k_1) P_\mathrm{L}(k_2)
  \\ \times
  \hat{\Gamma}^{(1)}_{X_1l_1}(k_1)
  \hat{\Gamma}^{(1)}_{X_2l_2}(k_2)
  \sum_{l_1',l_2',l_1'',l_2''}
  (-1)^{l_1'+l_2'} \{l_1'\}\{l_2'\}
  \\ \times
  \left[l_1\,l_1'\,l_1''\right]_{m_1}^{\phantom{m_1}m_1'm_1''}
  \left[l_2\,l_2'\,l_2''\right]_{m_2}^{\phantom{m_2}m_2'm_2''}
  \left(l_1''\,l_2''\,l_3\right)_{m_1''m_2''m_3}
  \\ \times
  C_{l_1'm_1'}(\hat{\bm{k}}_1) C_{l_2'm_2'}(\hat{\bm{k}}_2)
  \hat{\Gamma}^{(2)\,l_3}_{X_3l_1''l_2''}(k_1,k_2).
  \label{eq:316}
\end{multline}
Substituting Eqs.~(\ref{eq:195}) and (\ref{eq:210}), the above
expression is given by invariant forms of renormalized bias functions,
$c^{(0)}_X$, $c^{(1)}_{X\,l}(k)$ and $c^{(2)\,l}_{Xl_1l_2}(k_1,k_2)$.

The sum of the products of three $3j$-symbols appearing in the above
expression is represented by a $9j$-symbol using the formulas of
Eqs.~(\ref{eq:428}) and (\ref{eq:444}) in
Appendix~\ref{app:3njSymbols}. Comparing the resulting equation with
Eq.~(\ref{eq:309}), the invariant bispectrum is derived, and the
result is given by
\begin{multline}
  \tilde{B}^{l_1l_2l_3;l_1'l_2';LL'}_{X_1X_2X_3}(k_1,k_2)
  = (-1)^{L+L'} 
  \sqrt{\frac{\{l_1'\}\{l_2'\}\{L\}\{L'\}}{\{l_1\}\{l_2\}\{l_3\}}}
  \\ \times
  \hat{\Gamma}^{(1)}_{X_1l_1}(k_1)
  \hat{\Gamma}^{(1)}_{X_2l_2}(k_2)
  P_\mathrm{L}(k_1) P_\mathrm{L}(k_2)
  \sum_{l_1'',l_2''} (-1)^{l_1''+l_2''}  
  \\ \times
  \begin{pmatrix}
    l_1 & l_1' & l_1'' \\
    0 & 0 & 0
  \end{pmatrix}
  \begin{pmatrix}
    l_2 & l_2' & l_2'' \\
    0 & 0 & 0
  \end{pmatrix}
  \begin{Bmatrix}
    l_1 & l_1' & l_1'' \\
    l_2 & l_2' & l_2'' \\
    L & L' & l_3
  \end{Bmatrix}
  \hat{\Gamma}^{(2)\,l_3}_{X_3l_1''l_2''}(k_1,k_2).
  \label{eq:317}
\end{multline}
This is one of the generic predictions of our theory. 

As a consistency check, let us see if the bispectrum of the scalarly
biased field in the lowest-order perturbation theory can be correctly
reproduced. Considering the case of $l_1=l_2=l_3=m_1=m_2=m_3=0$ and
$X_1=X_2=X_3\equiv X$ in Eq.~(\ref{eq:316}), and using
Eqs.~(\ref{eq:427}), (\ref{eq:433}) and (\ref{eq:395}), we
straightforwardly derive
\begin{multline}
  \tilde{B}^{(000)}_{XXX\,000}(\bm{k}_1,\bm{k}_2,\bm{k}_3)
  =
  P_\mathrm{L}(k_1) P_\mathrm{L}(k_2)
  \sum_l \frac{(-1)^l}{\sqrt{\{l\}}}
  \mathit{P}_l(\hat{\bm{k}}_1\cdot\hat{\bm{k}}_2)
  \\ \times
  \hat{\Gamma}^{(1)}_{X0}(k_1) \hat{\Gamma}^{(1)}_{X0}(k_2)
  \hat{\Gamma}^{(2)\,0}_{Xll}(k_1,k_2).
  \label{eq:318}
\end{multline}
The same result is also derived from Eqs.~(\ref{eq:309}) and
(\ref{eq:317}) with Eq.~(\ref{eq:251}). We substitute
Eqs.~(\ref{eq:196}) and (\ref{eq:210}) in the above equation,
and straightforward calculations derive the scalar bispectrum as
\begin{multline}
  B_X(\bm{k}_1,\bm{k}_2,\bm{k}_3) =
  B^{(000)}_{XXX\,000}(\bm{k}_1,\bm{k}_2,\bm{k}_3)
  \\
  =
    b_1(k_1) b_1(k_2)
  P_\mathrm{L}(k_1) P_\mathrm{L}(k_2)
  \Biggl\{
  b^\mathrm{L}_2(\bm{k}_1,\bm{k}_2)
  + b_1(k_1) +  b_1(k_2)
  \\
    - \frac{4}{7}
    + \left[
      \frac{k_2}{k_1} b_1(k_1) +
      \frac{k_1}{k_2} b_1(k_2)
    \right]
    \frac{\bm{k}_1\cdot\bm{k}_2}{k_1k_2}
    + \frac{4}{7}
    \left(\frac{\bm{k}_1\cdot\bm{k}_2}{k_1k_2}\right)^2
    \Biggr\}
    \\
    + \mathrm{cyc.},
  \label{eq:319}
\end{multline}
where $b_1(k) \equiv c^{(0)}_X + c^{(1)}_{X0}(k)$ corresponds to the
Eulerian bias factor and is the same as $b_X(k)$ defined in
Eq.~(\ref{eq:245}), and
\begin{equation}
  b^\mathrm{L}_2(\bm{k}_1,\bm{k}_2)
  \equiv c^{(2)}_{X00}(\bm{k}_1,\bm{k}_2)
  = \sum_l \frac{(-1)^l}{\sqrt{\{l\}}}
    \mathit{P}_l(\hat{\bm{k}}_1\cdot\hat{\bm{k}}_2)
    c^{(2)0}_{Xll}(k_1,k_2)
  \label{eq:320}
\end{equation}
corresponds to the Lagrangian (nonlocal) bias function of the second order
\cite{Matsubara:2011ck}. The more commonly known form of the bispectrum in
the lowest-order perturbation theory is represented by the Eulerian
bias parameters $b_1$, $b_2$, $b_{s^2}$
\cite{Matarrese:1997sk,McDonald:2009dh,Sheth:2012fc,Baldauf:2012hs}.
It is natural to take a normalization
$\langle F^\mathrm{L}_X\rangle = 1$ [cf., Eq.~(\ref{eq:50})] and we
have $c^{(0)}_X=1$ in this scalar case. On one hand, in the special
case of local bias, in which $b_1$ and $b^\mathrm{L}_2$ are constants
and do not depend on wave vectors, Eq.~(\ref{eq:319}) reduces to
\begin{multline}
  B_X(\bm{k}_1,\bm{k}_2,\bm{k}_3)
  =
  {b_1}^2
  P_\mathrm{L}(k_1) P_\mathrm{L}(k_2)
  \\ \times
  \left[
  b^\mathrm{L}_2 + 2b_1
    - \frac{4}{7}
    + b_1
    \left(
      \frac{k_2}{k_1} + \frac{k_1}{k_2}
      \right)
    \frac{\bm{k}_1\cdot\bm{k}_2}{k_1k_2}
    + \frac{4}{7}
    \left(\frac{\bm{k}_1\cdot\bm{k}_2}{k_1k_2}\right)^2
    \right]
    \\
    + \mathrm{cyc.}
  \label{eq:321}
\end{multline}
On the other hand, the prediction of scalar bispectrum in the standard
(Eulerian) perturbation theory is given by \cite{Baldauf:2012hs},
\begin{multline}
  B_X(\bm{k}_1,\bm{k}_2,\bm{k}_3)
  =
  {b_1}^2
  P_\mathrm{L}(k_1) P_\mathrm{L}(k_2)
  \\ \times
  \left\{
    b_2 + 2b_{s^2}
    \left[
      \left(\frac{\bm{k}_1\cdot\bm{k}_2}{k_1k_2}\right)^2
      - \frac{1}{3}
    \right]
    + b_1 F_2(\bm{k}_1,\bm{k}_2)
    \right\}
    \\
    + \mathrm{cyc.},
  \label{eq:322}
\end{multline}
where
\begin{equation}
  F_2(\bm{k}_1,\bm{k}_2) =
  \frac{10}{7} + 
  \left(
    \frac{k_2}{k_1} + \frac{k_1}{k_2}
  \right)
  \frac{\bm{k}_1\cdot\bm{k}_2}{k_1k_2}
  + \frac{4}{7}
  \left(\frac{\bm{k}_1\cdot\bm{k}_2}{k_1k_2}\right)^2
  \label{eq:323}
\end{equation}
is the second-order kernel\footnote{Our choice of the normalizations
  for kernel functions $F_n$ is different from those in many previous
  references, where our $F_n$ corresponds to $n!F_n$ in the latter,
  minimizing the occurrence of $n!$ in resulting expressions
  \cite{Matsubara:2011ck}. } of the Eulerian perturbation theory
\cite{Bernardeau:2001qr}. Comparing the coefficients of
Eqs.~(\ref{eq:321}) and (\ref{eq:322}), we find the relations between
Lagrangian local bias parameters and Eulerian local bias parameters as
\begin{equation}
  b_1 =  1 + b^\mathrm{L}_1, \quad
  b_2 =  b^\mathrm{L}_2 - \frac{8}{21} b^\mathrm{L}_1, \quad
  b_{s^2} =  - \frac{2}{7} b^\mathrm{L}_1.
    \label{eq:324}
\end{equation}
The above relations exactly agree with those which are found in the
literature \cite{Sheth:2012fc,Baldauf:2012hs}. It is also possible to include the
nonlocal Lagrangian bias $b^\mathrm{L}_{s^2}$ in
$b^\mathrm{L}_2(\bm{k}_1,\bm{k}_2)$, and derive consistent results
with those in the literature \cite{Sheth:2012fc,Baldauf:2012hs}.

\section{\label{sec:Semilocal}
  semilocal models of bias for tensor fields
}

So far the formulation is general enough, and we have not assumed any
model of bias for the tensor fields so far. The bias model is
naturally specified in Lagrangian space through the renormalized bias
functions $c^{(n)}_{Xlm}$ defined by Eq.~(\ref{eq:71}), which is
evaluated once a model of bias $F^\mathrm{L}_{Xlm}(\bm{q})$ is
analytically given as a functional of the linear density field
$\delta_\mathrm{L}(\bm{q})$. In this section, we consider a category
of bias models which are commonly adopted in models of cosmological
structure formation. We call this category ``semilocal models'' of
bias.

\subsection{The semilocal models of bias}

\subsubsection{Defining the semilocal models}

The concept of the semilocal models is somehow similar to the
EFTofLSS approach in perturbation theory of biased tracers
\cite{Mirbabayi:2014zca,Desjacques:2016bnm}, where the biased field is
given by a finite set of semilocal operators made of spatial and
temporal derivatives of the gravitational potential in the
perturbative expansions order by order. However, while the EFTofLSS
approach is based on phenomenologically perturbative expansions of the
biased field, our iPT approach does not assume the underlying
operators are perturbative, because we essentially use orthogonal
expansions instead of Taylor expansions to include the fully nonlinear
biasing into the cosmological perturbation theory through the
renormalized bias functions.

This category of semilocal models of bias is defined so that the
field $F^\mathrm{L}_{Xlm}(\bm{q})$ is given by functions (instead of
functionals) of the spatial derivatives of the gravitational potential at
the same position $\bm{q}$, smoothed with (generally, multiple numbers
of) window functions $W^{(a)}(k)$:
\begin{equation}
  \chi^{(a)}_{i_1i_2\cdots i_{L_a}}(\bm{q}) =
  \partial_{i_1}\partial_{i_2}\cdots\partial_{i_{L_a}}
  \psi^{(a)}(\bm{q}),
  \label{eq:325}
\end{equation}
where
\begin{equation}
  \psi^{(a)}(\bm{q}) =
  \int \frac{d^3k}{(2\pi)^3} e^{i\bm{k}\cdot\bm{q}}
  \delta_\mathrm{L}(\bm{k})\,
   k^{-L_a}W^{(a)}(k)
  \label{eq:326}
\end{equation}
is the smoothed linear density field with an isotropic window function
$k^{-L_a} W^{(a)}(k)$. The label ``$a$'' distinguishes different kinds
of the linear tensor field with a particular rank $L_a$, i.e., the
label $a$ uniquely specifies the rank $L_a$ and the form of the window
function $W^{(a)}(k)$. The window function usually contains a
parameter of smoothing radius $R_a$, which can take different values
for each window function $W^{(a)}$. While typical window functions
include the Gaussian window, $W^{(a)}(k) = \exp(-k^2{R_a}^2/2)$, and
the top-hat window, $W^{(a)}(k) = 3j_1(kR_a)/(kR_a)$, we do not assume
any specific form of the window function here. When only a single linear
tensor field of a fixed window function is considered, one can omit
the label $a$ in the above. We include the possibility of using
multiple numbers of fields and window functions in the general
formulation below.

For a concrete example of biasing using multiple numbers of fields and
window functions, see Ref.~\cite{Matsubara:2016wth}. For example, the
second derivatives of the (normalized and smoothed) linear
gravitational potential $\varphi$ correspond to
$\varphi = \Laplace^{-1} \delta_R = -\psi^{(a)}$ with a rank of
$L_a=2$, and
$\chi^{(a)}_{ij} = -\partial_i\partial_j\Laplace^{-1}\delta_R$ where
$\delta_R$ is a smoothed linear density field with a smoothing window
function $W^{(a)}(k) = W(kR)$, and $\Laplace^{-1}$ is the inverse
operator of the Laplacian $\Laplace = \partial_i\partial_i$.

The linear density contrast $\delta_\mathrm{L}$ is a real function in
configuration space, and thus we have
$\delta_\mathrm{L}^{*}(\bm{k}) = \delta_\mathrm{L}(-\bm{k})$ as shown
from Eq.~(\ref{eq:62}). If the derivative field
$\chi^{(a)}_{i_1\cdots i_{L_a}}(\bm{q})$ and thus $\psi^{(a)}(\bm{q})$
are real functions, the window $W^{(a)}(k)$ is a real function.

In the semilocal models of bias, the field value
$F^\mathrm{L}_{Xlm}(\bm{q})$ at a particular position is given by a
function of the field values $\chi^{(a)}_{i_1\cdots i_l}(\bm{q})$ at
the same position in Lagrangian space. This function is common among
all the positions. Therefore, without loss of generality, one can
consider the particular position at the origin, $\bm{q}=\bm{0}$, to
describe the relation in the semilocal model of bias, and thus the
field $F^\mathrm{L}_{Xlm}$ is a function of
\begin{equation}
  \chi^{(a)}_{i_1\cdots i_{L_a}}
  = i^{L_a}
  \int \frac{d^3k}{(2\pi)^3}\,
  \hat{k}_{i_1}\cdots\hat{k}_{i_{L_a}}\,
  \delta_\mathrm{L}(\bm{k}) W^{(a)}(k),
  \label{eq:327}
\end{equation}
where $\hat{k}_i = k_i/k$. The tensor field $F^\mathrm{L}_{Xlm}$ can
depend on multiple local tensors of various ranks, $\chi^{(a_0)}$,
$\chi^{(a_1)}_i$, $\chi^{(a_2)}_{ij}$, $\chi^{(a_3)}_{ijk}$, etc.

There is a caveat in the terminology here. In conventional
perturbation theory, the bias relations involving the nondiagonal part
of second-order derivatives of the gravitational potential,
$\partial_i\partial_j\varphi$, are often referred to as a ``nonlocal
bias.'' In the terminology of this paper and of iPT in general, this
type of bias also falls into the category of ``semilocal bias.''

\subsubsection{Irreducible decomposition of semilocal variables}

A local tensor can be decomposed by an irreducible spherical basis
according to the procedure explained in
Sec.~\ref{sec:SphericalBasis}. The tensor of Eq.~(\ref{eq:327}) is
decomposed into traceless tensor of rank $L_a,L_a-2,\ldots$, and each
component of traceless tensor is decomposed by a spherical basis, as
in the Eqs.~(\ref{eq:36})--(\ref{eq:43}). The decomposition is given
by
\begin{equation}
  \chi^{(a)}_{i_1\cdots i_l}
  = \chi^{(a,l)}_{i_1i_2\cdots i_l} + 
  \frac{l(l-1)}{2(2l-1)}
  \delta_{(i_1i_2} \chi^{(a,l-2)}_{i_3\cdots i_l)}
  + \cdots,
  \label{eq:328}
\end{equation}
where we simply denote $L_a=l$ in the above, and
$\chi^{(a,l-2)}_{i_3\cdots i_l}$ is the traceless part of the first
trace $\chi^{(a)}_{jji_3\cdots i_l}$ and so forth.

The decomposed traceless tensors are further decomposed by spherical
basis. Similarly to Eqs.~(\ref{eq:42}) and (\ref{eq:44}), we define
\begin{align}
  \chi^{(a,l)}_{i_1i_2\cdots i_l}
  &=
    A_l\,   
    \chi^{(a[l])}_{lm} \mathsf{Y}^{(m)}_{i_1i_2\cdots i_l},
  \label{eq:329}\\
  \chi^{(a,l-2)}_{i_3\cdots i_l}
  &=
      A_{l-2}\,
      \chi^{(a[l])}_{l-2,m}
  \mathsf{Y}^{(m)}_{i_3\cdots i_l},
  \label{eq:330}
\end{align}
and so forth.
In our simplified notation with $l=L_a$, the rank of the
linear tensor field $a$ is not obvious if we write, e.g.,
$\chi^{(a)}_{l-2,m}$ for the first trace of the original tensor, and
thus we instead employ a notations $\chi^{(a[l])}_{lm}$,
$\chi^{(a[l])}_{l-2,m}$, etc.~to remind the rank of the original field
$a$ is $l$ in the lower-rank trace parts of the tensor. Namely,
\begin{equation}
  \chi^{(a[l])}_{lm}
  = \left.\chi^{(a)}_{L_am}\right|_{L_a=l}, \quad
  \chi^{(a[l])}_{l-2,m}
  = \left.\chi^{(a)}_{L_a-2,m}\right|_{L_a=l},
  \label{eq:331}
\end{equation}
and so forth.

The relations of Eqs.~(\ref{eq:329}) and (\ref{eq:330}) are inverted
as well as Eqs.~(\ref{eq:43}) and (\ref{eq:45}), and further using
Eqs.~(\ref{eq:18}) and (\ref{eq:43}), the spherical tensors are given
by
\begin{align}
  \chi^{(a[l])}_{lm}
  &=
    \frac{1}{A_l}\,
  \chi^{(a)}_{i_1\cdots i_l}
  \mathsf{Y}^{(m)*}_{i_1\cdots i_l}
    \nonumber\\
  &=
      i^l
    \int \frac{d^3k}{(2\pi)^3}
    \delta_\mathrm{L}(\bm{k})
    C_{lm}(\hat{\bm{k}}) W^{(a)}(k),
    \label{eq:332}\\
  \chi^{(a[l])}_{l-2,m}
  &=
      \frac{1}{A_{l-2}}\,  
  \chi^{(a)}_{jji_3\cdots i_l}
    \mathsf{Y}^{(m)*}_{i_3\cdots i_l}
    \nonumber\\
  &= i^l
    \int \frac{d^3k}{(2\pi)^3}
    \delta_\mathrm{L}(\bm{k})
    C_{l-2,m}(\hat{\bm{k}})
    W^{(a)}(k),
    \label{eq:333}
\end{align}
and so forth, where $A_l$ is given by Eq.~(\ref{eq:19}), and
$C_{lm}(\hat{\bm{k}})$ is the spherical harmonics with Racah's
normalization. In general, we have
\begin{align}
  \chi^{(a[l])}_{l-2p,m}
  &=
      \frac{1}{A_{l-2p}}\,
    \chi^{(a)}_{j_1j_1\cdots j_pj_pi_{2p+1}\cdots i_l}
    \mathsf{Y}^{(m)*}_{i_{2p+1}\cdots i_l}
    \nonumber\\
  &=
      i^l
    \int \frac{d^3k}{(2\pi)^3}
    \delta_\mathrm{L}(\bm{k})
    C_{l-2p,m}(\hat{\bm{k}})
    W^{(a)}(k),
    \label{eq:334}
\end{align}
where $p=0,1,\ldots,[l/2]$. Changing the labels of ranks by
$l \rightarrow L$ and $l-2p \rightarrow l$, Eq.~(\ref{eq:334}) is
equivalently represented by
\begin{equation}
  \chi^{(a[L])}_{lm}
  =
    i^L
  \int \frac{d^3k}{(2\pi)^3}
  \delta_\mathrm{L}(\bm{k})
  C_{lm}(\hat{\bm{k}})
  W^{(a)}(k),
  \label{eq:335}
\end{equation}
where $l=L,L-2$, \ldots, (0 or 1) for a linear tensor field $a$ of
rank $L$. The smallest value of $l$ in the last equation is 0 if $L$
is even, and 1 if $L$ is odd. When the rank $L_a$ of the original
tensor $\chi^{(a)}_{i_1\cdots i_{L_a}}$ is obvious, one can employ a
simplified notation $\chi^{(a)}_{lm}$ instead of $\chi^{(a[L])}_{lm}$,
where $l = L, L-2,\ldots$. In Eq.~(\ref{eq:335}), $L-l$ is an even
number, and the complex conjugate of the variable is given by
\begin{equation}
  \chi^{(a)*}_{lm} = g_{(l)}^{mm'} \chi^{(a)}_{lm'},
  \label{eq:336}
\end{equation}
where a simplified notation $\chi^{(a)}_{lm}$ is adopted to represent
the original variable, $\chi^{(a[L_a])}_{lm}$.

\subsubsection{Simple examples}

For a simple example, we consider the second derivatives of the
normalized potential $\varphi=\Laplace^{-1}\delta_R$, smoothed with a
window function $W(kR)$. In this case, we have
\begin{equation}
  \chi^{(\varphi)}_{ij} =
  - \partial_i\partial_j\varphi
  = - \int \frac{d^3k}{(2\pi)^3}\, \hat{k}_i \hat{k}_j
  \delta_\mathrm{L}(\bm{k}) W(kR).
  \label{eq:337}
\end{equation}
The general variables in Eqs.~(\ref{eq:326}) and (\ref{eq:327})
correspond to $\psi^{(a)}=-\varphi$ and $W^{(\varphi)}(k) = W(kR)$.
The second-rank tensor is decomposed into irreducible components as
\begin{align}
  \chi^{(\varphi)}_{ij}
  &= -
  \left(
    \partial_i\partial_j - \frac{\delta_{ij}}{3} \Laplace
  \right)
  \varphi -
    \frac{\delta_{ij}}{3} \Laplace \varphi
    \nonumber \\
  &=
    \sqrt{\frac{2}{3}}\,\chi^{(\varphi)}_{2m}\, \mathsf{Y}^{(m)}_{ij}
    + \frac{\delta_{ij}}{3}\, \chi^{(\varphi)}_{00}\, \mathsf{Y}^{(0)},
  \label{eq:338}
\end{align}
where, corresponding to Eqs.~(\ref{eq:332}) and (\ref{eq:333}), we
have
\begin{align}
  \chi^{(\varphi[2])}_{2m}
  &=
  - \int \frac{d^3k}{(2\pi)^3}
    \delta_\mathrm{L}(\bm{k})
    C_{2m}(\hat{\bm{k}})
    W(kR),
    \label{eq:339}\\
  \chi^{(\varphi[2])}_{00}
  &=
  -
    \int \frac{d^3k}{(2\pi)^3}
    \delta_\mathrm{L}(\bm{k})
    C_{00}(\hat{\bm{k}})
    W(kR).
  \label{eq:340}
\end{align}
The last variable is apparently given by
\begin{equation}
  \chi^{(\varphi[2])}_{00} =
  -
  \int \frac{d^3k}{(2\pi)^3}
  \delta_\mathrm{L}(\bm{k}) W(kR)
  \equiv
  -\,
  \delta_R,
  \label{eq:341}
\end{equation}
where $\delta_R$ is the smoothed density field with a smoothing
function $W(kR)$.

Similarly, the second derivatives of the smoothed density field are
given by
\begin{equation}
  \chi^{(\delta)}_{ij} =
  \partial_i\partial_j\delta_R
  = - \int \frac{d^3k}{(2\pi)^3}\,k^2\, \hat{k}_i \hat{k}_j
  \delta_\mathrm{L}(\bm{k}) W(kR).
  \label{eq:342}
\end{equation}
The general variables in Eqs.~(\ref{eq:326}) and (\ref{eq:327})
correspond to $\psi^{(a)}=\delta_R$ and
$W^{(\delta)}(k) = k^2 W(kR)$. The second-rank tensor is decomposed
into irreducible components as
\begin{align}
  \chi^{(\delta)}_{ij} =
  \partial_i\partial_j \delta_R
  &=
    \left(
    \partial_i\partial_j - \frac{\delta_{ij}}{3} \Laplace
    \right) \delta_R
    + \frac{\delta_{ij}}{3} \Laplace \delta_R
    \nonumber \\
  &=
    \sqrt{\frac{2}{3}}\chi^{(\delta[2])}_{2m} \mathsf{Y}^{(m)}_{ij}
    + \frac{\delta_{ij}}{3} \chi^{(\delta[2])}_{00} \mathsf{Y}^{(0)},
  \label{eq:343}
\end{align}
where
\begin{align}
  \chi^{(\delta[2])}_{2m}
  &=
  - \int \frac{d^3k}{(2\pi)^3}
  \delta_\mathrm{L}(\bm{k})
    C_{2m}(\hat{\bm{k}})\,
    k^2 W(kR),
    \label{eq:344}\\
  \chi^{(\delta[2])}_{00}
  &=
      -
  \int \frac{d^3k}{(2\pi)^3}
  \delta_\mathrm{L}(\bm{k})
    C_{00}(\hat{\bm{k}})\,
    k^2 W(kR),
  \label{eq:345}
\end{align}
and the last variable is apparently given by
\begin{equation}
  \chi^{(\delta[2])}_{00} =
      -
  \int \frac{d^3k}{(2\pi)^3}
  \delta_\mathrm{L}(\bm{k})\,k^2 W(kR)
  =
      \Laplace\delta_R.
  \label{eq:346}
\end{equation}

\subsection{Renormalized bias parameters}

\subsubsection{Renormalized bias functions in the semilocal models}

In the semilocal models of bias, the tensor field
$F^\mathrm{L}_{Xlm}$ is generally considered as a function of various
irreducible tensors
\begin{equation}
  F^\mathrm{L}_{Xlm} = 
  F^\mathrm{L}_{Xlm}\left(\left\{\chi^{(a)}_{l'm'}\right\}\right)
  \label{eq:347}
\end{equation}
at every position in (Lagrangian) configuration space. In this case,
the functional derivative in the definition of renormalized bias
functions, Eq.~(\ref{eq:71}), can be replaced by (with a simplified
notation $\tilde{\delta}_\mathrm{L} \rightarrow \delta_\mathrm{L}$ in
Fourier space)
\begin{equation}
  (2\pi)^3 \frac{\delta}{\delta\delta_\mathrm{L}(\bm{k})}
  \rightarrow
  \sum_{a}
    i^{L_a}
  W^{(a)}(k)
  \sum_{l=L_a,L_a-2,\ldots}
  C_{lm}(\hat{\bm{k}})
  \frac{\partial}{\partial \chi^{(a)}_{lm}},
  \label{eq:348}
\end{equation}
where the integer $L_a$ is the original rank of the linear tensor
$\chi^{(a)}_{i_1\cdots i_{L_a}}$. The summation over $m$ in
Eq.~(\ref{eq:348}) is implicitly assumed according to the Einstein
summation convention, just as in the rest of this paper. Therefore,
the renormalized bias functions of Eq.~(\ref{eq:71}) are given by
\begin{multline}
  c^{(n)}_{Xlm}(\bm{k}_1,\ldots\bm{k}_n)
  =
    (-i)^l
  \sum_{a_1,\ldots,a_n}
    i^{L_{a_1}+\cdots +L_{a_n}}
  \\ \times
  W^{(a_1)}(k_1) \cdots W^{(a_n)}(k_n)
  \sum_{l_1,\ldots,l_n}
  C_{l_1m_1}(\hat{\bm{k}}_1)
  \cdots C_{l_n,m_n}(\hat{\bm{k}}_n)
  \\ \times
  \left\langle
  \frac{\partial^n F^\mathrm{L}_{Xlm}}
  {\partial \chi^{(a_1)}_{l_1m_1} \cdots \partial
    \chi^{(a_n)}_{l_nm_n}}
  \right\rangle,
  \label{eq:349}
\end{multline}
where integers $l_i$ ($i=1,\ldots,n$) run over
$l_i = L_{a_i}, L_{a_i}-2, \ldots$, (0 or 1). Comparing the above
equation with Eq.~(\ref{eq:108}), we have
\begin{multline}
  c^{(n)\,l;l_1\cdots l_n}_{Xm;m_1\cdots m_n}(k_1,\ldots,k_n)
  =
    (-i)^l
  \sum_{a_1,\ldots,a_n}
    i^{L_{a_1}+\cdots +L_{a_n}}
  \\ \times
  W^{(a_1)}(k_1) \cdots W^{(a_n)}(k_n)\,
  g^{(l_1)}_{m_1m_1'} \cdots g^{(l_n)}_{m_nm_n'}
  \\ \times
  \left\langle
  \frac{\partial^n F^\mathrm{L}_{Xlm}}
  {\partial \chi^{(a_1)}_{l_1m_1'} \cdots \partial
    \chi^{(a_n)}_{l_nm_n'}}
  \right\rangle.
  \label{eq:350}
\end{multline}

The renormalized bias functions in the form of Eq.~(\ref{eq:349}) are
evaluated for a given model of the biased tensor field, once the
underlying statistics of the fields $\chi^{(a)}_{lm}$ are specified.
These fields just linearly depend on the linear density contrast
$\delta_\mathrm{L}$ through Eqs.~(\ref{eq:332})--(\ref{eq:335}) and
therefore the statistics are straightforwardly given by those of
linear density field.

Using Eqs.~(\ref{eq:75}) and (\ref{eq:335}), and the orthonormality
relation of the spherical harmonics, Eq.~(\ref{eq:397}), the
covariance of the fields (at the same position, the same applies
hereafter) can be straightforwardly calculated and is given by
\begin{equation}
  \left\langle
    \chi^{(a)}_{lm} \chi^{(b)}_{l'm'}
  \right\rangle
  =
  \delta_{ll'} \gamma^{ab}_{(l)} g^{(l)}_{mm'},
  \label{eq:351}
\end{equation}
where
\begin{equation}
  \gamma^{ab}_{(l)} \equiv
  \frac{i^{L_a+L_b}(-1)^l}{2l+1}
  \int \frac{k^2dk}{2\pi^2}
  W^{(a)}(k) W^{(b)}(k)
  P_\mathrm{L}(k).
  \label{eq:352}
\end{equation}

One can regard the set of parameters $\gamma^{ab}_{(l)}$ as matrix
elements of a matrix $\bm{\gamma}_{(l)}$, whose components are given
by $[\bm{\gamma}_{(l)}]_{ab} = \gamma^{ab}_{(l)}$. We denote the
matrix elements of the inverse matrix as
\begin{equation}
  \gamma^{(l)}_{ab} \equiv
  \left[
    \bm{\gamma}_{(l)}^{-1}
  \right]_{ab}.
  \label{eq:353}
\end{equation}
When one considers a set of variables $\chi^{(a)}_{lm}$ as a vector
with the set of indices $(a,l,m)$, the inverse of the covariance
matrix of Eq.~(\ref{eq:351}) is given by
$\delta_{ll'} \gamma^{(l)}_{ab} g_{(l)}^{mm'}$ with our notations.

In Eq.~(\ref{eq:351}), the integer of rank-$l$ should satisfy a
condition that integers $L_a-l$ and $L_b-l$ are non-negative even
numbers, where $L_a$ and $L_b$ are original ranks of tensors.
Therefore, the integer $L_a+L_b$ is also an even number, and
$l \leq \mathrm{min}(L_a,L_b)$. The matrix $\bm{\gamma}_{(l)}$
composed from matrix elements of Eq.~(\ref{eq:352}) is a real
symmetric matrix when the constraint above is satisfied, and so is the
inverse matrix, i.e.,
\begin{equation}
  \gamma^{ab}_{(l)} = \gamma^{ab*}_{(l)} = \gamma^{ba}_{(l)}, \quad
  \gamma_{ab}^{(l)} = \gamma_{ab}^{(l)*} = \gamma_{ba}^{(l)},
  \label{eq:354}
\end{equation}
for physically meaningful set of indices in Eq.~(\ref{eq:351}). When
the initial density field $\delta_\mathrm{L}$ is a Gaussian random
field, the two-point covariance of Eq.~(\ref{eq:351}) contains all the
information for the statistical distribution of the variables. In this
case, the distribution function is given by a multivariate Gaussian
distribution function,
\begin{multline}
  P_\mathrm{G}\left(\left\{\chi^{(a)}_{lm}\right\}\right)
  = \frac{1}{\sqrt{(2\pi)^N \det \bm{\gamma}_{(l)}}}
  \\ \times
  \exp\left[
    -\frac{1}{2} \sum_l
      \gamma^{(l)}_{ab}\, g_{(l)}^{mm'}
    \chi^{(a)}_{lm} \chi^{(b)}_{lm'}
  \right],
  \label{eq:355}
\end{multline}
where the Einstein summation convention is applied also to the indices
$a$, $b$ as well as azimuthal indices $m$, $m'$ and thus the summation
over these indices is implicitly assumed in the exponent. While the
variables $\chi^{(a)}_{lm}$ are complex numbers, the exponent of
Eq.~(\ref{eq:355}) is a real number, when the original derivative
field $\chi^{(a)}_{i_1\cdots i_{L_a}}$ is real. This property is
readily shown by noting that each term in the summation is represented
by $\sum_m \chi^{(a)*}_{lm}\gamma^{(l)}_{ab} \chi^{(b)}_{lm}$, and the
matrix elements $\gamma^{ab}_{(l)}$ make a real symmetric matrix (real
Hermitian matrix) due to Eq.~(\ref{eq:354}), and any quadratic form of
a real symmetric matrix is a real number.

\subsubsection{Defining the renormalized bias parameters}

When the initial density fluctuations are Gaussian, the expectation
value in Eq.~(\ref{eq:349}) is calculated by multivariate Gaussian
integrals with the distribution function of Eq.~(\ref{eq:355}), whose
evaluation is analytically possible in many cases when the semilocal
bias function of Eq.~(\ref{eq:347}) is given by an analytic function
of a finite number of variables. When there is a small non-Gaussianity
in the initial condition, and the linear density field is not exactly
a random Gaussian field, one can evaluate non-Gaussian corrections to
the Gaussian distribution function of Eq.~(\ref{eq:355}). The
procedure of deriving the non-Gaussian corrections is found in
Refs.~\cite{Matsubara:1995wd,Matsubara:2020lyv}, which is
straightforward to apply in this case. However, for the illustrative
examples below of this paper, we do not need to include the
non-Gaussian corrections in the evaluations of renormalized bias
functions.

The expectation value on the rhs of Eqs.~(\ref{eq:349}) and
(\ref{eq:350}) is represented by rotationally invariant variables,
just in the case of the renormalized bias function
$c^{(n)l;l_1\cdots l_n}_{Xm;m_1\cdots m_n}(k_1,\ldots,k_n)$, while the
arguments of $k_i$ are not present here. Similarly to
Eqs.~(\ref{eq:126}), (\ref{eq:130}), (\ref{eq:140}), and
(\ref{eq:151}), we have
\begin{align}
  \left\langle
  \frac{\partial F^\mathrm{L}_{Xlm}}{\partial\chi^{(a_1)}_{l_1m_1}}
  \right\rangle
  &=
    \frac{(-i)^{l+L_{a_1}}}{\sqrt{2l+1}}
    \delta_{ll_1} \delta_m^{m_1} b^{(1:a_1)}_{Xl},
    \label{eq:356}\\
  \left\langle
    \frac{\partial^2 F^\mathrm{L}_{Xlm}}
    {\partial\chi^{(a_1)}_{l_1m_1}\partial\chi^{(a_2)}_{l_2m_2}}
  \right\rangle
  &=
      i^{l-(L_{a_1}+L_{a_2})}
    \left(l\,l_1\,l_2\right)_m^{\phantom{m}m_1m_2}
    b^{(2:a_1a_2)}_{Xl;l_1l_2},
  \label{eq:357}\\
  \left\langle
  \frac{\partial^3 F^\mathrm{L}_{Xlm}}
  {\partial\chi^{(a_1)}_{l_1m_1}
  \partial\chi^{(a_2)}_{l_2m_2}
  \partial\chi^{(a_3)}_{l_3m_3}}
  \right\rangle
  &=
      i^{l-(L_{a_1}+L_{a_2}+L_{a_3})}
    \sum_L (-1)^L\sqrt{\{L\}}
    \nonumber \\
  & \hspace{-2pc} \times
    \left(l\,l_1\,L\right)_m^{\phantom{m}m_1M}
    \left(L\,l_2\,l_3\right)_M^{\phantom{M}m_2m_3}
    b^{(3:a_1a_2a_3)L}_{Xl;l_1l_2l_3},
  \label{eq:358}
\end{align}
and
\begin{multline}
  \left\langle
    \frac{\partial^3 F^\mathrm{L}_{Xlm}}
    {\partial\chi^{(a_1)}_{l_1m_1}\cdots
      \partial\chi^{(a_n)}_{l_nm_n}}
  \right\rangle
    =
      i^{l-(L_{a_1}+\cdots+L_{a_n})}
    \sum_{L_2,\ldots,L_{n-1}} (-1)^{L_2+\cdots L_{n-1}}
    \\ \times
    \sqrt{\{L_2\}\cdots\{L_{n-1}\}}
    \left(l\,l_1\,L_2\right)_m^{\phantom{m}m_1M_2}
    \\ \times
    \left(L_2\,l_2\,L_3\right)_{M_2}^{\phantom{M_2}m_2M_3}
    \cdots
    \left(L_{n-2}\,l_{n-2}\,L_{n-1}\right)_{M_{n-2}}^{\phantom{M_{n-2}}m_2M_{n-1}}
    \\ \times
    \left(L_{n-1}\,l_{n-1}\,l_n\right)_{M_{n-1}}^{\phantom{M_{n-1}}m_{n-1}m_n}
     b^{(n:a_1\cdots a_n)L_2\cdots L_{n-1}}_{Xl;l_1\cdots l_n}.
  \label{eq:359}
\end{multline}
The invariant constants $b^{(1;a_1)}_{Xl}$,
$b^{(2;a_1a_2)}_{Xl;l_1l_2}$, $b^{(3;a_1a_3)L}_{Xl;l_1l_2l_3}$ etc.~are
parameters that characterize the semilocal models of bias. 

Therefore, comparing Eq.~(\ref{eq:350}) with Eqs.~(\ref{eq:126}),
(\ref{eq:130}), (\ref{eq:140}) and (\ref{eq:151}), invariant
functions of the renormalized bias functions are given by
\begin{align}
    c^{(1)}_{Xl}(k)
  &
    = \sum_a
    b^{(1:a)}_{Xl}W^{(a)}(k),
  \label{eq:360}\\
    c^{(2)\,l}_{Xl_1l_2}(k_1,k_2)
  &
    = \sum_{a_1,a_2}
    b^{(2:a_1a_2)}_{Xl;l_1l_2}
    W^{(a_1)}(k_1) W^{(a_2)}(k_2),
  \label{eq:361}\\
    c^{(3)\,l;L}_{Xl;l_1l_2l_3}(k_1,k_2,k_3)
  &
    =
    \sum_{a_1,a_2,a_3}
    b^{(3:a_1a_2a_3)L}_{Xl;l_1l_2l_3}
    \nonumber \\
  & \qquad \times
    W^{(a_1)}(k_1) W^{(a_2)}(k_2) W^{(a_3)}(k_3),
  \label{eq:362}
\end{align}
and
\begin{multline}
    c^{(n)\,l;L_2\cdots L_{n-1}}_{Xl_1\cdots l_n}(k_1,\ldots,k_n)
    =
    \sum_{a_1,\ldots,a_n}
    b^{(n:a_1\cdots a_n)L_2\cdots L_{n-1}}_{Xl;l_1\cdots l_n}
  \\ \times
    W^{(a_1)}(k_1) \cdots W^{(a_n)}(k_n).
  \label{eq:363}
\end{multline} 
Thus, the renormalized bias functions are given by superpositions of
products of window functions with constant parameters which are
determined by a given semilocal model of bias. We naturally call
these parameters as the ``renormalized bias parameters.'' The scale
dependencies in the invariant renormalized functions are all contained
in the window functions which are fixed by the construction of the
semilocal model with a finite number of scale-independent parameters.
These properties simplify the calculations of loop corrections in our
applications. As the invariant renormalized functions and the window
functions $W^{(a)}(k)$ are both real, the renormalized bias parameters
are also real parameters.

Corresponding to the interchange symmetries of the renormalized bias
functions with orders greater than 2, Eqs.~(\ref{eq:139}),
(\ref{eq:149}), (\ref{eq:150}), (\ref{eq:158}), and (\ref{eq:159}),
the same symmetries for the constant coefficients of bias are given by
\begin{equation}
  b^{(2:a_2a_1)}_{Xl;l_2l_1}
  = (-1)^{l+l_1+l_2}
  b^{(2:a_1a_2)}_{Xl;l_1l_2}
  \label{eq:364}
\end{equation}
for the second-order bias,
\begin{align}
  b^{(3:a_1a_3a_2)L}_{Xl;l_ll_3l_2}
  &= (-1)^{l_2+l_3+L}
  b^{(3:a_1a_2a_3)L}_{Xl;l_ll_2l_3},
  \label{eq:365}\\
  b^{(3:a_2a_1a_3)L}_{Xl;l_2l_1l_3}
  &= (-1)^{l_1+l_2}
  \sum_{L'} (2L'+1)
  \begin{Bmatrix}
    l_1 & l & L \\
    l_2 & l_3 & L'
  \end{Bmatrix}
  b^{(3:a_1a_2a_3)L'}_{Xl;l_ll_2l_3}
  \label{eq:366}
\end{align}
for the third-order bias, and
\begin{align}
  &
    b^{(n:a_1\cdots a_{n-2}a_na_{n-1})L_2\cdots L_{n-2}L_{n-1}}_{Xl;l_l\cdots l_{n-2}l_nl_{n-1}}
  = (-1)^{l_{n-1}+l_n+L_{n-1}}
  b^{(n)L_2\cdots L_{n-1}}_{Xl;l_l\cdots l_n},
  \label{eq:367}\\
  &
    b^{(n:a_1\cdots a_{i-1}a_{i+1}a_ia_{i+2}\cdots a_n);L_2\cdots L_{i-1}L_{i+1}L_iL_{i+2}\cdots L_{n-1}}
    _{Xl_1\cdots l_{i-1}l_{i+1}l_il_{i+2}\cdots l_n}
    \nonumber\\
  &\qquad
  = (-1)^{l_i+l_{i+1}}
  \sum_{L'} (2L'+1)
  \begin{Bmatrix}
    l_i & L_i & L_{i+1} \\
    l_{i+1} & L_{i+2} & L'
  \end{Bmatrix}
    \nonumber\\
  &\hspace{8pc} \times
  b^{(n:a_1\cdots a_n);L_2\cdots L_iL'L_{i+2}\cdots L_{n-1}}_{Xl;l_1\cdots l_n}
  \label{eq:368}
\end{align}
for higher-order bias with $n>3$, where $i=1,\ldots,n-2$.  

\subsubsection{An example: The scale-dependent bias in tensor fields}

As a specific example, there is a loop integral in the scale-dependent
bias of Eq.~(\ref{eq:298}). Applying an expression
Eq.~(\ref{eq:361}) of semilocal models, the integral is given by
\begin{equation}
  \int \frac{p^2dp}{2\pi^2}
  c^{(2)l}_{Xl_1l_2}(p,p) P_\mathrm{L}(p)
  =
  \sum_{a_1,a_2} b^{(2:a_1a_2)}_{Xl;l_1l_2} \sigma^{(a_1a_2)},
  \label{eq:369}
\end{equation}
where
\begin{equation}
  \sigma^{(a_1a_2)} \equiv
  \int \frac{k^2dk}{2\pi^2} P_\mathrm{L}(k)
  W^{(a_1)}(k) W^{(a_2)}(k).
  \label{eq:370}
\end{equation}
Thus Eq.~(\ref{eq:298}) reduces to
\begin{equation}
  \Delta b_{Xl}(k) =
  \frac{2f_\mathrm{NL}^{(l)}}{\mathcal{M}(k)}
    \frac{1}{\sqrt{\{l\}}}
  \sum_{l_1,l_2}
  (-1)^{l_2}
  \begin{pmatrix}
    l_1 & l_2 & l \\
    0 & 0 & 0
  \end{pmatrix}
  \sum_{a_1,a_2} b^{(2:a_1a_2)}_{Xl;l_1l_2} \sigma^{(a_1a_2)},
  \label{eq:371}
\end{equation}
for $l=\mathrm{even}$ and $\Delta b_{Xl}(k)=0$ for $l=\mathrm{odd}$. 

For a simple case of scalar bias with the high-mass limit of the halo
model, comparison of Eq.~(\ref{eq:302}) with Eq.~(\ref{eq:361})
shows that only a term with
$b^{(2)\,0}_{\mathrm{h}\delta\delta;00} = \sigma^{-2}
\delta_\mathrm{c} b^\mathrm{L}_1$ survives, and one can readily see
the result of Eq.~(\ref{eq:303}) is reproduced as a consistency
check.

If we consider a simple model that the tensor bias is a local function
of only the second-order derivatives of the gravitational potential
$\partial_i\partial_j\varphi$, the window function is given by
$W^{(\varphi)}(k) = W(kR)$ as shown in Eq.~(\ref{eq:337}). We can
omit the label $(\varphi)$ in this case, because the linear tensor
field consists of only a single tensor. Thus Eq.~(\ref{eq:370})
simply reduces to
\begin{equation}
  \sigma^2 = \int \frac{k^2dk}{2\pi^2}
  P_\mathrm{L}(k) W^2(kR),
  \label{eq:372}
\end{equation}
which is the variance of the smoothed linear density field. The ranks
$l_1$ and $l_2$ take only values of 0 and 2 in the summation of
Eq.~(\ref{eq:371}) in this case, and due to the $3j$-symbol
in Eq.~(\ref{eq:371}), only the cases of $l=0,2,4$ are nonzero.
Substituting concrete numbers of the $3j$-symbols, Eq.~(\ref{eq:371})
is explicitly expanded in a finite number of terms and the result is
given by
\begin{multline}
  \Delta b_{Xl}(k) = 
  \frac{2\sigma^2}{\mathcal{M}(k)}
  \Biggl[
    \delta_{l0}
    f^{(0)}_\mathrm{NL}
    \left(
      b^{(2)}_{0;00} + \frac{1}{\sqrt{5}} b^{(2)}_{0;22}
    \right)
  \\
  +
    \frac{2}{5}\delta_{l2}
    f^{(2)}_\mathrm{NL}
    \left(
      b^{(2)}_{2;02}
      - \frac{1}{\sqrt{14}} b^{(2)}_{2;22}
    \right)
    \\
    +
      \frac{1}{3}
      \sqrt{\frac{2}{35}}\,
      \delta_{l4} f^{(4)}_\mathrm{NL}\,
    b^{(2)}_{4;22}
  \Biggr],
  \label{eq:373}
\end{multline}
where we use an interchange symmetry $b^{(2)}_{2;20} = b^{(2)}_{2;02}$
derived from Eq.~(\ref{eq:364}). Therefore, if the tensor bias is
modeled as a local function of only a second-rank linear tensor field,
the scale-dependent bias of the tensor field is nonzero only when the
rank of the biased tensor is 0, 2, or 4. Because of the rotational
symmetry, only a small number of bias parameters, $b^{(2)}_{0;00}$,
$b^{(2)}_{0;22}$, $b^{(2)}_{2;02}$, $b^{(2)}_{2;22}$, and
$b^{(2)}_{4;22}$ appear in this particular semilocal model.

The above examples probably do not sufficiently exhibit the practical
merits of semilocal models of bias. Their advantages are more
apparently shown in the calculation of loop corrections in the
higher-order perturbation theory. The present formalism with the full
use of the spherical basis is quite compatible with the FFT-PT
framework \cite{Schmittfull:2016jsw,Schmittfull:2016yqx} or the
FAST-PT framework \cite{McEwen:2016fjn,Fang:2016wcf}, which
dramatically reduces the dimensionality of multidimensional integrals
in the calculation of loop corrections in the higher-order
perturbation theory. Technical details of calculating the loop
corrections in the present formalism will be considered in a separate
paper of the series, Paper~II \cite{PaperII}.

\subsection{Relations to the conventional approach of bias
  renormalization}

On one hand, the renormalized bias functions in the iPT are fully
renormalized from the first place \cite{Matsubara:2011ck}, and we do
not need to renormalize the bias parameters order by order in
perturbation theory. On the other hand, the bias parameters in
conventional approaches in the literature of perturbation theory
should be renormalized order by order. These conventional approaches
consider only semilocal models of bias in the sense of our
terminology above. While our approach of iPT can include generally
nonlocal bias characterized by a hierarchy of renormalized bias
functions, our approach with semilocal bias models described above is
related to and consistent with other approaches.

In order to illustrate the relation, let us consider the simplest
example, the local-in-matter density (LIMD) model of bias in
Lagrangian space \cite{McDonald:2009dh,Desjacques:2016bnm}. For
simplicity, we ignore gravitational evolutions and assume that the
initial density fluctuations are Gaussian. In this toy model, the
density contrast of biased objects $X$ in Lagrangian space is given by
\begin{multline}
  \delta_X^\mathrm{L}(\bm{q}) = a_1 \delta_R(\bm{q})
  + \frac{a_2}{2!}
  \left\{
    \left[\delta_R(\bm{q})\right]^2 - \sigma^2
  \right\}
  \\
  + \frac{a_3}{3!} \left[\delta_R(\bm{q})\right]^3
  + \frac{a_4}{4!}
  \left\{
    \left[\delta_R(\bm{q})\right]^4 - 3 \sigma^2
  \right\}
  + \cdots,
  \label{eq:374}
\end{multline}
where $\delta_R(\bm{q})$ is the smoothed linear density field with
smoothing radius $R$, $\sigma^2 \equiv \langle \delta_R^2 \rangle$ is
the variance, and $a_n$ are bare bias parameters in the LIMD model.
The correlation function of the biased field is straightforwardly
calculated as
\begin{multline}
  \xi_X^\mathrm{L}(q) =
  \left( {a_1}^2 + a_1 a_3 \sigma^2 + \cdots \right) \xi_0(q)
  \\
  + \frac{1}{2}
  \left( {a_2}^2 + a_2 a_4 \sigma^2 + \cdots \right)
  \left[\xi_0(q)\right]^2 + \cdots,
  \label{eq:375}
\end{multline}
where
$\xi_0(|\bm{q}_1-\bm{q}_2|) =
\langle\delta_R(\bm{q}_1)\delta_R(\bm{q}_2)\rangle$ is the correlation
function of smoothed initial density field.

In the conventional procedure of bias renormalization, the resulting
coefficients on the rhs of the above equation correspond to the
renormalized bias parameters, as defined by
\begin{equation}
  b^\mathrm{L}_1 = a_1 + \frac{a_3\sigma^2}{2} + \cdots, \quad
  b^\mathrm{L}_2 = a_2 + \frac{a_4\sigma^2}{2} + \cdots,
  \label{eq:376}
\end{equation}
so that Eq.~(\ref{eq:375}) can be represented by a simple form,
\begin{equation}
  \xi^\mathrm{L}_X(q) =
  (b^\mathrm{L}_1)^2 \xi_0(q)
  +\frac{1}{2}(b^\mathrm{L}_2)^2 \left[\xi_0(q)\right]^2
  + \cdots.
  \label{eq:377}
\end{equation}
On large scales where the correlation function $\xi_0$ is small
enough, the above form is properly a perturbative series of
expansion, and the renormalized parameters
$b^\mathrm{L}_1,b^\mathrm{L}_2,\ldots$ correspond to the quantities
that can be determined by observations.

While the bias parameters should be renormalized order by order in
the conventional procedure explained above, the renormalized bias
functions in iPT are renormalized from the beginning. The
renormalized bias functions in iPT exactly correspond to
renormalized bias parameters to all orders in the semilocal models
of bias in the conventional procedure. One can show this property
model by model. For an illustrative example, we consider the LIMD
model of Eq.~(\ref{eq:374}). In our notation, the density field
corresponds to a scalar field $F^\mathrm{L}_{Xlm}(\bm{q})$ with
$l=m=0$, so that
$F^\mathrm{L}_{X00}(\bm{q}) = 1 + \delta^\mathrm{L}_X(\bm{q})$, as
understood by, e.g., comparing Eqs.~(\ref{eq:57}) and (\ref{eq:60}).
The Fourier transform of Eq.~(\ref{eq:374}) is given by
\begin{multline}
  \tilde{F}^\mathrm{L}_{X00}(\bm{k})
  = \sum_{n=1}^\infty \frac{a_n}{n!}
  \int \frac{d^3k_1}{(2\pi)^3} \cdots \frac{d^3k_n}{(2\pi)^3}
  \delta_\mathrm{D}^3(\bm{k}_1+\cdots +\bm{k}_n-\bm{k})
  \\ \times
  \delta_\mathrm{L}(\bm{k}_1) \cdots
  \delta_\mathrm{L}(\bm{k}_n)
  W(k_1R) \cdots W(k_nR),
  \label{eq:378}
\end{multline}
where we exclude the zero mode and assume $\bm{k} \ne \bm{0}$.
Substituting the above equation into the definition of the
renormalized bias functions, Eqs.~(\ref{eq:70}) or (\ref{eq:71}) with
$l=m=0$, and using a standard formula of Gaussian moments,
$\langle (\delta_R)^p \rangle = \sigma^2 (p-1)!!$ where $p$ is a
non-negative integer, we straightforwardly derive
\begin{equation}
  c^{(n)}_{X00}(\bm{k}_1,\ldots,\bm{k}_n)
  = W(k_1R) \cdots W(k_nR)
  \sum_{m=0}^\infty \frac{a_{2m+n}\sigma^{2m}}{(2m-1)!!}.
  \label{eq:379}
\end{equation}
Using this bias functions, the prediction of iPT for the correlation
function in Lagrangian space (neglecting the gravitational
evolution) is given by \cite{Matsubara:2011ck}
\begin{equation}
  \xi^\mathrm{L}_X(q) =
  \sum_{n=1}^\infty \frac{(b^\mathrm{L}_n)^n}{n!}
  \left[\xi_0(q)\right]^n,
  \label{eq:380}
\end{equation}
where
\begin{equation}
  b^\mathrm{L}_n =
  \sum_{m=0}^\infty \frac{a_{2m+n}\sigma^{2m}}{(2m-1)!!}.
  \label{eq:381}
\end{equation}
The last parameters are consistent with the renormalized bias
parameters in the conventional theory, Eq.~(\ref{eq:376}). We note
that the renormalized bias parameters are directly derived to full
orders with infinite terms at once, and we do not need to derive them
order by order in iPT. In the notations of semilocal bias in this
section, $\chi^{(a)}_{lm} = \delta_{l0}\delta_{m0} \delta_R$ with
$L_a=0$ and $W^{(a)}(k) = W(kR)$ in Eq.~(\ref{eq:332}), and the
renormalized bias parameters in our formalism, generally derived
from Eqs.~(\ref{eq:359}) and (\ref{eq:363}), simply correspond to
\begin{equation}
  b^{(1:a)}_{X0} = b^\mathrm{L}_1, \quad
  b^{(2:aa)}_{X0;00} = b^\mathrm{L}_2,\quad
  b^{(n:a\cdots a)0\cdots 0}_{X0;0\cdots 0} = b^\mathrm{L}_n
  \label{eq:382}
\end{equation}
in the LIMD model for scalar density fields.

The above example shows that the renormalized bias functions in
semilocal models exactly correspond to the renormalized bias
parameters in conventional theory of bias renormalization. With the
iPT, the bias parameters are renormalized from the first place of
introducing renormalized bias functions, and we do not need
order-by-order renormalizations of parameters as in the case of
conventional perturbation theory. The main reason is that the bias
relations between the biased field and mass density field are not
expanded into perturbative series in iPT, and are treated
nonperturbatively throughout the calculation. While the dynamics of
gravitational evolution for the mass density field are treated
perturbatively, the bias relations are not. On large scales where
amplitude of the correlation is small, e.g., $\xi_0(q) \ll 1$ in
Eq.~(\ref{eq:381}), the expansion scheme is well defined. In the
above, the LIMD model is employed just for an illustration, and one
can extend the arguments to more general cases with more complicated
models of bias and arbitrary ranks of tensors.

\section{\label{sec:Conclusions}
  Summary and Conclusions
}

\begin{figure*}
\centering
\includegraphics[width=30pc]{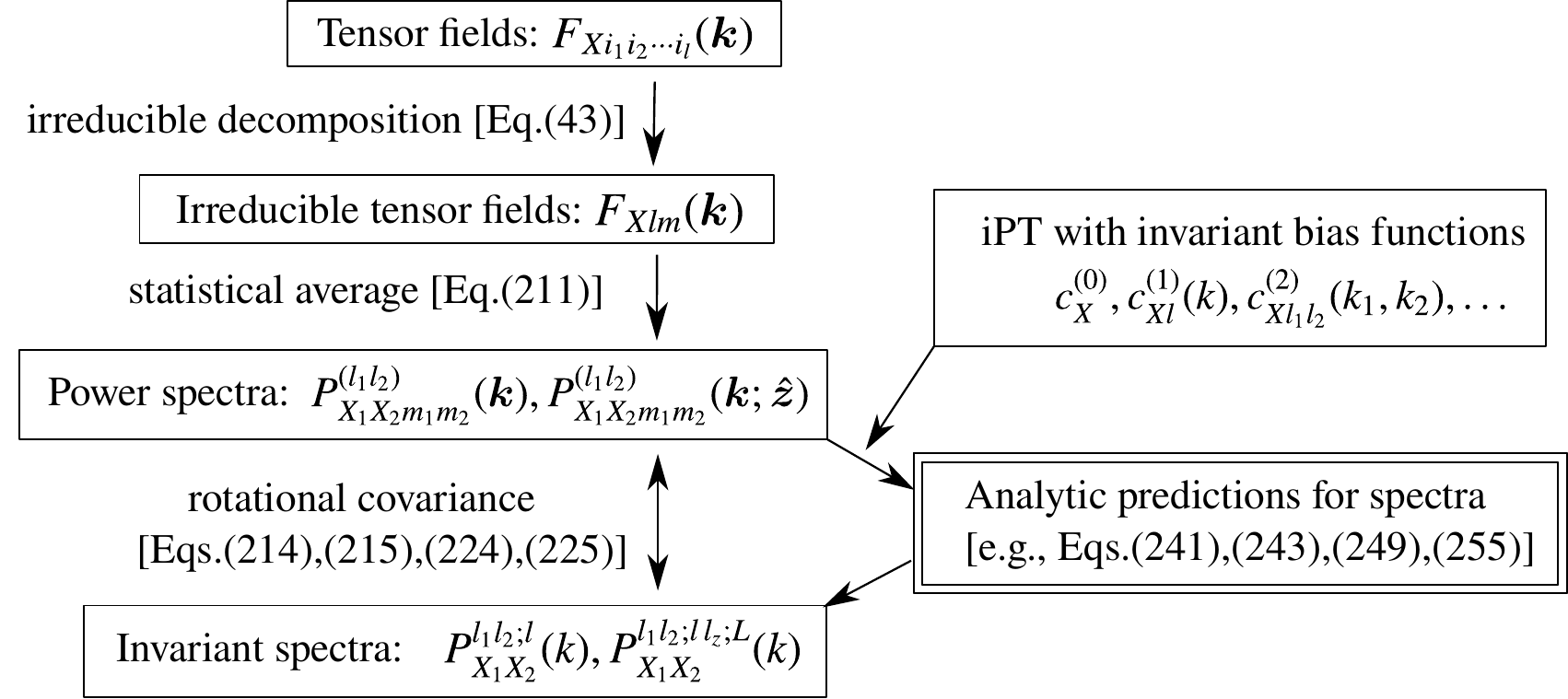}
\caption{\label{fig:1} An illustration of the procedure to derive
  analytic predictions for the power spectra of tensor fields in this
  paper. Some important equations are indicated. For other statistics
  such as correlation functions and bispectra, the procedures are
  similar. }
\end{figure*}

In this paper, the formalism iPT is generalized to calculate
statistics of generally tensor-valued objects, using the nonlinear
perturbation theory. Higher-rank tensors are conveniently decomposed
into spherical tensors which are irreducible representations of the
three-dimensional rotation group SO(3), and mathematical techniques
developed in the theory of angular momentum in quantum mechanics are
effectively applied.

The fundamental formalism of iPT can be recycled without essential
modification, and the only difference is that the biased field is
colored by tensor values. Because the original formalism of iPT is
supposed to be applied to scalar fields, the rotational symmetry in
this original version of the theory is relatively trivial. When the
theory is generalized to include tensor-valued fields, the statistical
properties of biasing are largely constrained by symmetries. In
particular, the renormalized bias functions, which are ones of the
important ingredients in the iPT, can be represented by a set of
invariant functions under rotations of the coordinates system.
Explicit constructions of these invariants are derived and presented
in this paper. These invariants can also be seen as coefficients of
angular expansions by polypolar spherical harmonics.

The iPT offers a systematic way of calculating the propagators of
nonlinear perturbation theory, both in real space and in redshift
space. The methodology of deriving propagators of tensor-valued fields
is generally explained in detail and several examples of lower-order
propagators are explicitly derived in this paper. For an illustrative
purpose, these examples of propagators are applied to simple cases of
predicting the correlation statistics, including the linear power
spectra and correlation functions, the lowest-order power spectrum
with primordial non-Gaussianity, the bispectrum in the tree-level
approximation, for generally tensor fields. So as to check
consistencies in the examples, we confirm that each derived formula
reproduces the known result for scalar fields, just substituting
$l=m=0$ in the general formulas. In Fig.~\ref{fig:1}, an example of
the procedure to derive analytic predictions of iPT for power spectra
of tensor fields in this paper is illustrated with relevant equations
in this paper. The correlation functions and bispectra are also
derived in similar procedures.

In the last section, the concept of semilocal models of bias is
introduced. The formalism of iPT does not have to assume any model of
bias, and all the uncertainties of bias are included in the
renormalized bias functions, which are determined by a nonlocal
functional of the biased field in terms of the linear density field,
and have infinite degrees of freedom in general. The semilocal models
of bias are defined so that the bias functional should be a function
of a finite number of variables derived from the linear density field
in Lagrangian space. As a result, the biasing is modeled by a finite
number of parameters in this category of models. Almost all the
existing models of bias introduced in the literature fall into this
category, including the halo model, the peak theory of bias, excursion
set peaks, EFTofLSS approach, and so forth. In this paper, only the
formal definition of the semilocal models of bias is described, and
rotationally invariant parameters in the models are identified. Their
applications to individual models of concrete targets will be
addressed in future work.

In this paper, the power spectrum and bispectrum are only evaluated in
a lowest-order, or tree-level approximation. Some further techniques
are required in calculating higher-order loop corrections, and they
are described in a subsequent paper of the series, Paper~II
\cite{PaperII}. In realistic observations, we can only observe
projected components of tensors onto the two-dimensional sky.
Predictions of the statistics of three-dimensional tensors given in
this paper can be transformed to those of projected tensors. Explicit
formulations of this transformation are given in another subsequent
paper of the series, Paper~III \cite{PaperIII}. While we assume the
distant-observer approximation in these first three papers of the
series, Paper~IV \cite{PaperIV} in the series gives a full-sky
formulation without the last approximation.

We hope that the
present formalism offers a way to carve out the future of applying the
cosmological perturbation theory to the analysis of observations in
the era of precision cosmology.

\begin{acknowledgments}
  I thank Y.~Urakawa, K.~Kogai, K.~Akitsu, and A.~Taruya for useful
  discussions. This work was supported by JSPS KAKENHI Grants
  No.~JP19K03835 and No.~21H03403.
\end{acknowledgments}


\appendix
\onecolumngrid

\section{\label{app:SphericalBasis}
  Spherical tensor basis
}

The spherical tensors $\mathsf{Y}^{(m)}_{i_1\cdots i_l}$ are the basis
to expand symmetric tensors in the Cartesian coordinates system to the
irreducible tensors on the spherical basis. In this Appendix, the
expressions of the spherical tensor basis for ranks $l=0,1,2,3,4$ are
explicitly given in terms of the spherical bases $\mathbf{e}^0$ and
$\mathbf{e}^\pm$, whose components are denoted by
${\mathrm{e}^m}_i = [\mathbf{e}^m]_i$ with $m=0,\pm$. The Cartesian
indices of the spherical tensor basis are totally symmetric,
$\mathsf{Y}^{(m)}_{i_1\cdots i_l} = \mathsf{Y}^{(m)}_{(i_1\cdots
  i_l)}$, where the round brackets indicate the symmetrization with
respect to the indices inside the brackets, i.e.,
\begin{equation}
  A_{(i_1\cdots i_n)} \equiv
  \frac{1}{n!}
  \sum_{\sigma \in \mathcal{S}_n}
  A_{i_{\sigma(1)}\cdots i_{\sigma(n)}},
  \label{eq:383}
\end{equation}
and $\mathcal{S}_n$ is the symmetric group of order $n$. The
following results are straightforwardly derived from the explicit form
of spherical harmonics $Y_{lm}(\theta,\phi)$, which uniquely determine
the form of spherical basis tensors $\mathsf{Y}^{(m)}_{i_1\cdots i_l}$
to satisfy Eq.~(\ref{eq:16}), i.e.,
\begin{equation}
  Y_{lm}(\theta,\phi) = \sqrt{\frac{(2l+1)!!}{4\pi\,l!}}\,
  \mathsf{Y}^{(m)*}_{i_1i_2\cdots i_l} n_{i_1} n_{i_2}\cdots n_{i_l}.
\label{eq:384}
\end{equation}
The procedure to derive the results are given as follows: First, we
represent the complex conjugate of the spherical harmonics
$Y_{lm}^*(\theta,\phi)$ in terms of the components of
$\bm{n} = (\sin\theta\cos\phi, \sin\theta\sin\phi, \cos\theta) =
(n_1,n_2,n_3)$, and the resulting expressions only contain factors of
$n_1\mp i n_2$ and $n_3$, because of the structure of spherical
harmonics. Second, we substitute
$n_1\mp in_2 \rightarrow \mp \sqrt{2}\,{\mathbf{e}^\pm}_i n_i$ and
$n_3\rightarrow {\mathbf{e}^0}_i n_i$ [cf., Eq.~(\ref{eq:2})] inq
the expression. Comparing the results with Eq.~(\ref{eq:384}), the
explicit forms of spherical basis tensors are uniquely and obviously
determined. We give the explicit expressions up to $l\leq 4$ below for
the reader's convenience.

Spherical basis tensors of rank-0 and rank-1:
\begin{equation}
  \mathsf{Y}^{(0)} = 1, \quad
  \mathsf{Y}^{(m)}_i = {\mathrm{e}^m}_i.
  \label{eq:385}
\end{equation}

Spherical basis tensors of rank-2:
\begin{equation}
  \mathsf{Y}^{(0)}_{ij} = \sqrt{\frac{3}{2}}
  \left(
    {\mathrm{e}^0}_i\,{\mathrm{e}^0}_j - \frac{1}{3}\delta_{ij}
  \right),
  \qquad
    \mathsf{Y}^{(\pm 1)}_{ij} = \sqrt{2}\,
    {\mathrm{e}^0}_{(i}\,{\mathrm{e}^\pm}_{j)},
    \qquad
    \mathsf{Y}^{(\pm 2)}_{ij} =
    {\mathrm{e}^\pm}_i\,{\mathrm{e}^\pm}_j.
  \label{eq:386}
\end{equation}

Spherical tensor basis of rank-3:
\begin{align}
  \mathsf{Y}^{(0)}_{ijk}
  &= \sqrt{\frac{5}{2}}
    \left[
    {\mathrm{e}^0}_i\,{\mathrm{e}^0}_j\,{\mathrm{e}^0}_k
    - \frac{3}{5} \delta_{(ij}{\mathrm{e}^0}_{k)}
    \right],
  &
     \mathsf{Y}^{(\pm 1)}_{ijk}
     &= \frac{\sqrt{15}}{2}
    \left[
    {\mathrm{e}^0}_{(i}\,{\mathrm{e}^0}_j\,{\mathrm{e}^\pm}_{k)}
    - \frac{1}{5} \delta_{(ij}{\mathrm{e}^\pm}_{k)}
    \right],
    \label{eq:387}\\
  \mathsf{Y}^{(\pm 2)}_{ijk}
  &= \sqrt{3}\,
    {\mathrm{e}^0}_{(i}\,{\mathrm{e}^\pm}_j\,{\mathrm{e}^\pm}_{k)},
  &
     \mathsf{Y}^{(\pm 3)}_{ijk}
  &=
    {\mathrm{e}^\pm}_{i}\,{\mathrm{e}^\pm}_j\,{\mathrm{e}^\pm}_k.
\label{eq:388}
\end{align}

Spherical tensor basis of rank-4:
\begin{align}
  \mathsf{Y}^{(0)}_{ijkl}
  &= \frac{1}{2}\sqrt{\frac{35}{2}}
    \left[
    {\mathrm{e}^0}_i\,{\mathrm{e}^0}_j\,{\mathrm{e}^0}_k\,{\mathrm{e}^0}_l
    - \frac{6}{7} \delta_{(ij}{\mathrm{e}^0}_k{\mathrm{e}^0}_{l)}
    + \frac{3}{35} \delta_{(ij} \delta_{kl)}
    \right],
  &
    \mathsf{Y}^{(\pm 1)}_{ijkl}
  &= \sqrt{7}
    \left[
    {\mathrm{e}^0}_{(i}\,{\mathrm{e}^0}_j\,{\mathrm{e}^0}_k\,{\mathrm{e}^\pm}_{l)}
    - \frac{3}{7} \delta_{(ij} {\mathrm{e}^0}_k\, {\mathrm{e}^\pm}_{l)}
    \right],
    \label{eq:389}\\
  \mathsf{Y}^{(\pm 2)}_{ijkl}
  &= \sqrt{7}
    \left[
    {\mathrm{e}^0}_{(i}\,{\mathrm{e}^0}_j\,{\mathrm{e}^\pm}_k\,{\mathrm{e}^\pm}_{l)}
    - \frac{1}{7} \delta_{(ij} {\mathrm{e}^\pm}_k\, {\mathrm{e}^\pm}_{l)}
    \right],
  &
     \mathsf{Y}^{(\pm 3)}_{ijkl}
  &= 2\,
    {\mathrm{e}^0}_{(i}\,{\mathrm{e}^\pm}_j\,{\mathrm{e}^\pm}_k\,
    {\mathrm{e}^\pm}_{l)},
\label{eq:390}\\
  \mathsf{Y}^{(\pm 4)}_{ijkl}
  &=
    {\mathrm{e}^\pm}_{i}\,{\mathrm{e}^\pm}_j\,{\mathrm{e}^\pm}_k
    \,{\mathrm{e}^\pm}_{l}.
  \label{eq:391}
\end{align}

\section{\label{app:SphericalHarmonics}
Formulas for spherical harmonics and polypolar spherical harmonics}

In this Appendix, various formulas regarding the spherical harmonics
are summarized, which are repeatedly used in the main text. As we
mainly use Racah's normalization of the spherical harmonics in this
paper, the corresponding formulas look different from those found in
standard textbooks. Accordingly bipolar, tripolar, and generally
polypolar spherical harmonics are also defined in convenient
normalizations used in this paper. Most of the formulas in this
Appendix are derived from those found in standard literature
\cite{Edmonds:1955fi,Khersonskii:1988krb}.

While the spherical harmonics are frequently represented by
$Y_l^m(\theta,\phi)$ in the literature, this function transforms as a
covariant tensor with respect to the coordinate rotation in the
spherical basis introduced in Sec.~\ref{sec:SphericalBasis}, thus we
instead prefer the notation $Y_{lm}(\theta,\phi)$ with the lower
position of the azimuthal index $m$. For simplicity, the spherical
harmonics are frequently represented by
$Y_{lm}(\bm{n}) = Y_{lm}(\theta,\phi)$, where the angular coordinates
are represented by a unit vector $\bm{n}$ whose spherical coordinates
are $(\theta,\phi)$. We also use the simplified notation that the
integration over the angular coordinates
$\int \sin\theta\,d\theta\,d\phi \cdots$ is represented by
$\int d^2n\cdots$. The standard normalization of the spherical
harmonics $Y_{lm}(\bm{n})$ is given by Eqs.~(\ref{eq:14}) and
(\ref{eq:15}). However, as described above, we intensively use the
Racah's normalization of the harmonics, defined by
Eq.~(\ref{eq:17}), i.e.,
\begin{equation}
  C_{lm}(\bm{n}) \equiv \sqrt{\frac{4\pi}{2l+1}} Y_{lm}(\bm{n})
  = \sqrt{\frac{(l-m)!}{(l+m)!}} P_l^m(\cos\theta)\,e^{im\phi},
  \label{eq:392}
\end{equation}
where $P_l^m(x)$ is the associated Legendre polynomial. 

The parity symmetry of the spherical harmonics is given by
\begin{equation}
  C_{lm}(-\bm{n}) = (-1)^l C_{lm}(\bm{n}).
  \label{eq:393}
\end{equation}
The complex conjugate is given by
\begin{equation}
  C_{lm}^*(\bm{n}) = g_{(l)}^{mm'} C_{lm'}(\bm{n}) = (-1)^m C_{l,-m}(\bm{n}),
  \label{eq:394}
\end{equation}
where $g^{mm'}_{(l)} = (-1)^m \delta_{m,-m'}$ is the metric tensor in
the spherical basis and the summation over $m'$ is implicitly assumed
when the same azimuthal index appears as a pair of lower and upper
indices, just as in the Einstein summation convention. An addition
theorem is given by
\begin{equation}
  g_{(l)}^{mm'} C_{lm}(\bm{n}_1) C_{lm'}(\bm{n}_2)
  = \mathit{P}_l(\bm{n}_1\cdot\bm{n}_2),
  \label{eq:395}
\end{equation}
where $P_l(x)$ is the Legendre polynomial of order $l$.
The Rayleigh expansion or plane-wave expansion is given by
\begin{equation}
  e^{\pm i\bm{k}\cdot\bm{x}} =
  \sum_{l=0}^\infty (\pm i)^l (2l+1) j_l(kx)
  g_{(l)}^{mm'} C_{lm}(\hat{\bm{k}}) C_{lm'}(\hat{\bm{x}}),
  \label{eq:396}
\end{equation}
where $k=|\bm{k}|$, $\hat{\bm{k}}=\bm{k}/k$, $x=|\bm{x}|$, and
$\hat{\bm{x}}=\bm{x}/x$. The orthonormality relation is given by
\begin{equation}
  \int \frac{d^2n}{4\pi}\,
  C_{lm}(\bm{n}) C_{l'm'}(\bm{n}) =
  \frac{\delta_{ll'}}{2l+1} g^{(l)}_{mm'},
  \label{eq:397}
\end{equation}
where $g^{(l)}_{mm'} = (-1)^m \delta_{m,-m'}$ is the inverse of
$g_{(l)}^{mm'}$, whose matrix elements are the same as
$g_{(l)}^{mm'}$. Combining Eqs.~(\ref{eq:396}) and (\ref{eq:397}), we
have
\begin{equation}
  \int \frac{d^2\hat{k}}{4\pi}\,
  e^{\pm i\bm{k}\cdot\bm{x}} C_{lm}(\hat{\bm{k}})
  = (\pm i)^l
    j_l(kx) C_{lm}(\hat{\bm{x}}).
  \label{eq:398}
\end{equation}
A product of two spherical harmonics with the same direction is given
by
\begin{equation}
  C_{l_1m_1}(\bm{n}) C_{l_2m_2}(\bm{n})
  = \sum_{l,m} (2l+1)
  \begin{pmatrix}
    l_1 & l_2 & l \\
    0 & 0 & 0
  \end{pmatrix}
  \begin{pmatrix}
    l_1 & l_2 & l \\
    m_1 & m_2 & m
  \end{pmatrix}
  C_{lm}^*(\bm{n}),
  \label{eq:399}
\end{equation}
where the last two factors but one are Wigner's $3j$-symbols.

The bipolar spherical harmonics in our notation are defined by
\begin{equation}
  X^{l_1l_2}_{lm}(\bm{n}_1,\bm{n}_2)
  = \left(l\,l_1\,l_2\right)_{m}^{\phantom{m}m_1m_2}
    C_{l_1m_1}(\bm{n}_1) C_{l_2m_2}(\bm{n}_2),
  \label{eq:400}
\end{equation}
where
\begin{equation}
  \left(l\,l_1\,l_2\right)_m^{\phantom{m}m_1m_2} =
  g_{(l_1)}^{m_1m_1'} g_{(l_2)}^{m_2m_2'}
  \begin{pmatrix}
    l & l_1 & l_2 \\
    m & m_1' & m_2'
  \end{pmatrix}
  = (-1)^{m_1+m_2}
  \begin{pmatrix}
    l & l_1 & l_2 \\
    m & -m_1 & -m_2
  \end{pmatrix}
  \label{eq:401}
\end{equation}
is a $3j$-symbol with two azimuthal indices raised by the metric
tensor of the spherical basis. The relation to the bipolar spherical
harmonics introduced in standard textbooks \cite{Khersonskii:1988krb},
$\{ Y_{l_1}(\bm{n}_1)\otimes Y_{l_2}(\bm{n}_2) \}_{lm}$, is given by
\begin{equation}
  \left\{
    Y_{l_1}(\bm{n}_1)\otimes Y_{l_2}(\bm{n}_2)
  \right\}_{lm}
  = \frac{(-1)^l}{4\pi}\sqrt{(2l+1)(2l_1+1)(2l_2+1)}\,
  X^{l_1l_2}_{lm}(\bm{n}_1,\bm{n}_2).
  \label{eq:402}
\end{equation}
The parity symmetry of the bipolar spherical harmonics is given by
\begin{equation}
  X^{l_1l_2}_{lm}(-\bm{n}_1,-\bm{n}_2)
  =(-1)^{l_1+l_2}
  X^{l_1l_2}_{lm}(\bm{n}_1,\bm{n}_2).
  \label{eq:403}
\end{equation}
The complex conjugate of the bipolar spherical harmonics is given by
\begin{equation}
  X^{l_1l_2*}_{lm}(\bm{n}_1,\bm{n}_2)
  =(-1)^{l+l_1+l_2} g_{(l)}^{mm'}
  X^{l_1l_2}_{lm'}(\bm{n}_1,\bm{n}_2).
  \label{eq:404}
\end{equation}
The orthonormality relation for the
bipolar spherical harmonics is given by
\begin{equation}
  \int \frac{d^2n_1}{4\pi} \frac{d^2n_2}{4\pi}
  X^{l_1l_2}_{lm}(\bm{n}_1,\bm{n}_2)
  X^{l_1'l_2'}_{l'm'}(\bm{n}_1,\bm{n}_2)
  = \frac{(-1)^{l+l_1+l_2}\delta_{ll'} \delta_{l_1l_1'} \delta_{l_2l_2'}
    \delta^\triangle_{l_1l_2l}}{(2l+1)(2l_1+1)(2l_2+1)} 
  g^{(l)}_{mm'},
  \label{eq:405}
\end{equation}
where
\begin{equation}
  \delta^\triangle_{l\,l_1l_2} =
  \begin{cases}
    1, & \mathrm{if\ \ } |l_1-l_2| \leq l \leq l_1+l_2, \\
    0, & \mathrm{otherwise},
  \end{cases}
  \label{eq:406}
\end{equation}
i.e., $\delta^\triangle_{l_1l_2l_3}$ is unity when the set of three
numbers $(l_1,l_2,l_3)$ satisfies the triangular condition, and is
zero otherwise. A product of two bipolar spherical harmonics with the
same set of directions is given by
\begin{multline}
  X^{l_1l_2}_{lm}(\bm{n}_1,\bm{n}_2)
  X^{l_1'l_2'}_{l'm'}(\bm{n}_1,\bm{n}_2)
  =
  \sum_{l''} (-1)^{l''}(2l''+1)
  \left(l\,l'\,l''\right)_{mm'}^{\phantom{mm'}m''}
  \\ \times
  \sum_{l_1'',l_2''} (-1)^{l_1''+l_2''} (2l_1''+1)(2l_2''+1)
  \begin{pmatrix}
    l_1 & l_1' & l_1'' \\
    0 & 0 & 0
  \end{pmatrix}
  \begin{pmatrix}
    l_2 & l_2' & l_2'' \\
    0 & 0 & 0
  \end{pmatrix}
  \begin{Bmatrix}
    l & l' & l'' \\
    l_1 & l_1' & l_1'' \\
    l_2 & l_2' & l_2''
  \end{Bmatrix}
  X^{l_1''l_2''}_{l''m''}(\bm{n}_1,\bm{n}_2),
  \label{eq:407}
\end{multline}
where the last factor but one is Wigner's $9j$-symbol. 

The tripolar spherical harmonics are defined by
\begin{equation}
  X^{l_1l_2l_3}_{L;lm}(\bm{n}_1,\bm{n}_2,\bm{n}_3)
  = (-1)^L \sqrt{2L+1}
  \left(l\,l_1\,L\right)_{m}^{\phantom{m}m_1M}
  \left(L\,l_2\,l_3\right)_{M}^{\phantom{M}m_2m_3}
  C_{l_1m_1}(\bm{n}_1) C_{l_2m_2}(\bm{n}_2)
  C_{l_3m_3}(\bm{n}_3).
  \label{eq:408}
\end{equation}
The relation to the tripolar spherical harmonics introduced in
standard textbooks \cite{Khersonskii:1988krb} is given by
\begin{equation}
  \left\{
    Y_{l_1}(\bm{n}_1)\otimes
    \left\{Y_{l_2}(\bm{n}_2)
      \otimes Y_{l_3}(\bm{n}_3)
    \right\}_L
  \right\}_{lm}
  = \frac{(-1)^l}{(4\pi)^{3/2}}\sqrt{(2l+1)(2l_1+1)(2l_2+1)(2l_3+1)}\,
  X^{l_1l_2l_3}_{L;lm}(\bm{n}_1,\bm{n}_2,\bm{n}_3).
  \label{eq:409}
\end{equation}
The parity symmetry of the tripolar spherical harmonics is given by
\begin{equation}
  X^{l_1l_2l_3}_{lm}(-\bm{n}_1,-\bm{n}_2,-\bm{n}_3)
  =(-1)^{l_1+l_2+l_3}
  X^{l_1l_2l_3}_{lm}(\bm{n}_1,\bm{n}_2,\bm{n}_3).
  \label{eq:410}
\end{equation}
The complex conjugate of the bipolar spherical harmonics is given by
\begin{equation}
  X^{l_1l_2l_3*}_{lm}(\bm{n}_1,\bm{n}_2,\bm{n}_3)
  =(-1)^{l+l_1+l_2+l_3} g_{(l)}^{mm'}
  X^{l_1l_2l_3}_{lm'}(\bm{n}_1,\bm{n}_2,\bm{n}_3).
  \label{eq:411}
\end{equation}
The orthonormality relation for the tripolar spherical harmonics is
given by
\begin{equation}
  \int \frac{d^2n_1}{4\pi} \frac{d^2n_2}{4\pi} \frac{d^2n_3}{4\pi}
  X^{l_1l_2l_3}_{L;lm}(\bm{n}_1,\bm{n}_2,\bm{n}_3)
  X^{l_1'l_2'l_3'}_{L';l'm'}(\bm{n}_1,\bm{n}_2,\bm{n}_3)
  = \frac{(-1)^{l+l_1+l_2+l_3} \delta_{ll'} \delta_{l_1l_1'} \delta_{l_2l_2'}
    \delta_{l_3l_3'} \delta_{LL'} \delta^\triangle_{l\,l_1L}
    \delta^\triangle_{Ll_2l_3}}{(2l+1)(2l_1+1)(2l_2+1)(2l_3+1)} 
    g^{(l)}_{mm'}.
  \label{eq:412}
\end{equation}
A product of two tripolar spherical harmonics with the same set of
directions is given by
\begin{multline}
  X^{l_1l_2l_3}_{L;lm}(\bm{n}_1,\bm{n}_2,\bm{n}_3)
  X^{l_1'l_2'l_3'}_{L';l'm'}(\bm{n}_1,\bm{n}_2,\bm{n}_3)
  = \sqrt{(2L+1)(2L'+1)}
  \sum_{l''} (-1)^{l''} (2l''+1)
  \left(l\,l'\,l''\right)_{mm'}^{\phantom{mm'}m''}
  \\ \times
  \sum_{l_1'',l_2'',l_3''} (-1)^{l_1''+l_2''+l_3''}
  (2l_1''+1)(2l_2''+1)(2l_3''+1)
  \begin{pmatrix}
    l_1 & l_1' & l_1'' \\
    0 & 0 & 0
  \end{pmatrix}
  \begin{pmatrix}
    l_2 & l_2' & l_2'' \\
    0 & 0 & 0
  \end{pmatrix}
  \begin{pmatrix}
    l_3 & l_3' & l_3'' \\
    0 & 0 & 0
  \end{pmatrix}
  \\ \times
  \sum_{L''} \sqrt{2L''+1}
  \begin{Bmatrix}
    l & l' & l'' \\
    l_1 & l_1' & l_1'' \\
    L & L' & L''
  \end{Bmatrix}
  \begin{Bmatrix}
    L & L' & L'' \\
    l_2 & l_2' & l_2'' \\
    l_3 & l_3' & l_3''
  \end{Bmatrix}
  X^{l_1''l_2''l_3''}_{L'';l''m''}(\bm{n}_1,\bm{n}_2,\bm{n}_3).
  \label{eq:413}
\end{multline}

While they are not found in standard textbooks, we generalize the
bipolar and tripolar spherical harmonics to higher-order counterparts,
and define polypolar spherical harmonics of arbitrary order by
\begin{multline}
  X^{l_1l_2\cdots l_n}_{L_2L_3\cdots L_{n-1};lm}
  (\bm{n}_1,\bm{n}_2,\ldots,\bm{n}_n)
  = (-1)^{L_2+\cdots +L_{n-1}}
  \sqrt{(2L_2+1)(2L_3+1)\cdots(2L_{n-1}+1)}
  \left(l\,l_1\,L_2\right)_{m}^{\phantom{m}m_1M_2}
  \\ \times
  \left(L_2\,l_2\,L_3\right)_{M_2}^{\phantom{M_2}m_2M_3} \cdots
  \left(L_{n-2}\,l_{n-2}\,L_{n-1}\right)_{M_{n-2}}^{\phantom{M_{n-2}}m_{n-2}M_{n-1}}
  \left(L_{n-1}\,l_{n-1}\,l_n\right)_{M_{n-1}}^{\phantom{M_{n-1}}m_{n-1}m_n}
  C_{l_1m_1}(\bm{n}_1) C_{l_2m_2}(\bm{n}_2) \cdots
  C_{l_nm_n}(\bm{n}_n).
  \label{eq:414}
\end{multline}
They satisfy recursive relations,
\begin{equation}
  X^{l_1l_2\cdots l_n}_{L_2L_3\cdots L_{n-1};lm}(\bm{n}_1,\bm{n}_2,\ldots,\bm{n}_n)
  = (-1)^{L_2} \sqrt{(2L_2+1)}
  \left(l\,l_1\,L_2\right)_{m}^{\phantom{m}m_1M_2}
  C_{l_1m_1}(\bm{n}_1)
  X^{l_2\cdots l_n}_{L_3\cdots L_{n-1};L_2M_2}(\bm{n}_2,\bm{n}_3,\ldots,\bm{n}_n).
  \label{eq:415}
\end{equation}
The parity symmetry of the polypolar spherical harmonics is given by
\begin{equation}
  X^{l_1l_2\cdots l_n}_{L_2L_3\cdots L_{n-1};lm}(-\bm{n}_1,-\bm{n}_2,\ldots,-\bm{n}_n)
  =(-1)^{l_1+l_2+\cdots+l_n}
  X^{l_1l_2\cdots l_n}_{L_2L_3\cdots L_{n-1};lm}(\bm{n}_1,\bm{n}_2,\ldots,\bm{n}_n).
  \label{eq:416}
\end{equation}
The complex conjugate of the bipolar spherical harmonics is given by
\begin{equation}
  X^{l_1l_2\cdots l_n*}_{L_2L_3\cdots L_{n-1};lm}(\bm{n}_1,\bm{n}_2,\ldots,\bm{n}_n)
  =(-1)^{l+l_1+l_2+\cdots+l_n} g_{(l)}^{mm'}
  X^{l_1l_2\cdots l_n}_{L_2L_3\cdots L_{n-1};lm'}(\bm{n}_1,\bm{n}_2,\ldots,\bm{n}_n).
  \label{eq:417}
\end{equation}
The orthonormality relation for the polypolar spherical harmonics in
general is straightforwardly derived from the orthonormality of the
$3j$-symbols, and the result is given by
\begin{multline}
  \int \frac{d^2n_1}{4\pi} \cdots \frac{d^2n_n}{4\pi}\,
  X^{l_1l_2\cdots l_n}_{L_2L_3\cdots
    L_{n-1};lm}(\bm{n}_1,\bm{n}_2,\ldots,\bm{n}_n) 
  X^{l_1'l_2'\cdots l_n'}_{L_2'L_3'\cdots
    L_{n-1}';l'm'}(\bm{n}_1,\bm{n}_2,\ldots,\bm{n}_n) 
    \\
    = \frac{(-1)^{l+l_1+\cdots +l_n} \delta_{ll'} \delta_{l_1l_1'}
      \cdots \delta_{l_nl_n'} \delta_{L_2L_2'} \cdots \delta_{L_{n-1}L_{n-1}'}
    \delta^\triangle_{l\,l_1L_2} \delta^\triangle_{L_2l_2L_3} \cdots
    \delta^\triangle_{L_{n-2}l_{n-2}L_{n-1}}
    \delta^\triangle_{L_{n-1}l_{n-1}l_n}}{(2l+1)(2l_1+1)(2l_2+1)\cdots (2l_n+1)} 
     g^{(l)}_{mm'}.
  \label{eq:418}
\end{multline}
A product of polypolar harmonics with the same set of directions is
given by
\begin{multline}
  X^{l_1l_2\cdots l_n}_{L_2L_3\cdots
    L_{n-1};lm}(\bm{n}_1,\bm{n}_2,\ldots,\bm{n}_n) 
  X^{l_1'l_2'\cdots l_n'}_{L_2'L_3'\cdots
    L_{n-1}';l'm'}(\bm{n}_1,\bm{n}_2,\ldots,\bm{n}_n) 
  =
  \sqrt{(2L_2+1)\cdots(2L_{n-1}+1)(2L_2'+1)\cdots(2L_{n-1}'+1)}
  \\ \times
  \sum_{l''} (-1)^{l''}(2l''+1)
  \left(l\,l'\,l''\right)_{mm'}^{\phantom{mm'}m''}
  \sum_{l_1'',\ldots,l_n''} (-1)^{l_1''+\cdots+l_n''}
  (2l_1''+1)\cdots (2l_n''+1)
  \begin{pmatrix}
    l_1 & l_1' & l_1'' \\
    0 & 0 & 0
  \end{pmatrix}
  \cdots
  \begin{pmatrix}
    l_n & l_n' & l_n'' \\
    0 & 0 & 0
  \end{pmatrix}
  \\ \times
  \sum_{L_2'',\ldots,L_{n-1}''} \sqrt{(2L_2''+1)\cdots (2L_{n-1}''+1)}
  \begin{Bmatrix}
    l & l' & l'' \\
    l_1 & l_1' & l_1'' \\
    L_2 & L_2' & L_2''
  \end{Bmatrix}
  \begin{Bmatrix}
    L_2 & L_2' & L_2'' \\
    l_2 & l_2' & l_2'' \\
    L_3 & L_3' & L_3''
  \end{Bmatrix}
  \cdots
  \begin{Bmatrix}
    L_{n-2} & L_{n-2}' & L_{n-2}'' \\
    l_{n-2} & l_{n-2}' & l_{n-2}'' \\
    L_{n-1} & L_{n-1}' & L_{n-1}''
  \end{Bmatrix}
  \begin{Bmatrix}
    L_{n-1} & L_{n-1}' & L_{n-1}'' \\
    l_{n-1} & l_{n-1}' & l_{n-1}'' \\
    l_n & l_n' & l_n''
  \end{Bmatrix}
  \\ \times
  X^{l_1''l_2''\cdots l_n''}_{L_2''L_3''\cdots
    L_{n-1}'';l''m''}(\bm{n}_1,\bm{n}_2,\ldots,\bm{n}_n).
  \label{eq:419}
\end{multline}
The last result can be derived by applying the sum rules of
Eqs.~(\ref{eq:444}) and (\ref{eq:445}) in
Appendix~\ref{app:3njSymbols}.

\section{\label{app:3njSymbols}
The $\bm{3j}$-, $\bm{6j}$- and $\bm{9j}$-symbols}

Wigner's $3j$-, $6j$- and $9j$-symbols are the coefficients in adding
and recoupling of angular momenta in quantum mechanics. These
coefficients are frequently used in the main text of this paper,
because they are useful as well in our formulation, regarding
rotational symmetry of the Universe in a statistical sense. We
summarize some of the formulas in this Appendix for the readers'
convenience. All the formulas below can be found in, or derived from,
Refs.~\cite{Edmonds:1955fi,Khersonskii:1988krb}. In the following, we
assume $l, l_1,l_2,\ldots$ are non-negative integers (excluding the
possibility of half-integers) as well as azimuthal indices, and we
assume $(-1)^l = (-1)^{-l}$, $(-1)^m = (-1)^{-m}$.

In this paper, we employ a simplified notation for the $3j$-symbol,
\begin{equation}
  \left(l_1\,l_2\,l_3\right)_{m_1m_2m_3} \equiv
    \begin{pmatrix}
      l_1 & l_2 & l_3 \\
      m_1 & m_2 & m_3
    \end{pmatrix},
  \label{eq:420}
\end{equation}
where the $3j$-symbol is nonzero only when $m_1+m_2+m_3=0$ in the
above, and azimuthal indices $m$'s are raised or lowered by spherical
metric tensors,
$g^{(l)}_{mm'} = (-1)^m\delta_{m,-m'} = g_{(l)}^{mm'}$, as exemplified
in Eq.~(\ref{eq:401}). In particular, the symmetries of the
$3j$-symbol indicates
\begin{equation}
  \left(l_1\,l_2\,l_3\right)^{m_1m_2m_3} =
    \begin{pmatrix}
      l_1 & l_2 & l_3 \\
      -m_1 & -m_2 & -m_3
    \end{pmatrix} =
    (-1)^{l_1+l_2+l_3}
    \begin{pmatrix}
      l_1 & l_2 & l_3 \\
      m_1 & m_2 & m_3
    \end{pmatrix} =
  (-1)^{l_1+l_2+l_3}
    \left(l_1\,l_2\,l_3\right)_{m_1m_2m_3},
  \label{eq:421}
\end{equation}
since $m_1+m_2+m_3=0$. Even permutations of the $3j$-symbols have the
same value,
\begin{equation}
  \left(l_1\,l_2\,l_3\right)_{m_1m_2m_3} =
  \left(l_2\,l_3\,l_1\right)_{m_2m_3m_1} =
  \left(l_3\,l_1\,l_2\right)_{m_3m_1m_2},
  \label{eq:422}
\end{equation}
while odd permutations are given by
\begin{equation}
  \left(l_2\,l_1\,l_3\right)_{m_2m_1m_3} =
  \left(l_1\,l_3\,l_2\right)_{m_1m_3m_2}
  =
  \left(l_3\,l_2\,l_1\right)_{m_3m_2m_1} =
  (-1)^{l_1+l_2+l_3} \left(l_1\,l_2\,l_3\right)_{m_1m_2m_3}.
  \label{eq:423}
\end{equation}
The above equalities hold even when some of the azimuthal indices are
raised on both sides of the equations. 

The orthogonality relations of the $3j$-symbol are represented by
\begin{align}
  &
    \left(l\,l_1\,l_2\right)_{mm_1m_2}
    \left(l'\,l_1\,l_2\right)_{m'}^{\phantom{m'}m_1m_2}
    =
    \frac{(-1)^{l+l_1+l_2} \delta_{ll'}}{2l+1}
    g^{(l)}_{mm'} \delta^\triangle_{l\,l_1l_2},
    \label{eq:424}\\
  &
    \sum_l (-1)^l (2l+1)
    \left(l\,l_1\,l_2\right)_{mm_1m_2}
    \left(l\,l_1\,l_2\right)^{m}_{\phantom{m}m_1'm_2'}
    = (-1)^{l_1+l_2} g^{(l_1)}_{m_1m_1'} g^{(l_2)}_{m_2m_2'},
    \label{eq:425}
\end{align}
where $\delta^\triangle_{l_1l_2l_3}=1$ only when $(l_1,l_2,l_3)$
satisfies the triangle inequality, and is zero otherwise. From
Eq.~(\ref{eq:424}), we have a derived formula,
\begin{equation}
    \left(l_1\,l_2\,l_3\right)_{m_1m_2m_3}
    \left(l_1\,l_2\,l_3\right)^{m_1m_2m_3}
    =
    (-1)^{l_1+l_2+l_3}
    \delta^\triangle_{l_1l_2l_3}.
  \label{eq:426}
\end{equation}
When one or two of the $l$'s in the $3j$-symbols are zero, we have
\begin{equation}
  \left(l\,0\,0\right)_{m00}
  = \delta_{l0} \delta_{m0}, \qquad
  \left(l\,l'\,0\right)_{mm'0}
  = \frac{(-1)^{l}\delta_{ll'}}{\sqrt{2l+1}} g^{(l)}_{mm'}.
  \label{eq:427}
\end{equation}

In this paper, the Gaunt integral with our notation is denoted by
\begin{equation}
  \left[l_1\,l_2\,l_3\right]_{m_1m_2m_3}
  \equiv
  \int \frac{d^2n}{4\pi}\,
  C_{l_1m_1}(\hat{\bm{n}})
  C_{l_2m_2}(\hat{\bm{n}})
  C_{l_3m_3}(\hat{\bm{n}})
  =
  \begin{pmatrix}
    l_1 & l_2 & l_3 \\
    0 & 0 & 0
  \end{pmatrix}
  \begin{pmatrix}
    l_1 & l_2 & l_3 \\
    m_1 & m_2 & m_3
  \end{pmatrix},
  \label{eq:428}
\end{equation}
and azimuthal indices are raised and lowered by the spherical metric
$g_{(l)}^{mm'}$, $g^{(l)}_{mm'}$ as in the case of $3j$-symbol.
A $3j$-symbol with vanishing $m$'s are nonzero only when the sum of
$l$'s are even, i.e.,
\begin{equation}
  \begin{pmatrix}
    l_1 & l_2 & l_3 \\
    0 & 0 & 0
  \end{pmatrix}
  = 0, \qquad \mathrm{when}\ \ l_1+l_2+l_3=\mathrm{odd}.
  \label{eq:429}
\end{equation}
Correspondingly, the Gaunt integrals are nonzero only when
$l_1+l_2+l_3=\mathrm{even}$ in the above equation. The covariant
version of the Gaunt integral has the same components as the
contravariant one:
\begin{equation}
  \left[l_1\,l_2\,l_3\right]_{m_1m_2m_3} =
  \left[l_1\,l_2\,l_3\right]^{m_1m_2m_3},
  \label{eq:430}
\end{equation}
because of a symmetry of the $3j$-symbols, Eq.~(\ref{eq:421}). 
The Gaunt integrals are totally symmetric under the permutation,
\begin{equation}
  \left[l_1\,l_2\,l_3\right]_{m_1m_2m_3} =
  \left[l_2\,l_3\,l_1\right]_{m_2m_3m_1} =
  \left[l_3\,l_1\,l_2\right]_{m_3m_1m_2}
  =
  \left[l_2\,l_1\,l_3\right]_{m_2m_1m_3} =
  \left[l_1\,l_3\,l_2\right]_{m_1m_3m_2} =
  \left[l_3\,l_2\,l_1\right]_{m_3m_2m_1}.
  \label{eq:431}
\end{equation}
An orthogonality relation of the $3j$-symbols, Eq.~(\ref{eq:424}),
indicates the orthogonality relation of the Gaunt integral:
\begin{equation}
  \left[l\,l_1\,l_2\right]_{mm_1m_2}
  \left[l'\,l_1\,l_2\right]_{m'}^{\phantom{m'}m_1m_2}
  = \frac{1}{2l+1}
  \begin{pmatrix}
    l & l_1 & l_2 \\
    0 & 0 & 0
  \end{pmatrix}^2
  \delta_{ll'}
  g^{(l)}_{mm'}.
  \label{eq:432}
\end{equation}
When one or two of the $l$'s in the Gaunt integral are zero, we have
\begin{equation}
  \left[l\,0\,0\right]_{m00}
  = \delta_{l0} \delta_{m0}, \qquad
  \left[l\,l'\,0\right]_{mm'0}
  = \frac{\delta_{ll'}}{2l+1} g^{(l)}_{mm'}.
  \label{eq:433}
\end{equation}

The Wigner's $6j$-symbol is associated with the recoupling of three
angular momenta, and defined by
\begin{equation}
  \begin{Bmatrix}
    l_1 & l_2 & l_3 \\
    l_4 & l_5 & l_6
  \end{Bmatrix}
  =
  (-1)^{l_1 + l_2 + l_3 + l_4 + l_5 + l_6}
  \left(l_1\,l_2\,l_3\right)^{m_1m_2m_3}
  \left(l_1\,l_5\,l_6\right)_{m_1}^{\phantom{m_1}m_5m_6}
  \left(l_4\,l_2\,l_6\right)^{m_4}_{\phantom{m_4}m_2m_6}
  \left(l_4\,l_5\,l_3\right)_{m_4m_5m_3}.
  \label{eq:434}
\end{equation}
or equivalently, 
\begin{equation}
  \begin{Bmatrix}
    l_1 & l_2 & l_3 \\
    l_4 & l_5 & l_6
  \end{Bmatrix}
  \left(l_1\,l_2\,l_3\right)_{m_1m_2m_3}
  = (-1)^{l_1 + l_2 + l_3}
  \left(l_1\,l_5\,l_6\right)_{m_1m_5m_6}
  \left(l_4\,l_2\,l_6\right)_{m_4m_2}^{\phantom{m_4m_2}m_6}
  \left(l_4\,l_5\,l_3\right)^{m_4m_5}_{\phantom{m_4m_5}m_3}.
  \label{eq:435}
\end{equation}
The $6j$-symbols are invariant against any permutation of the columns:
\begin{equation}
  \begin{Bmatrix}
    l_1 & l_2 & l_3 \\
    l_4 & l_5 & l_6
  \end{Bmatrix}
  =
  \begin{Bmatrix}
    l_2 & l_3 & l_1 \\
    l_5 & l_6 & l_4
  \end{Bmatrix}
  =
  \begin{Bmatrix}
    l_3 & l_1 & l_2 \\
    l_6 & l_4 & l_5
  \end{Bmatrix}
  =
  \begin{Bmatrix}
    l_2 & l_1 & l_3 \\
    l_5 & l_4 & l_6
  \end{Bmatrix}
  =
  \begin{Bmatrix}
    l_1 & l_3 & l_2 \\
    l_4 & l_6 & l_5
  \end{Bmatrix}
  =
  \begin{Bmatrix}
    l_3 & l_2 & l_1 \\
    l_6 & l_5 & l_4
  \end{Bmatrix},
  \label{eq:436}
\end{equation}
and also invariant against the interchange of the upper and lower
arguments in each of any two columns,
\begin{equation}
  \begin{Bmatrix}
    l_1 & l_2 & l_3 \\
    l_4 & l_5 & l_6
  \end{Bmatrix}
  =
  \begin{Bmatrix}
    l_1 & l_5 & l_6 \\
    l_4 & l_2 & l_3
  \end{Bmatrix}
  =
  \begin{Bmatrix}
    l_4 & l_2 & l_6 \\
    l_1 & l_5 & l_3
  \end{Bmatrix}
  =
  \begin{Bmatrix}
    l_4 & l_5 & l_3 \\
    l_1 & l_2 & l_6
  \end{Bmatrix}.
  \label{eq:437}
\end{equation}

The Wigner's $9j$-symbol is associated with the recoupling of four
angular momenta, and defined by
\begin{equation}
  \begin{Bmatrix}
    l_1 & l_2 & l_3 \\
    l_4 & l_5 & l_6 \\
    l_7 & l_8 & l_9
  \end{Bmatrix}
  =
  \left(l_1\,l_2\,l_3\right)^{m_1m_2m_3}
  \left(l_4\,l_5\,l_6\right)^{m_4m_5m_6}
  \left(l_7\,l_8\,l_9\right)^{m_7m_8m_9}
  \left(l_1\,l_4\,l_7\right)_{m_1m_4m_7}
  \left(l_2\,l_5\,l_8\right)_{m_2m_5m_8}
  \left(l_3\,l_6\,l_9\right)_{m_3m_6m_9}.
  \label{eq:438}
\end{equation}
The $9j$-symbols are invariant under transposition,
\begin{equation}
  \begin{Bmatrix}
    l_1 & l_2 & l_3 \\
    l_4 & l_5 & l_6 \\
    l_7 & l_8 & l_9
  \end{Bmatrix} =
  \begin{Bmatrix}
    l_1 & l_4 & l_7 \\
    l_2 & l_5 & l_8 \\
    l_3 & l_6 & l_9
  \end{Bmatrix}.
  \label{eq:439}
\end{equation}
and also invariant against cyclic (even) permutations of the columns
or lows:
\begin{equation}
  \begin{Bmatrix}
    l_1 & l_2 & l_3 \\
    l_4 & l_5 & l_6 \\
    l_7 & l_8 & l_9
  \end{Bmatrix} =
  \begin{Bmatrix}
    l_2 & l_3 & l_1 \\
    l_5 & l_6 & l_4 \\
    l_8 & l_9 & l_7
  \end{Bmatrix} =
  \begin{Bmatrix}
    l_3 & l_1 & l_2 \\
    l_6 & l_4 & l_5 \\
    l_9 & l_7 & l_8
  \end{Bmatrix} =
  \begin{Bmatrix}
    l_7 & l_8 & l_9 \\
    l_1 & l_2 & l_3 \\
    l_4 & l_5 & l_6
  \end{Bmatrix} =
  \begin{Bmatrix}
    l_4 & l_5 & l_6 \\
    l_7 & l_8 & l_9 \\
    l_1 & l_2 & l_3
  \end{Bmatrix}.
  \label{eq:440}
\end{equation}
They change the sign by noncyclic (odd) permutations of the columns
or lows, if the sum of all the arguments is odd:
\begin{equation}
  (-1)^R
  \begin{Bmatrix}
    l_1 & l_2 & l_3 \\
    l_4 & l_5 & l_6 \\
    l_7 & l_8 & l_9
  \end{Bmatrix} =
  \begin{Bmatrix}
    l_2 & l_1 & l_3 \\
    l_5 & l_4 & l_6 \\
    l_8 & l_7 & l_9
  \end{Bmatrix} =
  \begin{Bmatrix}
    l_3 & l_2 & l_1 \\
    l_6 & l_5 & l_4 \\
    l_9 & l_8 & l_7
  \end{Bmatrix} =
  \begin{Bmatrix}
    l_1 & l_3 & l_2 \\
    l_4 & l_6 & l_5 \\
    l_7 & l_9 & l_8
  \end{Bmatrix} =
  \begin{Bmatrix}
    l_4 & l_5 & l_6 \\
    l_1 & l_2 & l_3 \\
    l_7 & l_8 & l_9
  \end{Bmatrix} =
  \begin{Bmatrix}
    l_7 & l_8 & l_9 \\
    l_4 & l_5 & l_6 \\
    l_1 & l_2 & l_3
  \end{Bmatrix} =
  \begin{Bmatrix}
    l_1 & l_2 & l_3 \\
    l_7 & l_8 & l_9 \\
    l_4 & l_5 & l_6
  \end{Bmatrix},
  \label{eq:441}
\end{equation}
where
\begin{equation}
  R = l_1 + l_2 +  l_3 + l_4 +  l_5 + l_6 +  l_7 + l_8 +  l_9. 
  \label{eq:442}
\end{equation}

Many formulas among $3j$-, $6j$-, and $9j$-symbols, including
orthogonality relations and sum rules, can be found in the standard
literature mentioned above. In particular, we use the following
sum rules in this paper for products of $3j$-symbols:
\begin{equation}
  \left(l_1\,l_2\,L\right)_{m_1m_2}^{\phantom{m_1m_2}M}
  \left(L\,l_3\,l_4\right)_{Mm_3m_4}
  = (-1)^{l_2+l_3} \sum_{L'} (2L'+1)
  \left(l_1\,l_3\,L'\right)_{m_1m_3}^{\phantom{m_1m_3}M'}
  \left(L'\,l_2\,l_4\right)_{M'm_2m_4}
  \begin{Bmatrix}
    l_1 & l_2 & L \\
    l_4 & l_3 & L'
  \end{Bmatrix},
  \label{eq:443}
\end{equation}
\begin{equation}
  \left(L\,l_1\,l_2\right)_{M}^{\phantom{M}m_1m_2}
  \left(L'\,l_1'\,l_2'\right)_{M'}^{\phantom{M'}m_1'm_2'}
  \left(l_2'\,l_2\,l_2''\right)_{m_2'm_2}^{\phantom{m_2'm_2}m_2''}
  = \sum_{l_1'',L''} (2l_1''+1)(2L''+1)
  \left(L\,L'\,L''\right)_{MM'}^{\phantom{MM'}M''}
  \left(L''\,l_2''\,l_1''\right)_{M''}^{\phantom{m''}m_2''m_1''}
  \left(l_1\,l_1'\,l_1''\right)^{m_1m_1'}_{\phantom{m_1m_1'}m_1''}
  \begin{Bmatrix}
    l_1 & l_1' & l_1'' \\
    l_2 & l_2' & l_2'' \\
    L & L' & L''
  \end{Bmatrix},
  \label{eq:444}
\end{equation}
and
\begin{equation}
  \left(L\,l_1\,l_2\right)_{M}^{\phantom{M}m_1m_2}
  \left(L'\,l_1'\,l_2'\right)_{M'}^{\phantom{M'}m_1'm_2'}
  \left(l_1'\,l_1\,l_1''\right)_{m_1'm_1}^{\phantom{m_1'm_1}m_1''}
  \left(l_2'\,l_2\,l_2''\right)_{m_2'm_2}^{\phantom{m_2'm_2}m_2''}
  = \sum_{L''} (2L''+1)
  \left(L\,L'\,L''\right)_{MM'}^{\phantom{MM'}M''}
  \left(L''\,l_2''\,l_1''\right)_{M''}^{\phantom{m''}m_2''m_1''}
  \begin{Bmatrix}
    l_1 & l_1' & l_1'' \\
    l_2 & l_2' & l_2'' \\
    L & L' & L''
  \end{Bmatrix}.
  \label{eq:445}
\end{equation}


\twocolumngrid
\renewcommand{\apj}{Astrophys.~J. }
\newcommand{\aap}{Astron.~Astrophys. }
\newcommand{\aj}{Astron.~J. }
\newcommand{\apjl}{Astrophys.~J.~Lett. }
\newcommand{\apjs}{Astrophys.~J.~Suppl.~Ser. }
\newcommand{\apss}{Astrophys.~Space Sci. }
\newcommand{\cqg}{Class.~Quant.~Grav. }
\newcommand{\jcap}{J.~Cosmol.~Astropart.~Phys. }
\newcommand{\mnras}{Mon.~Not.~R.~Astron.~Soc. }
\newcommand{\mpla}{Mod.~Phys.~Lett.~A }
\newcommand{\pasj}{Publ.~Astron.~Soc.~Japan }
\newcommand{\physrep}{Phys.~Rep. }
\newcommand{\ptp}{Progr.~Theor.~Phys. }
\newcommand{\ptep}{Prog.~Theor.~Exp.~Phys. }
\newcommand{\jetp}{JETP }
\newcommand{\jhep}{Journal of High Energy Physics}


\end{document}